\documentclass[11pt,a4paper]{article}

\usepackage{amsmath,amsfonts,amssymb,mathrsfs,mathabx}
\usepackage[dvips]{graphicx}
\usepackage{tikz}
\usepackage{comment}

\usepackage{booktabs}
\usepackage[footnotesize,bf]{caption}
\usepackage{hyperref}

\hypersetup{
    unicode=true,          % non-Latin characters in Acrobat's bookmarks
    pdftoolbar=false,        % show Acrobat's toolbar?
    pdfmenubar=false,        % show Acrobat's menu?
    pdffitwindow=false,     % window fit to page when opened
    pdfstartview={FitH},    % fits the width of the page to the window
    pdftitle={Classical irregular block, N=2 pure gauge theory},    % title
    pdfauthor={M. Piatek, A. R. Pietrykowski},     % author
    pdfsubject={Integrable systems},   % subject of the document
    pdfcreator={M. Piatek, A. R. Pietrykowski},   % creator of the document
    pdfproducer={M. Piatek, A. R. Pietrykowski}, % producer of the document
    pdfkeywords={Integrable systems} {Supersymetry}
                {Seiberg-Witten theory} {AGT correspondence} {Mathieu equation}, % list of
    pdfnewwindow=true,      % links in new window
    colorlinks=true,       % false: boxed links; true: colored links
    linkcolor=black,          % color of internal links
    citecolor=blue,        % color of links to bibliography
    filecolor=blue,      % color of file links
    urlcolor=blue           % color of external links
}

\def \a {\alpha}
\def \b {\beta}

\def \d {\delta}
\def \e {\epsilon}

\def \vf{\varphi}
\def \g {\gamma}

\def \i {\iota}

\def \w {\omega}

\def \r {\rho}

\def \z {\zeta}

\def \n{\nu}

\def \La{\Lambda}
\def \pt{\partial}

\newcommand \ord[1]{\mathcal{O}({#1})}
\newcommand \cbr[1]{\left({#1}\right)}
\newcommand \sbr[1]{\left[{#1}\right]}
\newcommand \pbr[1]{\left\{{#1}\right\}}

\newlength{\intwidth}

\DeclareRobustCommand{\vpiint}
   {\mathop{%
      \text{%
        \settowidth{\intwidth}{$\iint$}%
        \makebox[0pt][l]{\makebox[\intwidth]{$-\!-$}}%
        $\iint$}}}

\newcommand\bigforall{\mbox{\huge $\mathsurround0pt\forall$}}

\numberwithin{equation}{section}

\newcommand{\preprintsize}{
      \headheight=5pt                              % header space
     \topmargin= 0 cm \headsep=0.1cm
     \oddsidemargin= 0cm
      \evensidemargin= 0cm  % adjust left
      \textheight = 23truecm \textwidth=16truecm      %
}

\preprintsize

\begin{document}

\begin{center}
{\sf\LARGE Classical irregular block, ${\cal N}=2$ pure gauge theory
\\[8pt]
and Mathieu equation}
\end{center}

\begin{center}
{\large {\sf Marcin Piatek}$^{\,a,\,c,\;}$}\footnote{\href{mailto:piatek@fermi.fiz.univ.szczecin.pl}{e-mail: piatek@fermi.fiz.univ.szczecin.pl}}
\hskip 1.0cm
{\large {\sf Artur R. Pietrykowski}$^{\,b,\,c,\;}$}\footnote{\href{mailto:pietrie@theor.jinr.ru}{e-mail: pietrie@theor.jinr.ru}}

\vskip 4mm
${}^{a}$
Institute of Physics, University of Szczecin\\
ul. Wielkopolska 15, 70-451 Szczecin, Poland

\vskip 4mm
${}^{b}$
Institute of Theoretical Physics\\
University of Wroc{\l}aw\\
pl. M. Borna, 950-204 Wroc{\l}aw, Poland

\vskip 4mm
${}^{c}$
Bogoliubov Laboratory of Theoretical Physics,\\
Joint Institute for Nuclear Research, 141980 Dubna, Russia
\end{center}

\vskip .5cm

\noindent
\begin{abstract}\noindent
Combining the semiclassical/Nekrasov--Shatashvili limit of the AGT conjecture 
and the Bethe/gauge correspondence results in a triple correspondence 
which identifies classical conformal blocks with twisted superpotentials and
then with Yang--Yang functions. In this paper the triple correspondence is studied
in the simplest, yet not completely understood case of pure $SU(2)$ super-Yang--Mills gauge theory. 
A missing element of that correspondence is identified with the classical irregular block.
Explicit tests provide a convincing evidence that such a function exists.
In particular, it has been shown that the classical irregular block can be
recovered from classical blocks on the torus and sphere in suitably defined decoupling
limits of classical external conformal weights. These limits are ``classical analogues'' of known 
decoupling limits for corresponding quantum blocks.
An exact correspondence between the classical irregular block
and the $SU(2)$ gauge theory twisted superpotential has been obtained as 
a result of another consistency check. The latter 
determines the spectrum of the 2-particle periodic Toda (sin-Gordon) Hamiltonian
in accord with the Bethe/gauge correspondence.
An analogue of this statement is found entirely within $2d$ CFT. 
Namely, considering the classical limit of the null vector decoupling equation for the 
degenerate irregular block a celebrated Mathieu's equation is obtained with an eigenvalue
determined by the classical irregular block. As it has been checked this result 
reproduces a well known weak coupling expansion of Mathieu's eigenvalue. 
Finally, yet another new formulae for Mathieu's eigenvalue relating 
the latter to a solution of certain Bethe-like equation are found.
\end{abstract}

\newpage
{\small \hrule \tableofcontents \vskip .5cm\hrule}

\section{Introduction}
Studying the problem of the oscillations of an elliptical membrane
E.~Mathieu \cite{Mathieu:1868} obtained the following two ordinary
differential equations with real coefficients \cite{Whittaker:1996,McLachlan:1947}:\footnote{
In the present paper we adopt the notation from \cite{MuellerKirsten:2006}.}
\begin{equation}\label{Mathieu1}
\frac{{\rm d}^2 \psi}{{\rm d}x^2}+\left(\lambda - 2h^2 \cos 2x \right)\psi \;=\; 0,
\;\;\;\;\;\;\;\;\;\;\;\;\;\;x\in\mathbb{R}
\end{equation}
and
\begin{equation}
\label{modifiedM}
\frac{{\rm d}^2 \psi}{{\rm d}x^2}-\left(\lambda - 2h^2 \cosh 2x \right)\psi \;=\; 0
\;\;\;\;\;{\rm i.e.}\;\;\;\;\;
\frac{{\rm d}^2 \psi}{{\rm d}(ix)^2}+\left(\lambda - 2h^2 \cos 2ix \right)\psi\;=\;0.
\end{equation}
In honor of their originator the eqs.~(\ref{Mathieu1}),
(\ref{modifiedM}) are now called Mathieu and modified Mathieu equations respectively.
As has been explicitly written down the modified Mathieu eq.~(\ref{modifiedM}) may be derived
from eq.~(\ref{Mathieu1}) by writing $ix$ for $x$ and vice versa.

For certain $\lambda$ and $h^2$ there exists a general solution $\psi(x)$ of the
eq.~(\ref{Mathieu1}) and a {\it Floquet exponent} $\nu$ such that
$$
\psi(x+\pi)\;=\;{\rm e}^{i\pi\nu}\psi(x).
$$
If $\psi_{+}(x)$ is the solution of the Mathieu equation satisfying the initial
conditions $\psi_{+}(0)={\rm const.}=a$ and $\psi_{+}^{\prime}(0)=0$, the parameter $\nu$
can be determined from the relation \cite{MuellerKirsten:2006}:
\begin{equation}\label{nu}
\cos\pi\nu \;=\;\frac{\psi_{+}(\pi)}{a}.
\end{equation}
Thus the Floquet exponent is determined by the value at $x=\pi$ of the solution
$\psi_{+}(x)$  which is even around $x=0$.
To the lowest order ($h^2=0$) the even solution around $x=0$ is
$\psi_{+}^{(0)}(x)=a\cos\sqrt{\lambda}x$.
Hence, from (\ref{nu}) for $h^2=0$ we have $\nu=\sqrt{\lambda}$, and more in general
$$
\nu^2 \;=\; \lambda + {\cal O}\left(h^2\right).
$$
One can derive various terms of this expansion perturbatively. For instance, in the weak
coupling regime, for small $h^2$, the eigenvalue $\lambda$ as a function of
$\nu$ and $h^2$ explicitly reads as follows \cite{MuellerKirsten:2006}:
\begin{eqnarray}\label{lambda}
\lambda &=& \nu^2 +
\frac{h^4}{2 \left(\nu ^2-1\right)}+
\frac{\left(5 \nu ^2+7\right)h^8}{32 \left(\nu ^2-4\right) \left(\nu ^2-1\right)^3}\nonumber
\\[8pt]
&+&
\frac{\left(9 \nu ^4+58 \nu ^2+29\right)h^{12}}{64 \left(\nu ^2-9\right) \left(\nu ^2-4\right) \left(\nu^2-1\right)^5}
+\mathcal{O}\left(h^{16}\right).
\end{eqnarray}
One sees that this expansion cannot hold for integer values of $\nu$. These cases
have to be dealt with separately, cf.~\cite{MuellerKirsten:2006}.

The solutions of the (modified) Mathieu equation govern a vast number of problems
in physics: (i)~the propagation of electromagnetic waves along elliptical cylinder,
(ii)~the vibrations of a membrane in the shape of an ellipse with a rigid boundary
(Mathieu's problem) \cite{Whittaker:1996}, (iii)~the motion of a rod fixed on one end
and being under periodic tension at the other end \cite{McLachlan:1947},
(iv)~the motion of particles in electromagnetic traps \cite{Paul:1990zz}, (v)~the inverted
pendulum and the quantum pendulum, (vi)~the wave scattering by D-brane \cite{Gubser:1998iu},
(vii)~fluctuations of scalar field about a D3-brane \cite{Manvelyan:2000yv,Park:2000iy},
(viii)~reheating process in inflationary models \cite{Lachapelle}, (ix)~determination
of the mass spectrum of a scalar field in a world with latticized
and circular continuum space \cite{Cho:2003rp}, are just some of them.

Eq.~(\ref{Mathieu1}) can be looked at as a one-dimensional Schr\"{o}dinger
equation ${\sf H}\psi=E\psi$ with the Hamiltonian (the Mathieu operator)
${\sf H}=-{\rm d}^2/{\rm d}x^2 + 2h^2\cos2x$ and the energy eigenvalue $E=\lambda$.
The Mathieu operator ${\sf H}$ belongs to the class of Schr\"{o}dinger operators with
periodic potentials \cite{ReedSimon} which are of special importance in solid state physics.

Mathieu's equation has recently emerged in a fascinating context, namely in the studies of the
interrelationships between quantum integrable systems (QIS), ${\cal N}=2$ super-Yang--Mills (SYM) theories and
two-dimensional conformal field theory ($2d$ CFT). In order to spell out
aims of the present work let us discuss certain aspects of these investigations in detail.
First, recall that eq.~(\ref{Mathieu1})
with $\lambda=8u\hbar^{-2}$, $h^2=4\hat\Lambda^2\hbar^{-2}$, $2x=\varphi$, i.e.:
\begin{equation}\label{qsG}
\left[-\frac{\hbar^2}{2}\frac{\partial^2}{\partial\varphi^2}+\hat\Lambda^2\cos\varphi\right]\psi(\varphi)
\;=\;u\,\psi(\varphi)
\end{equation}
is a Schr\"{o}dinger equation for the quantum one-dimensional sine-Gordon
system defined by the Lagrangian ${\cal L}=\frac{1}{2}\dot{\varphi}^2-\hat\Lambda^2\cos\varphi$.
In this context quantities $\hat\Lambda$, $\varphi$, $u$ can be complex valued.\footnote{In other words we are dealing
here with quantized integrable model defined on the complex plane which is not hermitian quantum-mechanical system.}

Secondly, it is a well known fact that the (classical) sine-Gordon model
encodes an information about the ${\cal N}=2$ $SU(2)$ pure
gauge Seiberg--Witten theory \cite{Gorsky:1995zq}. More precisely,
the so-called Bohr--Sommerfeld periods
$$
\Pi(\Gamma)\;=\;\oint\limits_{\Gamma}\sqrt{2(u-\hat\Lambda^2\cos\varphi)}\,{\rm d}\varphi
\;=:\;\oint\limits_{\Gamma}P_{0}(\varphi)\,{\rm d}\varphi
$$
for two complementary contours $\Gamma=A, B$ encircling the two turning points $\hat\Lambda^2\cos\pm\varphi_0$
define the Seiberg--Witten system \cite{Seiberg:1994rs}:
$a=\Pi(A)$,
$\partial {\cal F}(a)/\partial a=\Pi(B)$.\footnote{Cf.~appendix \ref{app_WKB}.}
Here, $a$ is a modulus and ${\cal F}(a)$ denotes the Seiberg--Witten prepotential determining
the low energy effective dynamics of the ${\cal N}=2$ supersymmetric $SU(2)$ pure
gauge theory.

As has been observed in \cite{Mironov:Morozov:2010} (see also \cite{He:2010zzc,Maruyoshi:2010})
the above statement has its ``quantum analogue''
or ``quantum generalization''  which can be formulated as follows. Namely,
the monodromies\footnote{Here,
$P(\varphi, \hbar)=P_{0}(\varphi)+P_{1}(\varphi)\hbar+P_{2}(\varphi)\hbar^2+\ldots$.}
$$
\widetilde{\Pi}(\Gamma)\;=\;\oint\limits_{\Gamma}P(\varphi, \hbar)\,{\rm d}\varphi
$$
of the exact WKB solution
$$
\psi(\varphi)\;=\;\exp\left\lbrace \frac{i}{\hbar}\int\limits^{\varphi}
P(\rho, \hbar)\,{\rm d}\rho\right\rbrace
$$
to the eq.~(\ref{qsG}) define the Nekrasov--Shatashvili system \cite{NS:2009}:
$a=\widetilde{\Pi}(A)$,
$\partial {\cal W}(\hat\Lambda, a,\hbar)/\partial a=\widetilde{\Pi}(B)$.\footnote{
For $SU(N)$ generalization of this result, see \cite{Mironov:2009dv}.}
Here, ${\cal W}(\hat\Lambda, a, \hbar)={\cal W}_{\rm pert}(\hat\Lambda, a, \hbar)
+{\cal W}_{\rm inst}(\hat\Lambda, a, \hbar)$
is the {\it effective twisted superpotential} of the
$2d$ $SU(2)$ pure gauge ($\Omega$-deformed) SYM theory defined in \cite{NS:2009}
as the following (Nekrasov--Shatashvili) limit
$$
{\cal W}(\hat\Lambda, a, \hbar)\;\equiv\;
\lim_{\epsilon_2\to 0}\epsilon_2\log{\cal Z}(\hat\Lambda, a, \epsilon_{1}\!\!=\hbar,\epsilon_2)
$$
of the Nekrasov partition function
${\cal Z}(\hat\Lambda, a, \epsilon_1, \epsilon_2)={\cal Z}_{\rm pert}(\hat\Lambda, a, \epsilon_1, \epsilon_2)
{\cal Z}_{\rm inst}(\hat\Lambda, a, \epsilon_1, \epsilon_2)$ \cite{Nekrasov:Okounkov:2003,Nekrasov:2002qd}.

Twisted superpotentials determine the low energy effective dynamics of the two-dimensional SYM theories
restricted to the $\Omega$-background. These quantities play
also a pivotal role in the so-called Bethe/gauge correspondence
\cite{NS:2009,Nekrasov:2009uh,Nekrasov:2009ui,Nekrasov:2013xda}
that maps supersymmetric vacua of the ${\cal N} = 2$ theories to Bethe states of quantum integrable systems.
A result of that duality is that the twisted superpotentials are identified
with the {\it Yang--Yang functions} \cite{NS:2009}
which describe the spectrum of the corresponding quantum integrable systems.\footnote{The Yang--Yang functions
are potentials for Bethe equations.}

Twisted superpotentials occur also in the context related to the AGT correspondence \cite{Alday:2009aq}.
The AGT conjecture states that the Liouville field theory (LFT) correlators on the
Riemann surface $C_{g,n}$ with genus $g$ and $n$ punctures can be identified with the
partition functions of a class $T_{g,n}$ of four-dimensional ${\cal N}=2$ supersymmetric $SU(2)$ quiver gauge
theories. A significant part of the AGT conjecture is an exact correspondence between the Virasoro
blocks on $C_{g,n}$ and the instanton sectors of the Nekrasov partition functions of the gauge theories
$T_{g,n}$. Soon after its discovery, the AGT hypothesis has been extended to the $SU(N)$-gauge
theories/conformal Toda correspondence \cite{Wyllard:2009hg}. The AGT duality works at the level of the quantum Liouville
field theory. At this point arises a question, what happens if we proceed to the classical limit
of the Liouville theory. This is the limit in which the central charge and the external and intermediate
conformal weights of LFT correlators tend to infinity in such a way that their ratios are fixed. It is
commonly believed that such limit exists and the Liouville correlation functions, and in particular, conformal
blocks behave in this limit exponentially.
It turns out that the semiclassical limit of the LFT correlation functions corresponds to the
Nekrasov--Shatashvili limit ($\epsilon_2\to 0$, $\epsilon_1\!=\!{\rm const.}$)
of the Nekrasov partition functions. A consequence of that correspondence is
that the instanton parts of the effective twisted superpotentials can be identified with {\it classical conformal blocks}.

Hence, joining together classical/Nekrasov--Shatashvili limit of the AGT duality
and the Bethe/gauge correspondence one thus gets the triple correspondence
which links the classical blocks to the twisted superpotentials and
then to the Yang--Yang functions. The simplest, although not yet completely understood
examples of that correspondence are listed below:
\medskip
\begin{center}
\begin{tabular}{|r|l|c|}
\hline\hline
\;{\sf classical}\;{\sf block}\; & \;{\sf twisted}\;{\sf superpotential}\; & \;{\sf Yang--Yang}\;{\sf function}\;\\ \hline\hline
on\;4-punctured\;sphere &  $SU(2)\;{\rm N_f}=4$         & $SL(2)$-type\;Gaudin\;model              \\ \hline
on\;1-punctured\;torus  &  $SU(2)\;{\cal N}=2^*$        & 2-particle\;elliptic\;Calogero--Moser\;model \\ \hline
?\;\;\;\;\;\;\;\;\;\;\;\;\;\;\;\;&  $SU(2)$\;pure\;gauge                 & 2-particle\;periodic\;Toda\;chain       \\ \hline
\end{tabular}
\end{center}
\medskip

Our goals in this paper are twofold. First, we study the triple correspondence
in case where the SYM theory is the $SU(2)$ pure gauge theory (the third example above).\footnote{
For a discussion of the first example in the table above, see \cite{Teschner}.}
We identify the question mark in the table above as the {\it classical irregular block}
$f_{\rm irr}$. The latter is the classical limit of the quantum irregular
block ${\cal F}_{\rm irr}$ \cite{Gaiotto:2009}.

Motivations to study classical blocks were, for a long time, mainly confined
to applications in pure mathematics, in particular, to the celebrated
uniformization problem of Riemann surfaces \cite{Hempel,KRV}
which is closely related to the monodromy problem for certain ordinary differential equations
\cite{Zamolodchikov:1995aa,Hadasz:2006rb,Menotti:2014kra,Menotti:2013bka,Menotti:2012wq,Menotti:2011ws,Menotti:2010en,
Piatek:2013ifa,Ferrari:2012gc}.\footnote{In a somewhat different context, see also \cite{Bazhanov:2013cua}.}
The importance of the classical blocks is not only limited to
the uniformization theorem, but gives also information about the solution of the
Liouville equation on surfaces with punctures.
Recently, a mathematical application of classical blocks
emerged in the context of Painlev\'{e} VI equation \cite{Litvinov:2013sxa}.
Due to the recent discoveries the classical blocks are also relevant for physics.
Indeed, in addition to the correspondence discussed in the present paper
lately the classical blocks have been of use to studies of the entanglement entropy
within the ${\rm AdS}_3/2d\;{\rm CFT}$ holography \cite{Hartman:2013mia} and
the topological string theory \cite{KashaniPoor:2012wb}.

The second purpose of the present work is to discuss implications of the
aforementioned triple correspondence for the eigenvalue problem of Mathieu's operator.
More concretely, as has been observed in \cite{Mironov:Morozov:2010}
the eigenvalue $u$ in eq.~(\ref{qsG})
(or equivalently $\lambda$ in eq.~(\ref{Mathieu1}))
can be found from eq.~$a=\widetilde{\Pi}(A)$ as a series in $\hbar$ by applying the exact WKB method.
The result can be re-expressed as a logarithmic derivative of the twisted superpotential
${\cal W}(\hat\Lambda, a, \epsilon_1)$ w.r.t.~$\hat\Lambda$.\footnote{See appendix \ref{app_WKB}.}
The same ``should be visible'' on the
conformal field theory side. Indeed, Mathieu's equation occurs entirely within formalism of $2d$ CFT
as a classical limit of the null vector decoupling equation satisfied by the 3-point degenerate
irregular block ( = matrix element of certain primary degenerate chiral vertex operator between Gaiotto
states \cite{Gaiotto:2009}). As expected, the eigenvalue in this equation is given
by the logarithmic derivative of the classical irregular block $f_{\rm irr}$ w.r.t.~$\hat\Lambda$.
Concluding, these observations pave the way for working out new methods
for calculating  Mathieu's eigenvalues and eigenfunctions.
The second goal of this paper is to check this possibility.

Our studies of Mathieu's equation and the classical irregular block
are in particular motivated by recent results obtained by
one of the authors in \cite{Piatek:2013ifa}. There have been derived novel expressions for the so-called
accessory parameter ${\sf B}$ of the Lam\'{e} equation:\footnote{Here,
$\wp(z)$ is the Weierstrass elliptic function and $E_{2}(\tau)$ denotes
the second Eisenstein series.}
\begin{equation*}
\frac{{\rm d}^2\eta}{{\rm d}z^2}-\left[\kappa\,\wp(z)+{\sf B}\right]\eta\;=\;0.
\end{equation*}
In particular, it has been found that
\begin{equation}\label{B}
\frac{{\sf B}(\tau)}{4\pi^2}\;=\;q\frac{\rm d}{{\rm d}q}f_{\rm torus}(-\kappa, \delta_{*}; q)
+ \frac{\kappa}{12}\,E_{2}(\tau),
\end{equation}
where $q=\exp (2\pi i \tau)$ and $\tau$ is a torus modular parameter; $f_{\rm torus}(\,\cdot\,,\,\cdot\,; q)$
denotes the classical torus block.\footnote{In eq.~(\ref{B}) the classical torus block is evaluated
on the so-called saddle point intermediate classical weight $\delta_{*}=\frac{1}{4}+p^{2}_{*}$, where $p_{*}$
is a solution of the following equation ($p\in\mathbb{R}$):
$$
\frac{\partial}{\partial p}{\rm S}_{\rm L}^{(3)}\left(2p,-2i\sqrt{\kappa+\frac{1}{4}},2p\right)\;=\;
2\mathfrak{R}f_{\rm torus}\left(-\kappa,{\frac{1}{4}}+p^2; q\right)\Big|_{p=p_{*}}.
$$
${\rm S}_{\rm L}^{(3)}$ is known classical Liouville action on the Riemann sphere with three
hyperbolic singularities (holes), cf.~\cite{Piatek:2013ifa}.}
It is a well known fact that in a certain limit the Lam\'{e} equation becomes the Mathieu equation.
Hence, one may expect that the Lam\'{e} equation with the eigenvalue expressed in terms of
the classical torus block $f_{\rm torus}$ consistently reduces to the Mathieu equation with the eigenvalue
given by the classical irregular block $f_{\rm irr}$ if in such a limit
$f_{\rm torus}\to f_{\rm irr}.$
If this statement is true it will give a strong evidence that conjectured formula (\ref{B}) is correct.

The organization of the paper is as follows. In section \ref{sec2} we define the quantum
irregular block ${\cal F}_{\rm irr}$ related to the $SU(2)$ pure gauge Nakrasov instanton partition function,
in accordance with the so-called {\it non-conformal AGT relation} \cite{Gaiotto:2009}.
Inspired by the latter and the results of \cite{NS:2009} we then conjecture
that ${\cal F}_{\rm irr}$ exponentiates to the classical irregular block $f_{\rm irr}$ in the classical limit.
Indeed, for the low orders of expansion of
the quantum irregular block one can see that the classical limit of  ${\cal F}_{\rm irr}$
exists yielding consistent definition of the classical irregular block.
The latter corresponds to the twisted superpotential of the $2d$ ${\cal N}=2$ $SU(2)$ pure gauge theory.
In addition, we perform another consistency checks suggesting that the function $f_{\rm irr}$
really ``lives its own life''. In particular, we verify that classical blocks
on the 1-punctured torus $C_{1,1}$ and on the 4-punctured sphere $C_{0,4}$
reduce to $f_{\rm irr}$ in certain properly defined {\it decoupling limit}
of external classical weights. This limit is a classical analogue of known
decoupling limits for quantum blocks on $C_{1,1}$ and $C_{0,4}$.

In section \ref{sec3} (see also appendix~\ref{AppI}) 
we consider the classical limit of the null vector decoupling equation
satisfied by the 3-point degenerate irregular block \cite{Maruyoshi:2010} and find an expression
for the Mathieu eigenvalue. As has been already mentioned the latter is determined
by the classical irregular block. This formula yields the well known weak coupling
expansion (\ref{lambda}) of the eigenvalue of the Mathieu operator.

Section 4 is devoted to the derivation of the Mathieu eigenvalue from
the non-conformal AGT counterpart of the classical irregular block,
namely the twisted superpotential. The latter is obtained from the Nekrasov instanton partition
function for pure $SU(2)$ gauge theory as a zero limit in
one of the two deformation parameters $\epsilon_1$, $\epsilon_2$.
Since the Nekrasov partition function can be represented
by the sum over profiles of the Young diagrams \cite{Nekrasov:Okounkov:2003}
(see appendix \ref{app_instNekrasov} for details) the two deformation
parameters are associated with two edges of elementary box of anisotropic
Young diagrams. As a result the superpotential is obtained from the critical
Young diagram which is determined by a dominating contribution to the
partition function. The Mathieu eigenvalue can be thus found by means of the
Bethe/gauge correspondence postulated in ref. \cite{NS:2009}.

In section \ref{sec5} we present our conclusions. The problems that are still
open and the possible extensions of the present work are discussed.

In the appendix \ref{App0} are collected formulae for expansion coefficients 
of $2d$ CFT and gauge theory functions used in the main text.
In the appendix \ref{app_WKB} the Mathieu eigenvalue is obtained form the
exact WKB method. Appendices \ref{app_instNekrasov} and \ref{app_ids} contain supplementary
material to the section \ref{sec4}. Specifically, the Nekrasov partition function
is given in terms of the profiles of the Young diagrams.

\section{Quantum and classical irregular blocks}
\label{sec2}
\subsection{Quantum irregular block}
In order to define the quantum irregular block first we will need to introduce
the notion of the Gaiotto state. This is the vector $|\,\Delta,\Lambda^2\,\rangle$ defined by the
following conditions \cite{Gaiotto:2009,Maruyoshi:2010}:
\begin{eqnarray}\label{Gstate1}
L_{0}|\,\Delta,\Lambda^2\,\rangle &=&
\left(\Delta+\frac{\Lambda}{2}\frac{\partial}{\partial\Lambda}\right)|\,\Delta,\Lambda^2\,\rangle,
\\\label{Gstate2}
L_{1}|\,\Delta,\Lambda^2\,\rangle &=& \Lambda^2 |\,\Delta,\Lambda^2\,\rangle,
\\\label{Gstate3}
L_{n}|\,\Delta,\Lambda^2\,\rangle &=& 0 \;\;\;\;\;\;\;\forall\;n\geq 2.
\end{eqnarray}

In \cite{Marshakov:2009} it was shown that
the state $|\,\Delta,\Lambda^2\,\rangle$ which solves the Gaiotto constraint equations
has the following form
\begin{eqnarray}\label{Gsolution}
|\,\Delta,\Lambda^2\,\rangle &=& \sum\limits_{n=0}\Lambda^{2n}\,|
\,\Delta, n\,\rangle =\sum\limits_{n=0}\Lambda^{2n}\sum\limits_{|J|=n}
\left[G^{n}_{c,\Delta}\right]^{(1^n) J}L_{-J}|\,\Delta\,\rangle.
\end{eqnarray}
Above $\left[G^{n}_{c,\Delta}\right]^{IJ}$ denotes the inverse of the Gram matrix
$\left[G^{n}_{c,\Delta}\right]_{IJ}=\langle\,\Delta\,|L_{I}L_{-J}|\,\Delta\,\rangle$
in the standard basis
\begin{eqnarray}\label{basis}
|\,\Delta,J\,\rangle\;:=\;L_{-J}|\,\Delta\,\rangle\;=\;L_{-k_1}\ldots L_{-k_n}|\,\Delta\,\rangle,
&&
J=\left( k_1 \geq\ldots\geq k_n \geq 1\right),
\\
&&
|J|=k_n+\ldots +k_1 =n\nonumber
\end{eqnarray}
of the Verma module with the central charge $c$ and the highest weight $\Delta$.

The quantum irregular block is defined as the scalar product $\langle\,\Delta, \Lambda^2\,|
\,\Delta,\Lambda^2\,\rangle$ of the Gaiotto state. Hence, taking into account (\ref{Gsolution})
one gets\footnote{For an explicit computation of the first few coefficients
in (\ref{InProd}), see appendix \ref{App0}.}
\begin{eqnarray}\label{InProd}
\langle\,\Delta, \Lambda^2\,|\,\Delta,\Lambda^2\,\rangle &=&
\sum\limits_{n=0}\Lambda^{4n}\langle\,\Delta, n\,|\,\Delta, n\,\rangle
=\sum\limits_{n=0}\Lambda^{4n} \left[G^{n}_{c,\Delta}\right]^{(1^n) (1^n)}.
\end{eqnarray}

Due to the AGT relation the irregular block can be expressed through the
$SU(2)$ pure gauge Nekrasov instanton partition function $\mathcal{Z}_{\rm inst}^{\rm N_f=0, SU(2)}$.
Indeed, the following relation
\begin{eqnarray}\label{AGT}
{\cal F}_{c,\Delta}(\Lambda) &:=&
\langle\,\Delta, \Lambda^2\,|\,\Delta,\Lambda^2\,\rangle =
\sum\limits_{n=0}\Lambda^{4n} \left[G^{n}_{c,\Delta}\right]^{(1^n) (1^n)}\nonumber
\\
&=&\sum\limits_{n=0}\hat\Lambda^{4n}\,\mathcal{Z}_{n}(a,\epsilon_1,\epsilon_2)
=\mathcal{Z}_{\rm inst}^{\rm N_f=0, SU(2)}(\hat\Lambda,a,\epsilon_1,\epsilon_2)
\end{eqnarray}
holds for
\begin{equation}\label{para1}
\Lambda\;=\;\frac{\hat\Lambda}{\sqrt{\epsilon_1\epsilon_2}},
\;\;\;\;\;\;\;\;\;\;
\Delta \;=\; \frac{(\epsilon_1+\epsilon_2)^2 - 4a^2}{4\epsilon_1\epsilon_2},
\;\;\;\;\;\;\;\;\;\;
c\;=\;1+6\left(\frac{\epsilon_1+\epsilon_2}{\sqrt{\epsilon_1\epsilon_2}}\right)^2
\equiv 1+6Q^2,
\end{equation}
where
\begin{equation}
\label{para2}
Q\;=\;b+\frac{1}{b}\;\equiv\;\sqrt{\frac{\epsilon_2}{\epsilon_1}} + \sqrt{\frac{\epsilon_1}{\epsilon_2}}
\;\;\;\;\Leftrightarrow\;\;\;\; b\;=\;\sqrt{\frac{\epsilon_2}{\epsilon_1}}.
\end{equation}
For a non rigorous derivation of eq.~(\ref{AGT}), see \cite{Hadasz:2010xp}.
Rigorous proof can be found in ref.~\cite{Maulik:2012wi}.

\subsection{Classical irregular block}
\label{S22}
In \cite{NS:2009} it was observed that in the limit $\epsilon_2\to 0$ the Nekrasov partition functions
behave exponentially. In particular, for the instantonic sector we have
\begin{equation}\label{asymZ}
\mathcal{Z}_{\rm inst}^{\rm N_f=0, SU(2)}(\hat\Lambda,a,\epsilon_1,\epsilon_2)\;\stackrel{\epsilon_2\to 0}{\sim}\;
\exp\left\lbrace \frac{1}{\epsilon_2}\,\mathcal{W}_{\rm inst}^{\rm N_f=0, SU(2)}(\hat\Lambda,a,\epsilon_1)\right\rbrace.
\end{equation}
In other words, there exists the limit
\begin{equation}\label{W_Lambda_expansion}
\mathcal{W}_{\rm inst}^{\rm N_f=0, SU(2)}(\hat\Lambda,a,\epsilon_1)\;\equiv\;
\lim\limits_{\epsilon_2\to 0}\epsilon_2\log \mathcal{Z}_{\rm inst}^{\rm N_f=0, SU(2)}
= \sum\limits_{k=1}\hat\Lambda^{4k} \mathcal{W}_{k}(a,\epsilon_1)
\end{equation}
called the effective twisted superpotential.

Taking into account (\ref{AGT}), (\ref{para1}), (\ref{para2}) and (\ref{asymZ}) one can expect the
exponential behaviour of the irregular block in the limit $b\to 0$. Indeed,
let $b=\sqrt{\epsilon_2/\epsilon_1}$ and $\Lambda=\hat\Lambda/(\epsilon_1 b)$,
$c=1+6Q^2$, $\Delta=\frac{1}{b^2}\delta$, $\delta={\cal O}(b^0)$ then, we conjecture that
\begin{equation}\label{ClIrr}
{\cal F}_{c,\Delta}(\Lambda)\;=\;
\langle\,\Delta, \Lambda^2\,|\,\Delta,\Lambda^2\,\rangle
\;\stackrel{b\to 0}{\sim}\;
\exp\left\lbrace\frac{1}{b^2}f_{\delta}\left(\frac{\hat\Lambda}{\epsilon_1}\right)\right\rbrace.
\end{equation}
Equivalently, there exists the limit
\begin{eqnarray}\label{CLIrr2}
f_{\delta}\left(\frac{\hat\Lambda}{\epsilon_1}\right)
&=&
\lim\limits_{b\to 0} b^2 \log {\cal F}_{c, \Delta}\left(\frac{\hat\Lambda}{\epsilon_1 b}\right)
=
\lim\limits_{b\to 0} b^2 \log\left[1+\sum\limits_{n=1}\left(\frac{\hat\Lambda}{\epsilon_1 b}\right)^{4n}
\left[G^{n}_{c, \Delta}\right]^{(1^n)(1^n)} \right]\nonumber
\\
&=&
\sum\limits_{n=1}\left(\frac{\hat\Lambda}{\epsilon_1}\right)^{\!\!4n}\!\!f_{\delta}^n
\end{eqnarray}
called the classical irregular block.
It should be stressed that the
asymptotical behaviuor (\ref{ClIrr}) is a nontrivial statement concerning the quantum irregular block.
Although there is no proof of this property the existence of the classical irregular
block can be checked, first, by direct calculation. For instance, up to $n=3$ one finds
\begin{equation}\label{sec2_block_coeffs}
f_{\delta}^1 \;=\; \frac{1}{2 \delta },
\;\;\;\;\;\;\;\;\;\;\;\;\;
f_{\delta}^2 \;=\; \frac{5 \delta -3}{16 \delta ^3 (4 \delta +3)},
\;\;\;\;\;\;\;\;\;\;\;\;\;
f_{\delta}^3 \;=\; \frac{9 \delta ^2-19 \delta +6}{48 \delta ^5 \left(4 \delta ^2+11 \delta +6\right)}.
\end{equation}
Secondly, there exist two other equivalent ways to get the function
$f_{\delta}(\hat\Lambda/\epsilon_1)$.
As has been shown in ref.~\cite{Marshakov:2009}
the quantum irregular block ${\cal F}_{c,\Delta}(\Lambda)$ can be recovered from
the 4-point block on the sphere ${\cal F}_{c,\Delta}\!
\left[_{\Delta_{4}\;\Delta_{1}}^{\Delta_{3}\;\Delta_{2}}\right]\!(x)$ in a properly
defined decoupling limit of the external conformal weights $\Delta_i$,\footnote{Here and below,
$\Lambda=\hat\Lambda/(\epsilon_1 b)$.}
\begin{equation}\label{delim1}
{\tilde{\cal F}}_{c,\Delta}\!\left[_{\Delta_{4}\;\Delta_{1}}^{\Delta_{3}\;\Delta_{2}}\right]\!(x)
\;\xrightarrow[x\mu_1\mu_2\mu_3\mu_4=\hat\Lambda^4]{\mu_1,\mu_2,\mu_3,\mu_4\,\to\,\infty}\;
{\cal F}_{c,\Delta}(\Lambda),
\end{equation}
where ($\epsilon=\epsilon_1+\epsilon_2$):
\begin{eqnarray}\label{wagi4p}
\Delta_4 = \frac{\left(\epsilon+\mu
   _3-\mu _4\right) \left(\epsilon-\mu _3+\mu _4\right)}{4
   \epsilon _1 \epsilon _2},
&&
\Delta_3 = \frac{\left(\mu _3+\mu _4\right) \left(2\epsilon-\mu _3-\mu
   _4\right)}{4 \epsilon _1 \epsilon _2},\nonumber
\\
\Delta_2 = \frac{\left(\epsilon+\mu
   _1-\mu _2\right) \left(\epsilon-\mu _1+\mu _2\right)}{4
   \epsilon _1 \epsilon _2},
&&
\Delta_1 = \frac{\left(\mu _1+\mu _2\right) \left(2
   \epsilon-\mu _1-\mu
   _2\right)}{4 \epsilon _1 \epsilon _2}.
\end{eqnarray}
The same phenomenon occurs in the case of the 1-point block on the torus
${\cal F}_{c,\Delta}^{\tilde\Delta}(q)$ with
\begin{equation}\label{waga}
\tilde\Delta\;=\;[m \left(\epsilon _1+\epsilon _2-m\right)]/(\epsilon _1 \epsilon _2).
\end{equation}
The torus 1-point block yields ${\cal F}_{c,\Delta}(\Lambda)$
after a decoupling of the external weight $\tilde\Delta$ \cite{Alba:2009fp},
\begin{equation}\label{delim2}
\tilde{{\cal F}}_{c,\Delta}^{\tilde\Delta}(q)
\;\xrightarrow[qm^4=\hat\Lambda^4]{m\,\to\,\infty}\;
{\cal F}_{c,\Delta}(\Lambda).
\end{equation}
Our claim is that the decoupling limits (\ref{delim1}), (\ref{delim2}) work also
on the ``classical level'',
i.e. after taking the classical limit of the quantum conformal
blocks ${\cal F}_{c,\Delta}\!
\left[_{\Delta_{4}\;\Delta_{1}}^{\Delta_{3}\;\Delta_{2}}\right]\!(x)$ and
${\cal F}_{c,\Delta}^{\tilde\Delta}(q)$.
We describe this observation in detail in the next subsection.

\subsection{Decoupling limits }
As a starting point let us recall definitions of the 4-point block
on the sphere and the 1-point block on the torus.

Let $x$ be the modular parameter of the 4-punctured Riemann sphere then the $s$-channel
conformal block on $C_{0,4}$ is defined as the following formal $x$-expansion:
\begin{eqnarray*}
{\cal F}_{c,\Delta}\!\left[_{\Delta_{4}\;\Delta_{1}}^{\Delta_{3}\;\Delta_{2}}\right]\!(x)
&=& x^{\Delta-\Delta_{2}-\Delta_{1}}\left( 1 + \sum_{n=1}^\infty
{\cal
F}^{\,n}_{c,\Delta}\!\left[_{\Delta_{4}\;\Delta_{1}}^{\Delta_{3}\;\Delta_{2}}\right]
x^{n} \right)
\\
&=:& x^{\Delta-\Delta_{2}-\Delta_{1}}\,
{\tilde{\cal F}}_{c,\Delta}\!\left[_{\Delta_{4}\;\Delta_{1}}^{\Delta_{3}\;\Delta_{2}}\right]\!(x),
\\
{\cal F}^{\,n}_{c,\Delta}\!\left[_{\Delta_{4}\;\Delta_{1}}^{\Delta_{3}\;\Delta_{2}}\right]
&=&
\lim_{z\to 1} \sum\limits_{n=|I|=|J|}
\left\langle\,\Delta_4\,|\,V_{\Delta_3}(z)\,|\Delta,I\,\right\rangle
\;
\Big[G_{c,\Delta}\Big]^{IJ}
\;
\left\langle\,\Delta,J\,|\,V_{\Delta_2}(z)\,|\,\Delta_1\,\right\rangle.
\end{eqnarray*}

Let $q=\textrm{e}^{2\pi i \tau}$  be the elliptic variable on the torus
with modular parameter $\tau$ then
the conformal block on $C_{1,1}$ is given by the following formal $q$-series:
\begin{eqnarray*}
\label{torusblock} {\cal
F}_{c,\Delta}^{\tilde\Delta}(q)&=&q^{\Delta-\frac{c}{24}}
\left(1+\sum\limits_{n=1}^{\infty}{\cal
F}_{c,\Delta}^{\tilde\Delta,n}q^n \right)
\\
&=:&q^{\Delta-\frac{c}{24}}\,{\tilde{\cal F}}_{c,\Delta}^{\tilde\Delta}(q),
\\
\label{torusCoeff} \mathcal{F}^{\tilde\Delta,
n}_{c,\Delta}&=&\lim_{z\to 1}
\sum\limits_{n=|I|=|J|}
\left\langle\,\Delta, I\,|\,V_{\tilde\Delta}(z)\,|\Delta,J\,\right\rangle
\;\Big[ G_{c,\Delta}\Big]^{IJ}.
\end{eqnarray*}

Above appear the matrix elements  of the primary chiral vertex operator between
the basis states (\ref{basis}). In order to calculate them it is sufficient to know:
\begin{itemize}
\item[$(i)$] the covariance properties of the primary chiral vertex operator w.r.t.
the Virasoro algebra,
$$
\left[L_n , V_{\Delta}(z)\right] \;=\; z^{n}\left(z
\frac{{\rm d}}{{\rm d}z} + (n+1)\Delta
\right)V_{\Delta}(z)\,,\;\;\;\;\;\;\;\;\;\;n\in\mathbb{Z};
$$
\item[$(ii)$] the form of the normalized matrix element of the
primary chiral vertex operator,\footnote{The normalization condition takes the form
$\left\langle\,\Delta_i\,|\,V_{\Delta_{j}}(1)\,|\,\Delta_k\,\right\rangle=1$.}
$$
\left\langle\,\Delta_i\,|\,V_{\Delta_{j}}(z)\,|\,\Delta_k\,\right\rangle
\;=\;z^{\Delta_i -\Delta_{j}-\Delta_k}.
$$
\end{itemize}

Let us consider the classical limit of conformal blocks, i.e. the limit
in which the central charge $c=1+6(b+\frac{1}{b})^2$, external and intermediate conformal weights
tend to infinity in such a way that their ratios are fixed.
It is known (however not yet proved) that in such limit quantum conformal blocks exponentiate.
In particular, for conformal blocks on $C_{0,4}$ and $C_{1,1}$ we have
\begin{eqnarray}
\label{defccb}
{\tilde{\cal F}}_{\! c,\Delta}
\!\left[_{\Delta_{4}\;\Delta_{1}}^{\Delta_{3}\;\Delta_{2}}\right]
\!(x)
& \stackrel{b\,\to\,0}{\,\sim} &
\exp \left\{
\frac{1}{b^2}\,f_{\delta}
\!\left[_{\delta_{4}\;\delta_{1}}^{\delta_{3}\;\delta_{2}}\right]
\!(x)
\right\},
\\
\label{classT}
{\tilde{\cal F}}^{\tilde\Delta}_{\! c,\Delta}(q) &\stackrel{b\to 0}{\sim}&
\exp\left\lbrace\frac{1}{b^2}f^{\tilde\delta}_{\delta}(q)\right\rbrace.
\end{eqnarray}
Above it is assumed that the quantum conformal weights are {\it heavy}:
$$
\Delta_i \;=\; \frac{1}{b^2}\delta_i,
\;\;\;\;\;
\tilde\Delta \;=\; \frac{1}{b^2}\tilde\delta,
\;\;\;\;\;
\Delta \;=\; \frac{1}{b^2}\delta,
\;\;\;\;\;
\delta_i, \tilde\delta, \delta \;=\;{\cal O}(b^0).
$$
The functions:
\begin{eqnarray*}
f_{\delta}\!\left[_{\delta_{4}\;\delta_{1}}^{\delta_{3}\;\delta_{2}}\right]\!(x)
\;=\;
\sum_{n=1}^\infty f^{\,n}_{\delta}\!\left[_{\delta_{4}\;\delta_{1}}^{\delta_{3}\;\delta_{2}}\right]\,x^{n},
\;\;\;\;\;\;\;\;\;\;\;\;\;\;\;\;\;
f^{\tilde\delta}_{\delta}(q)
\;=\;
\sum\limits_{n=1}^{\infty}f^{\,\tilde\delta, n}_{\delta} q^n
\end{eqnarray*}
are known as the classical conformal blocks on the sphere \cite{Zamolodchikov:1995aa,Z1,Z2}
and on the torus \cite{Piatek:2013ifa} respectively. The coefficients
$f^{\,n}_{\delta}\!\left[_{\delta_{4}\;\delta_{1}}^{\delta_{3}\;\delta_{2}}\right]$
and $f^{\,\tilde\delta, n}_{\delta}$
can be found directly from the semiclassical asymptotics (\ref{defccb}), (\ref{classT})
and the power expansions of quantum blocks:
\begin{eqnarray*}
\sum_{n=1}^\infty f^{\,n}_{\delta}\!\left[_{\delta_{4}\;\delta_{1}}^{\delta_{3}\;\delta_{2}}\right]\,x^{n}
&=&
\lim\limits_{b \to 0} {b^2} \log\left(1 +
\sum_{n=1}^\infty {\cal
F}^{\,n}_{\!c,\Delta}\!\left[_{\Delta_{4}\;\Delta_{1}}^{\Delta_{3}\;\Delta_{2}}\right]
x^{n}\right),
\\[8pt]
\sum\limits_{n=1}^{\infty}f^{\,\tilde\delta, n}_{\delta} q^n
&=&
\lim\limits_{b\to 0}b^2\log
\left(1+\sum\limits_{n=1}^{\infty}{\cal F}_{c,\Delta}^{\tilde\Delta,n}\;q^n \right),
\end{eqnarray*}
see appendix \ref{App0}.

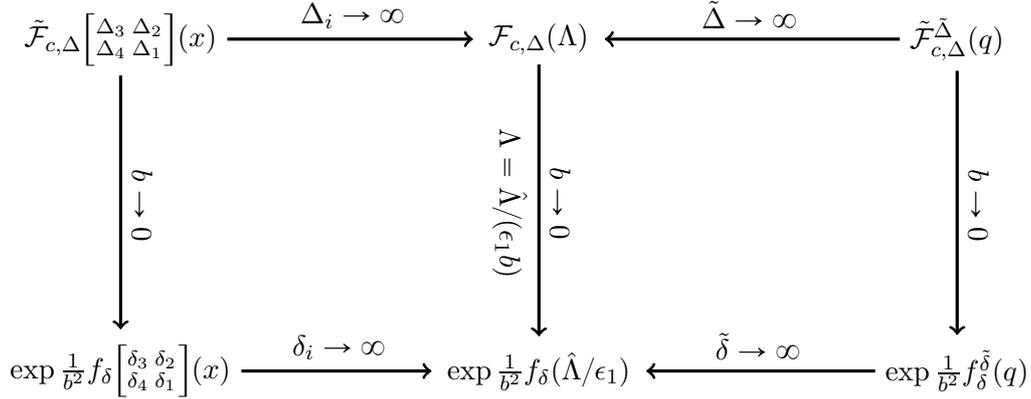
\begin{figure}[htb]
\centering
\begin{tikzpicture}[shorten >=2pt,scale=1.1]
\node (a) at (0,0) {${\tilde{\cal F}}_{c,\Delta}\!\left[_{\Delta_{4}\;\Delta_{1}}^{\Delta_{3}\;\Delta_{2}}\right]\!(x)$};
\node (b) at (5,0) {$\mathcal{F}_{c,\Delta}(\Lambda)$};
\node (c) at (10,0) {${\tilde{\cal F}}_{c,\Delta}^{\tilde\Delta}(q)$};
\node (d) at (0,-4) {$\exp\frac{1}{b^2} f_{\delta}\!\left[_{\delta_{4}\;\delta_{1}}^{\delta_{3}\;\delta_{2}}\right]\!(x)$};
\node (e) at (5,-4) {$\exp\frac{1}{b^2} f_{\delta}(\hat\Lambda/\epsilon_1)$};
\node (f) at (10,-4) {$\exp\frac{1}{b^2} f_{\delta}^{\tilde\delta}(q)$};
\path[very thick,->] (a) edge node[above] {$\Delta_i\rightarrow\infty$} node [below] {} (b)
				   edge node[above,rotate=-90] {$b\rightarrow 0$}  node [below,rotate=90] {} (d)
			   (d) edge node[above] {$\delta_i\rightarrow\infty$} node [below] {} (e)
               (c) edge node[above] {$\tilde\Delta\rightarrow\infty$} node[below] {} (b)
			   (c) edge node[above,rotate=-90] {$b\rightarrow 0$} node [below,rotate=-90] {} (f)
			   (f) edge node[above] {$\tilde\delta\rightarrow\infty$} node [below] {} (e)
               (b) edge node[above,rotate=-90] {$b\rightarrow 0$} node [below,rotate=-90] {$\Lambda=\hat\Lambda/(\epsilon_1 b)$} (e);
\end{tikzpicture}
\caption{Equivalent ways to get the classical irregular block
$f_{\delta}(\hat\Lambda/\epsilon_1)$
(above $c=1+6(b+\frac{1}{b})^2$ and
$(\Delta, \Delta_i, \tilde\Delta)=\frac{1}{b^2}(\delta,  \delta_i, \tilde\delta))$.\label{fig:1}}
\end{figure}

Let us observe
that the quantum external weights (\ref{wagi4p}), (\ref{waga}) are heavy
in the terminology of \cite{Zamolodchikov:1995aa}, i.e. exist the limits:
$$
\delta_i \;=\; \lim_{b\to 0}b^2\Delta_i=
\lim_{\epsilon_2\to 0}\frac{\epsilon_2}{\epsilon_1}\Delta_i,
\;\;\;\;\;\;\;\;\;\;\;\;\;\;\;\;
\tilde\delta \;=\; \lim_{b\to 0}b^2\tilde\Delta=
\lim_{\epsilon_2\to 0}\frac{\epsilon_2}{\epsilon_1}\tilde\Delta.
$$
The classical external weights explicitly read as follows
\begin{eqnarray}\label{Clwagi}
\delta_4 = \left[\epsilon _1^2-\left(\mu _3-\mu
   _4\right){}^2\right]/\left(4 \epsilon _1^2\right),
&&
\delta_3 = \left[\left(\mu _3+\mu _4\right) \left(2
   \epsilon _1-\mu _3-\mu _4\right)\right]/\left(4
   \epsilon _1^2\right),\nonumber
\\
\delta_2 = \left[\epsilon _1^2-\left(\mu _1-\mu
   _2\right){}^2\right]/\left(4 \epsilon _1^2\right),
&&
\delta_1 = \left[\left(\mu _1+\mu _2\right) \left(2
   \epsilon _1-\mu _1-\mu _2\right)\right]/\left(4
   \epsilon _1^2\right)
\end{eqnarray}
and
$$
\tilde\delta \;=\;\left[m \left(\epsilon _1-m\right)\right]/\epsilon _1^2.
$$
Hence, one can consider the $b\to 0$ limit of both sides
of the decoupling limits (\ref{delim1}), (\ref{delim2}).
Then, for the classical weights given by (\ref{Clwagi}) one can verify order by order that
\begin{eqnarray*}
f_{\delta}\!\left[_{\delta_{4}\;\delta_{1}}^{\delta_{3}\;\delta_{2}}\right]\!(x)
&\xrightarrow[x\mu_1\mu_2\mu_3\mu_4=\hat\Lambda^4]{\mu_1,\mu_2,\mu_3,\mu_4\,\to\,\infty}&
f_{\delta}(\hat\Lambda/\epsilon_1),
\\
f^{\tilde\delta}_{\delta}(q)
&\xrightarrow[qm^4=\hat\Lambda^4]{m\,\to\,\infty}&
f_{\delta}(\hat\Lambda/\epsilon_1).
\end{eqnarray*}
Calculations presented above can be visualized on the diagram (see fig.\ref{fig:1}).
Let us stress that the commutativity of this diagram lend a strong support to the
exponentiation hypothesis (\ref{ClIrr}), (\ref{defccb}), (\ref{classT}) of conformal blocks in the limit $b\to 0$.

As a final remark in this section let us observe that joining together
(\ref{AGT})-(\ref{para2}), (\ref{asymZ}) and (\ref{ClIrr}) one gets
\begin{equation}\label{classAGT}
f_{\delta}(\hat\Lambda/\epsilon_1) \;=\;\frac{1}{\epsilon_1}\,
\mathcal{W}_{\rm inst}^{\rm N_f=0, SU(2)}(\hat\Lambda,a,\epsilon_1),
\end{equation}
where
\begin{equation}\label{classDelta}
\delta\;=\;\lim\limits_{b\to 0}b^2\Delta \;=\;
\lim\limits_{\epsilon_2\to 0}\frac{\epsilon_2}{\epsilon_1}\,\Delta
\;=\; \frac{1}{4}-\frac{a^2}{\epsilon_{1}^{2}}.
\end{equation}
Indeed, to show the consistency of our approach one can use \eqref{sec2_block_coeffs} and formulae
from the appendix \ref{AppIII}, and check that
\begin{equation}
\label{coeffs_f_W_comparison}
\frac{1}{\epsilon_1}\,\mathcal{W}_{n}^{\rm N_f=0, SU(2)} \;=\;
\frac{1}{\epsilon_{1}^{4n}}\,f^{n}_{\frac{1}{4}-\frac{a^2}{\epsilon_{1}^{2}}},
\;\;\;\;\;\;\;\;\;\;
n=1,2,3,\ldots\;.
\end{equation}

\section{Mathieu equation in \texorpdfstring{$2d$}{2d} CFT}
\label{sec3}
\subsection{Null vector equation for the degenerate irregular block}
Let
$$
V_{+}(z)\;\equiv\;V_{\Delta_+}(z),
\qquad
\Delta_{+}\;=\;-\frac{1}{2}-\frac{3}{4}\,b^2
$$
denotes the degenerate primary chiral vertex operator \cite{Moore:1988qv}
$$
V_{+}(z) \;\equiv\; V^{\Delta'\,\Delta_+\,
\tilde\Delta}_{\;\infty \;\;\;z \;\;\;0}
\left(|\,\Delta_+\,\rangle\otimes \cdot\,\right) : {\cal
V}_{\tilde\Delta} \to {\cal V}_{\Delta'}
$$
acting between the Verma modules ${\cal V}_{\tilde\Delta}$ and ${\cal V}_{\Delta'}$.
We will assume that the conformal weights $\tilde\Delta$ and $\Delta'$ are related by the fusion rule, i.e.:
\begin{equation}\label{fusion}
\tilde\Delta\;=\;\Delta\left(\sigma-\frac{b}{4}\right),
\;\;\;\;\;\;\;\;
\Delta'\;=\;\Delta\left(\sigma+\frac{b}{4}\right),
\;\;\;{\rm where}\;\;\;
\Delta(\sigma)\;=\;\frac{Q^2}{4}-\sigma^2.
\end{equation}
Let us consider the descendant\footnote{Recall that \cite{Belavin:1984vu}
\begin{equation}\label{hatL}
\widehat{L}_{-k}(z)\;\equiv\;\frac{1}{2\pi i}\oint\limits_{C_z}dw (w-z)^{1-k}T(w).
\end{equation}}
$$
\label{zerowe2B}
\chi_{+}(z)\;=\;\left[\widehat{L}_{-2}(z) -
\frac{3}{2(2\Delta_{+} +1)}
\,\widehat{L}_{-1}^{\,2}(z)\right]V_{+}(z)
$$
of $V_{+}(z)$ which corresponds to the null vector appearing on the second
level of the Verma module ${\cal V}_{\Delta_+}$.
According to the null vector decoupling theorem \cite{FeiginFuchs}
(see also \cite{Teschner:2001rv}) the matrix element of $\chi_{+}(z)$,
between the states with the highest weights
obeying (\ref{fusion}), must vanish. In particular, vanishes the matrix element
\begin{eqnarray*}\label{NVDcondition}
\langle\,\Delta', \Lambda^2\,|
\,\chi_{+}(z)\,|\,\tilde\Delta,\Lambda^2\,\rangle\nonumber
&=& \langle\,\Delta', \Lambda^2\,|
\,\widehat{L}_{-2}(z)V_{+}(z)\,|\,\tilde\Delta,\Lambda^2\,\rangle
\\
&+&
\frac{1}{b^2}\,\langle\,\Delta', \Lambda^2\,|
\,\widehat{L}_{-1}^{2}(z)V_{+}(z)\,|\,\tilde\Delta,\Lambda^2\,\rangle\;=\;0,
\end{eqnarray*}
where $\langle\,\Delta', \Lambda^2\,|$ and $|\,\tilde\Delta,\Lambda^2\,\rangle$
are the Gaiotto states introduced in (\ref{Gstate1})-(\ref{Gstate3}).
The above null vector decoupling condition
can be converted to the following partial differential equation
for the degenerate irregular 3-point block $\Psi(\Lambda, z)=\langle\,\Delta', \Lambda^2\,|
V_{+}(z)|\,\tilde\Delta,\Lambda^2\,\rangle$ \cite{Maruyoshi:2010},
\begin{equation}\label{NVDeq}
\left[\frac{1}{b^2}\,z^2\frac{\partial^2}{\partial z^2}-
\frac{3z}{2}\frac{\partial}{\partial z}+\Lambda^2 \left(z+\frac{1}{z}\right)
+\frac{\Lambda}{4}\frac{\partial}{\partial\Lambda}+
\frac{\tilde\Delta+\Delta'-\Delta_+}{2}\right]\Psi(\Lambda, z)\;=\;0\,.
\end{equation}
Indeed, using the Ward identity
\begin{eqnarray}\label{MT2}
\langle\,\Delta', \Lambda^2\,|
\,T(w)V_{+}(z)\,|\,\tilde\Delta,\Lambda^2\,\rangle
&=&
\left[
\frac{z}{w(w-z)}\frac{\partial}{\partial z} + \frac{\Delta_+}{(w-z)^2}
+\left(\frac{\Lambda^{2}}{w}+\frac{\Lambda^2}{w^3}\right)\right.\nonumber
\\
&&\hspace{-35pt}+\left.
\frac{1}{2 w^2}\left(\frac{\Lambda}{2}\frac{\partial}{\partial \Lambda}
+\tilde\Delta+\Delta'-\Delta_{+}-z\frac{\partial}{\partial z}\right)
\right]\Psi(\Lambda, z),
\end{eqnarray}
where $T(w)=\sum_{n\in\mathbb{Z}}w^{-n-2}L_{n}$ is the holomorphic component
of the energy-momentum tensor, one can find
\begin{eqnarray}
\label{L2}
\langle\,\Delta', \Lambda^2\,|
\,\widehat{L}_{-2}(z)V_{+}(z)\,|\,\tilde\Delta,\Lambda^2\,\rangle
&=&
\left[
-\frac{1}{z}\frac{\partial}{\partial z}
+\left(\frac{\Lambda^{2}}{z}+\frac{\Lambda^2}{z^3}\right)\right.\nonumber
\\
&&\hspace{-35pt}+\left.
\frac{1}{2 z^2}\left(\frac{\Lambda}{2}\frac{\partial}{\partial \Lambda}
+\tilde\Delta+\Delta'-\Delta_{+}-z\frac{\partial}{\partial z}\right)
\right]\Psi(\Lambda, z),
\\
\label{L1}
\langle\,\Delta', \Lambda^2\,|
\,\widehat{L}_{-1}^{2}(z)V_{+}(z)\,|\,\tilde\Delta,\Lambda^2\,\rangle
&=& \frac{\partial^2}{\partial z^2}\Psi(\Lambda, z).
\end{eqnarray}
For a derivation of eqs.~(\ref{MT2}), (\ref{L2}), see appendix \ref{AppI}.

At this point, a few comments concerning $\Psi(\Lambda, z)$ are necessary.
First, let us note that from (\ref{Gsolution}) we have
\begin{eqnarray}
\Psi(\Lambda, z)&=&\langle\,\Delta', \Lambda^2\,|V_{+}(z)|\,\tilde\Delta,\Lambda^2\,\rangle
\;=\;z^{\Delta'-\Delta_+ -\tilde \Delta} \sum_{m,n\geq 0}\Lambda^{2(m+n)}z^{m-n}\nonumber
\\\label{matrix1}
&&\times
\sum_{|I|=m}\sum_{|J|=n} \left[G^{m}_{c,\Delta'}\right]^{(1^m) I}
\langle\,\Delta',I\, |\,V_{+}(1)\,|\,\tilde\Delta,J\,\rangle
\left[G^{n}_{c,\tilde\Delta}\right]^{J (1^n)}
\\[8pt]\label{matrix2}
&\equiv& z^{\kappa}\,\Phi(\Lambda,z),
\end{eqnarray}
where $\kappa\equiv\Delta'-\Delta_+ -\tilde \Delta$.

Let us observe that $\Phi(z,\Lambda)$ can be split into two parts, i.e. when
$m=n$ and $m\neq n$:
\begin{eqnarray}\label{Phi}
\Phi(\Lambda,z) &=& \Phi^{(m=n)}(\Lambda) + \Phi^{(m\neq n)}(\Lambda, z),
\\
\Phi^{(m=n)}(\Lambda) &=&\sum\limits_{n\geq 0}\Lambda^{4n}
\sum_{|I|=n}\left[G^{n}_{c,\Delta'}\right]^{(1^n) I}
\langle\,\Delta',I\, |\,V_{+}(1)\,|\,\tilde\Delta,I\,\rangle
\left[G^{n}_{c,\tilde\Delta}\right]^{I (1^n)},\nonumber
\\
\Phi^{(m\neq n)}(\Lambda, z)\!\!\!\! &=&
\!\!\!\!\!\!\!\!
\sum\limits_{\begin{array}{c}\scriptstyle m\neq n
\\[-7pt]\scriptstyle m,n\geq 0 \end{array}}
\!\!\!\!\!\Lambda^{2(m+n)}z^{m-n}\!\!\!\!\!
\sum\limits_{\begin{array}{c}\scriptstyle |I|=m,
\\[-7pt]\scriptstyle |J|=n \end{array}}
\left[G^{m}_{c,\Delta'}\right]^{(1^m) I}
\langle\,\Delta',I\, |\,V_{+}(1)\,|\,\tilde\Delta,J\,\rangle
\left[G^{n}_{c,\tilde\Delta}\right]^{J (1^n)}.\nonumber
\end{eqnarray}
Then, one can write
\begin{eqnarray}\label{Psi}
\Psi(\Lambda, z)&=& z^\kappa \; \exp\left\lbrace\log\left(
\Phi^{(m=n)}(\Lambda) + \Phi^{(m\neq n)}(\Lambda, z)\right)\right\rbrace
\\
&=& z^\kappa \;\exp\left\lbrace\log\Phi^{(m=n)}(\Lambda)+
\log\left(1+\frac{\Phi^{(m\neq n)}(\Lambda, z)}{\Phi^{(m=n)}(\Lambda)}\right)\right\rbrace
\;\equiv\;z^{\kappa}\,{\rm e}^{\phi_{1}(\Lambda)}\,{\rm e}^{\phi_{2}(\Lambda,z)}.\nonumber
\end{eqnarray}

Inserting (\ref{matrix2}) into the eq.~(\ref{NVDeq}) we get
\begin{eqnarray}\label{NVDeq2}
&&\left[\frac{1}{b^2}\,z^2\partial^2_z+\left(\frac{2\kappa}{b^2}-\frac{3}{2}\right)z\partial_z
+\frac{\Lambda}{4}\,\partial_\Lambda + \frac{\kappa(\kappa -1)}{b^2}
-\frac{3\kappa}{2}\right.\nonumber
\\
&&\hspace{145pt}\;\left.\,+\,\Lambda^2 \left(z+\frac{1}{z}\right)
+ \frac{\tilde\Delta+\Delta'-\Delta_+}{2}\right] \Phi(\Lambda,z)\;=\;0.
\end{eqnarray}

\subsection{Classical limit}
Now, we want to find the limit $b\to 0$ of the eq.~(\ref{NVDeq2}).
This firstly requires to rescale the parameter $\sigma$ in $\tilde\Delta$
and $\Delta'$, i.e. $\sigma=\xi/b$ and to express $\Lambda$ as
$\Lambda=\hat\Lambda/(\epsilon_1 b)$. After rescaling we have
\begin{eqnarray}\label{delta}
&&\Delta',\tilde\Delta\stackrel{b\to 0}{\sim}\frac{1}{b^2}\,\delta,
\;\;\;\;\;\;{\rm where}\;\;\;\;\;\;\delta\;=\;
\lim_{b\to 0}b^2\Delta'\;=\;\lim_{b\to 0}b^2\tilde\Delta\;=\;\frac{1}{4}-\xi^2,
\\[8pt]
&&
\tilde\Delta+\Delta'-\Delta_+\stackrel{b\to 0}{\sim}\frac{1}{b^2}
\,2\left(\frac{1}{4}-\xi^2\right)\;=\;\frac{1}{b^2}\,2\delta,
\\[8pt]
&&
\kappa\stackrel{b\to 0}{\longrightarrow}\frac{1}{2}-\xi,
\;\;\;\;\;\;\;\;\;\;\;
\kappa\left(\kappa-1\right)\stackrel{b\to 0}{\longrightarrow}
-\left(\frac{1}{4}-\xi^2\right)\;=\;-\delta.
\end{eqnarray}
Note, that $\Delta_+ \stackrel{b\to 0}{\sim} {\cal O}(b^0)$.

Secondly, one has to determine the behavior of the normalized degenerate irregular
block $\Phi=z^{-\kappa}\Psi$ when $b\to 0$.
For $\Lambda=\hat\Lambda/(\epsilon_1 b)$ and
$\Delta',\tilde\Delta\stackrel{b\to 0}{\sim}\frac{1}{b^2}\,\delta$
it is reasonable to expect, that
\begin{equation}\label{ClAsymp}
\Phi(\Lambda, z)\;=\;z^{-\kappa}\,\langle\,\Delta', \Lambda^2\,|\,
V_{+}(z)\,|\,\tilde\Delta,\Lambda^2\,\rangle \;\stackrel{b\to 0}{\sim}\;
v(\hat\Lambda/\epsilon_1, z)\,
\exp\left\lbrace\frac{1}{b^2}f_{\delta}(\hat\Lambda/\epsilon_1)\right\rbrace,
\end{equation}
where (cf.~(\ref{Psi}))
\begin{eqnarray}\label{v}
v(\hat\Lambda/\epsilon_1, z) &=&
\lim\limits_{b\to 0}{\rm e}^{\phi_{2}\left(\Lambda, z\right)}=
{\rm e}^{\lim_{b\to 0}\phi_{2}\left(\Lambda, z\right)},
\\
\label{f}
f_{\delta}(\hat\Lambda/\epsilon_1) &=& \lim\limits_{b\to 0}b^2 \phi_{1}\left(\Lambda\right)
= \lim\limits_{b\to 0}b^2\log\Phi^{(m=n)}\left(\Lambda\right).
\end{eqnarray}
Let us stress that the asymptotic (\ref{ClAsymp}) is a nontrivial statement concerning
the 3-point irregular block $\Phi(\Lambda, z)$. We have no rigorous proof of this conjecture.
However, the eq.~(\ref{ClAsymp}) can be well confirmed by direct calculations. Indeed,
one can check order by order that the limits (\ref{v}) and (\ref{f}) exist.
Moreover, the latter yields the classical irregular block.

Then, after substituting (\ref{ClAsymp}) into the eq.~(\ref{NVDeq2}), multiplying by $b^2$,
and taking the limit $b\to 0$ one gets\footnote{The key point here is the fact that
$\lim_{b\to 0}b^2\frac{\hat\Lambda}{4}\,\partial_{\hat\Lambda}v=0$.}
\begin{equation}\label{M1}
\left[\,z^2\partial^2_z+2(\tfrac{1}{2}-\xi)z\partial_z
+ \frac{\hat\Lambda^2}{\epsilon_1^2 } \left(z+\frac{1}{z}\right)
+\frac{\hat\Lambda}{4}\partial_{\hat\Lambda} f_{\delta}(\hat\Lambda/\epsilon_1)
\right]v(z)\;=\;0\,.
\end{equation}
In order to get the Mathieu equation we define a new function $\psi(z)$ related to the old
one by
$$
v(z)\;=\; z^{\xi}\,\psi(z).
$$
Now, what we obtain equals
\begin{equation}\label{NVDeq3}
\left[\,z^2\partial^2_z+z\partial_z
+ \frac{\hat\Lambda^2}{\epsilon_1^2 } \left(z+\frac{1}{z}\right)
+\frac{\hat\Lambda}{4}\partial_{\hat\Lambda} f_{\delta}(\hat\Lambda/\epsilon_1) - \xi^2
\right] \psi(z)\;=\;0\,.
\end{equation}
Since for $z = {\rm e}^w$ the derivatives transform as
$\left(z^2\partial^2_z  + z\partial_z \right)\psi(z)
=\partial^2_w \psi\left({\rm e}^w\right)$ then,
the eq.~(\ref{NVDeq3}) goes over to the form
\begin{equation}\label{NVDeq4}
\left[\frac{\mbox{d}^2}{\mbox{d}w^2} + 2\frac{\hat\Lambda^2}{\epsilon_1^2}\cosh(w)
+ \frac{\hat\Lambda}{4}\partial_{\hat\Lambda} f_{\delta}(\hat\Lambda/\epsilon_1) - \xi^2
\right] \psi\left({\rm e}^w\right) \;=\; 0 .
\end{equation}
Finally, the substitution $w=2ix$, $x\in\mathbb{R}$ in (\ref{NVDeq4}) yields
\begin{equation}\label{M2}
\left[-\frac{{\rm d}^2}{{\rm d}x^2} + 8\frac{\hat\Lambda^2}{\epsilon_{1}^{2}}\,\cos 2x
+\hat\Lambda\,\partial_{\hat\Lambda}f_{\delta}(\hat\Lambda/\epsilon_1)
-4\xi^2\right]\psi({\rm e}^{2ix})\;=\;0.
\end{equation}
Now, one can identify the parameters $\lambda$ and $h$ appearing in the Mathieu equation
(\ref{Mathieu1}) as follows
\begin{equation}\label{eigenvalue}\boxed{
\;\lambda \;=\; -\hat\Lambda\,\partial_{\hat\Lambda}f_{\delta}\left(\hat\Lambda/\epsilon_1\right)+
4\xi^2,
\;\;\;\;\;\;\;\;\;\;\;\;\;
h\;=\;\pm \frac{2\hat\Lambda}{\epsilon_1}.\;}
\end{equation}
Let us observe that using \eqref{sec2_block_coeffs} and postulating the following
relation
\begin{equation}\boxed{
\;\xi\;=\;\frac{\nu}{2}\;}
\end{equation}
between the parameter $\xi$ and the Floquet exponent $\nu$ one finds
\begin{eqnarray*}
\lambda & = &-\hat\Lambda\,\partial_{\hat\Lambda}
\left[\,\sum\limits_{n=1}\left(\hat\Lambda/\epsilon_1\right)^{\!\!4n}\!\!f_{\delta}^n\right]+
4\xi^2
\\[10pt]
&=&-\frac{4 h^4}{16}\,f_{\frac{1}{4}-\frac{\nu^2}{4}}^1-
\frac{8 h^8}{256}\,f_{\frac{1}{4}-\frac{\nu^2}{4}}^2 -
\frac{12 h^{12}}{4096}\,f_{\frac{1}{4}-\frac{\nu^2}{4}}^3 - \ldots + 4\left(\frac{\nu^2}{4}\right)
\\[10pt]
&&\hspace{-40pt}=\;\nu^2 +
\frac{h^4}{2 \left(\nu ^2-1\right)}+
\frac{\left(5 \nu ^2+7\right)h^8}{32 \left(\nu ^2-4\right) \left(\nu ^2-1\right)^3}+
\frac{\left(9 \nu ^4+58 \nu ^2+29\right)h^{12}}{64 \left(\nu ^2-9\right) \left(\nu ^2-4\right) \left(\nu^2-1\right)^5}
+\ldots\;.
\end{eqnarray*}
Hence, the formula (\ref{eigenvalue}) reproduces the expansion (\ref{lambda}).

On the other hand one can check that the formula (\ref{eigenvalue}) exactly matches
that obtained by means of the WKB method. Indeed, taking into account that (cf.~(\ref{classDelta}))
$$
\delta\;=\;\frac{1}{4}-\frac{a^2}{\epsilon_{1}^{2}}\;=\;\frac{1}{4}-\xi^2
\;\;\;\Longleftrightarrow\;\;\;\xi=\frac{a}{\epsilon_{1}}
$$
one can rewrite the eq.~(\ref{M2}) to the following Schr\"{o}dinger-like form:
\begin{equation}\label{SchEq}
\left[-\epsilon_{1}^{2}\frac{\textrm{d}^2}{\textrm{d}x^2}
+8\hat\Lambda^{2}\cos2x\right]\psi \;=\;{\sf E}\,\psi,
\end{equation}
where
\begin{equation}\label{EnergyWKB}
{\sf E}\;=\;\epsilon_{1}^{2}\,\lambda \;=\;4a^2-\epsilon_{1}^{2}\,\hat\Lambda\,
\partial_{\hat\Lambda}f_{\frac{1}{4}-\frac{a^2}{\epsilon_{1}^{2}}}(\hat\Lambda/\epsilon_1).
\end{equation}
The ``energy eigenvalue'' ${\sf E}$ in eq.~(\ref{SchEq}) can be computed by
making use of the WKB method. These calculations are performed in the appendix \ref{app_WKB}.
The result of the WKB calculations coincides with ${\sf E}$ given by eq.~(\ref{EnergyWKB}).
This check is yet another  example of an interesting
link between the semiclassical limit of conformal blocks and the WKB approximation.

\section{Mathieu eigenvalue from \texorpdfstring{$\mathcal{N}=2$}{N=2} gauge theory}
\label{sec4}
We saw in previous sections that the eigenvalue of the Mathieu operator is related to the
classical irregular conformal block. The latter appeared to be a limit of the quantum irregular conformal
block when $b^2\sim\hbar\to0$. The irregular quantum conformal block is found to be related \cite{Marshakov:2009} to the
Nekrasov's instanton partition function of the pure ($N_f = 0$)  $\mathcal{N}=2$ SYM. Nekrasov's partition
function in the zero limit of one of its parameters (in what follows we take it to be $\epsilon_2$)
yields an effective twisted superpotential. Therefore it is natural to expect
that thus established correspondence between the two theories at the quantum level also extends to the classical level. In the
preceding section we found a relation between the expansion parameters on both sides of the correspondence
$\mathcal{F}_{c,\Delta}(\Lambda) = \mathcal{Z}^{\rm N_f=0, SU(2)}_{\rm inst}(\hat\Lambda/\epsilon_1,a,b)$
and verified the agreement between expansion coefficients of classical irregular block
and twisted superpotential up to the third order.
In this section we address the derivation of the twisted superpotential $\mathcal{W}$ from
the representation of the instanton partition function
for $\mathcal{N} = 2$ pure gauge SYM in terms of profile
functions of the Young diagrams employed in
this context first by Nekrasov and Okounkov \cite{Nekrasov:Okounkov:2003}.

\subsection{Nekrasov--Shatashvili limit}

The Nekrasov's partition function for $\mathcal{N}=2$ pure SYM with $SU(N)$ symmetry
on $\Omega$-background relates instanton configurations on a moduli space
with partitions a graphical representation of whose are Young diagrams. This relationship
makes the mentioned gauge theory tantamount to the theory of random partitions.
There are five equivalent forms of the instanton partition function (see appendix \ref{app_instNekrasov}
for notation and more information about Nekrasov partition function). For our purpose
we use the following one
\begin{subequations}
\begin{equation}
\label{def_full_partition_func}
\mathcal{Z}^{\rm N_f=0, SU(N)}(\hat\Lambda,\mathbf{a},\e_1,\e_2)
= \exp\left\{-\sum_{\a,\b=1}\g_{\e_1 ,\e_2}(a_{\a}-a_\b;\hat\La)  \right\}
\mathcal{Z}_{\rm inst}^{\rm N_f=0, SU(N)}(\hat\Lambda,\mathbf{a},\e_1,\e_2),
\end{equation}
where the first factor on the right hand side is a perturbative part. The second is defined as follows
\begin{gather}
\label{def_z_inst}
\mathcal{Z}_{\rm inst}^{\rm N_f=0, SU(N)}(\hat\Lambda,\mathbf{a},\e_1,\e_2)
\equiv \sum_{k\geq 0 }
\cbr{\frac{\hat\La}{\e_1}}^{2N k} Z_{k}(\mathbf{a},\e_1,\e_2) ,
\\
Z_{k}(\mathbf{a},\e_1,\e_2)
\equiv \sum_{\mathbf{k}:|\mathbf{k}| = k }b^{-4Nk}
\prod_{(\a,i)\neq(\b,j)}\frac{\Gamma\cbr{\e_2^{-1}(x_{\a,i} - x_{\b,j} - \e_1 ) }
\Gamma\cbr{\e_2^{-1}(x^0_{\a,i}- x^0_{\b,j} ) } }
{\Gamma\cbr{\e_2^{-1}(x_{\a,i} - x_{\b,j} ) }\Gamma\cbr{\e_2^{-1}(x^0_{\a,i}- x^0_{\b,j} - \e_1 ) } } .
\end{gather}
\end{subequations}
The arguments of the Euler gamma functions are
defined as follows\footnote{More precisely, the point $x_{\a,i}$ is a coordinate of $i^{\rm th}$ column within a
$\a^{\rm th}$ Young diagram depicted in the Russian convention and projected onto the real axis.
The relation of the Russian convention of drawing a Young diagram
to e.g., the English one boils down to rotation of the latter by $135^\circ$ about its pivot point $a_\a$.
In this case the decline of columns coincides with the linear order of $\mathbb{R}$.
Since $\e_1\neq\e_2$ the boxes of a Young diagram are deformed to rectangles with edges $l_i = \sqrt{2}|\e_i|,\ i=1,2$.}
\begin{equation}
\label{points}
x_{\a,i} \equiv a_\a + \e_1(i-1) + \e_2 k_{\a i} ,
\qquad
x^0_{\a,i} \equiv a_\a + \e_1(i-1) .
\end{equation}
Thus, the instatnon partition function is expressed in terms of
partitions of an integer instanton number $k$ i.e., $k=|\mathbf{k}|=\sum |\Bbbk_\alpha|
=\sum\limits k_{\a,i}$, $\a = 1,\dots,N$ and
$i\in\mathbb{N}$, such that for any $i<j$ and fixed $\a$, $k_{\a,i}\geq k_{\a,j}\geq 0$.
This particular form of it is related
to the one, defined in terms of the profiles of the deformed Young diagrams. Namely,
\begin{multline}
\label{z_in_profile}
\mathcal{Z}^{\rm N_f=0, SU(N)}(\hat\Lambda,\mathbf{a},\e_1,\e_2)
\\
=
\sum_{f_{\mathbf{a},\mathbf{k}}\in \mathscr{P}(\mathbb{Y}^N) }
\exp\left\{-\frac{1}{4}\vpiint\limits_{\mathbb{R}^2}
\mbox{d} x\mbox{d} y\ f''_{\mathbf{a},\mathbf{k}}(x|\e_1,\e_2)\g_{\e_1 ,\e_2}(x-y;\hat\La)
 f''_{\mathbf{a},\mathbf{k}}(y|\e_1,\e_2)
\right\} ,
\end{multline}
where $\mathscr{P}(\mathbb{Y}^N)$ denotes the space of all profiles of $N$-tuple of Young diagrams.
The profile function is defined as follows\footnote{Note, that this definition of the profile function, when
defined for real argument, is meaningful if and only if both parameters $\e_1$ or $\e_2$ have opposite signs.}
\begin{multline}
\label{def_profile}
f_{\mathbf{a},\mathbf{k}}(x|\e_1,\e_2)
\equiv \sum_{\a =1}^N |x-a_\a|
+\sum_{\a =1}^N \sum_{i\geq 1}\Big(\left|x-x^0_{\a,i}-\e_1\right|
-\left|x - x_{\a,i}-\e_1\right|
\\
- \left|x- x^0_{\a,i}\right|+\left|x-x_{\a,i}\right|
\Big ) .
\end{multline}
In what follows we work with the instanton density rather then with the profile functions.
The linear density function of instantons at position $\mathbf{a}$ and configuration $\mathbf{k}$
is defined as
\begin{equation}
\label{def_inst_density}
\rho_{\mathbf{a},\mathbf{k}}(x|\e_1,\e_2)
\equiv f_{\mathbf{a},\mathbf{k}}(x|\e_1,\e_2)
-f_{\mathbf{a},\boldsymbol{\emptyset}}(x|\e_1,\e_2) ,
\end{equation}
where the second term is the empty profile given by the first term on the
right hand side of eq. \eqref{def_profile}.
The instanton density satisfies the following normalization condition
\begin{equation}
\label{inst_numb_density}
k = \frac{1}{-2 \e_1\e_2}
\int\limits_{\mathbb{R}}\mbox{d} x\, \rho_{\mathbf{a},\mathbf{k}}(x|\e_1,\e_2).
\end{equation}
The above equation shows that the density function stores the information
about both, the number of instantons and their configuration.
The partition function expressed in terms of profiles in eq. \eqref{z_in_profile}
can now be written in terms of instanton densities. Discarding the perturbative part which
takes the form of the multiplier in eq. \eqref{def_full_partition_func} the instatnon part reads
\begin{equation}
\label{z_inst_hamiltonian}
\mathcal{Z}_{\rm inst}^{\rm N_f=0, SU(N)}(\hat\Lambda,\mathbf{a},\e_1,\e_2)
= \sum_{\rho_{\mathbf{a},\mathbf{k}}\in \mathscr{R}(\mathbb{Y}^N) }
\exp\pbr{-H_{\rm inst}[\rho_{\mathbf{a},\mathbf{k}}](\e_1,\e_2,\hat\La)} ,
\end{equation}
where $\mathscr{R}(\mathbb{Y}^N)
{\tt\subset}\mathscr{P}(\mathbb{Y}^N)$.\footnote{$\mathscr{R}(\mathbb{Y}^N)$ is
a space of all functions over $N$-tuple of Young diagrams with a compact support.}
The Hamiltonian for instanton configurations structurally reads
\begin{multline}
\label{def_inst_hamilton}
H_{\rm inst}[\rho_{\mathbf{a},\mathbf{k}}](\e_1,\e_2,\hat\La)
\equiv \tfrac{1}{4} (f''_{\boldsymbol{\emptyset}} )^{\rm i}
\g_{\e_1,\e_2}(\hat\Lambda)_{\rm ij} (\rho''_{\mathbf{k}} )^{\rm j}
+ \tfrac{1}{4} (\rho''_{\mathbf{k}}  )^{\rm i }
\g_{\e_1,\e_2}(\hat\Lambda)_{\rm ij} (f''_{\boldsymbol{\emptyset}}  )^{\rm j}
\\
+ \tfrac{1}{4} (\rho''_{\mathbf{k}}  )^{\rm i }
\g_{\e_1,\e_2}(\hat\Lambda)_{\rm ij} (\rho''_{\mathbf{k}}  )^{\rm j}.
\end{multline}
In the above equation we introduced the following notation
\begin{equation}
\label{superdese_notation}
 (\rho''_{\mathbf{k}}  )^{\rm i}
\g_{\e_1,\e_2}(\hat\Lambda)_{\rm ij} (\rho''_{\mathbf{k}}  )^{\rm j}
\equiv  \vpiint\limits_{\mathbb{R}^2}
\mbox{d} x\mbox{d} y\ \r''_{\mathbf{a},\mathbf{k}}(x|\e_1,\e_2)
\g_{\e_1,\e_2}(x-y;\hat\La) \r''_{\mathbf{a},\mathbf{k}}(y|\e_1,\e_2).
\end{equation}
In what follows we are concerned with the form of the instanton partition
within the Nekrasov-Shatashvili limit i.e., when one of the deformation parameters tends to zero.
This limit of the Nekrasov partition function defines the effective twisted superpotential. Explicitly
\begin{equation}
\label{def_NS_limit}
\mathcal{W}^{\rm N_f=0, SU(N)}(\hat\Lambda,\mathbf{a},\epsilon_1)\;\equiv\;
-\lim_{\epsilon_2\to 0}
\epsilon_2\log \mathcal{Z}^{\rm N_f=0, SU(N)}(\hat\Lambda,\mathbf{a},\e_1,\e_2) ,
\end{equation}
where $\mathcal{W} = \mathcal{W}_{\rm pert} + \mathcal{W}_{\rm inst}$.
In the case under study this limit can be approached by taking
the thermodynamic limit with the number of instantons and simultaneously
squeezing the boxes in $\epsilon_2$ direction.
Let us consider the relation between the instanton number and
the density of instantons given in eq. \eqref{inst_numb_density}.
By definition this formula is satisfied by any partition of $k$.
Next, let us choose the one that corresponds to the colored Young
diagram with the highest column for $\a=1$ color index, i.e.,
$\mathbf{k}_0 \equiv\{ k_{\a,i} \}= \{\delta_{\a,1}\delta_{i,1} k\}$ and take the limit of
$\e_2 k$ while keeping the area under $\rho$ constant, i.e.,
\begin{equation}
\label{limit_e2}
\lim_{\substack{\e_2\to 0 \\ k\to \infty} }\e_2 k = -\frac{1}{2 \e_1}
\lim_{\substack{\e_2\to 0 \\ k\to \infty} }
\int\limits_{\mathbb{R}}\mbox{d} x\, \rho_{\mathbf{a},\mathbf{k}_0}(x|\e_1,\e_2)
= -\frac{1}{2 \e_1} \int\limits_{\mathbb{R}}\mbox{d} x\, \rho_{\mathbf{a},\boldsymbol{\w}_0}(x|\e_1)
= \omega .
\end{equation}
Within this limit columns of the diagram $\boldsymbol{\w}$ become
nonnegative real numbers and $\rho_{\mathbf{a},\boldsymbol{\w}}$
becomes a function of infinite many variables $\w_{\a,i}\in\mathbb{R}_{\leq 0}$.

With this picture in mind we would like to determine a squeezed colored Young diagram
upon which the weight in the partition function given in the deformed version of eq. \eqref{z_inst_hamiltonian}
yields a dominant contribution over the other summands. In order to find it,
let us expand the instanton Hamiltonian in $\e_2$ about zero, namely
\begin{subequations}
\begin{gather}
\label{hamilton_expansion}
H_{\rm inst}[\r_{\mathbf{a},\mathbf{k}}](\hat\Lambda,\e_1,\e_2)
= \sum_{g\geq 0}\e^{g-1}_2 c_g H^{(g)}_{\rm inst}[\r_{\mathbf{a},\boldsymbol{\w} }](\hat\Lambda,\e_1) ,
\\
\nonumber
c_0 = 1,\ c_1 = -\tfrac{1}{2},\ c_2 = \tfrac{1}{12},\ \dots ,
\end{gather}
where $H^{(g)}_{\rm inst}$ is defined in eq. \eqref{def_inst_hamilton} with the kernel $\g_{\e_1,\e_2}$ replaced
with the following one\footnote{$(s)_n \equiv \Gamma(n+s)/\Gamma(n)$.}
\begin{equation}
\label{gamma_expand_coeff}
\g^{(g)}_{\e_1}(x;\hat\Lambda)
= \e^{1-g}_1 \frac{\mbox{d}}{\mbox{d} s}\sbr{\cbr{\frac{\hat\Lambda}{\e_1}}^s
(s)_{g-1}\, \z\cbr{s +g-1,\tfrac{x}{\e_1}+1} }_{s=0}.
\end{equation}
\end{subequations}
>From explicit form of the expansion in eq. \eqref{hamilton_expansion} it is seen that
\begin{equation}
\label{def_zeroth_order}
H_{\rm inst}[\r_{\mathbf{a},\mathbf{k}}](\hat\Lambda,\e_1,\e_2)
\approx \frac{1}{\e_2} H^{(0)}_{\rm inst}[\r_{\mathbf{a},\boldsymbol{\w} }](\hat\Lambda,\epsilon_1)
\equiv \frac{1}{\e_2} \mathcal{H}[\r_{\mathbf{a},\boldsymbol{\w}}](\hat\Lambda,\epsilon_1).
\end{equation}
The squeezed Hamiltonian $\mathcal{H}[\r_{\mathbf{a},\boldsymbol{\w}}]$
takes the analogous form to the one in eq. \eqref{def_inst_hamilton}
with the following substitutions: $\r_{\mathbf{a},\mathbf{k}}(x|\e_1,\e_2) \to \r_{\mathbf{a},\boldsymbol{\w}}(x|\e_1) $,
$\g_{\e_1,\e_2}(x,\hat\La) \to \g_{\e_1}(x,\hat\La) \equiv \g^{(0)}_{\e_1}(x,\hat\La)$.
It can also be cast into the form that proves useful in what follows, namely\footnote{To find
this form the identity has been used:
$\g_{\e_1}(x;\hat\Lambda) = -\e_1\log\big(\hat\Lambda/\e_1\big)\z(-1,x/\e_1+1) + \g_{\e_1}(x;\e_1)$.}
\begin{equation}
\label{squeezed_hamiltonian_log_exposed}
\begin{split}
\mathcal{H}[\r_{\mathbf{a},\boldsymbol{\w}}](\hat\Lambda,\epsilon_1)
=& \frac{N}{\e_1}\log\left(\frac{\hat\Lambda}{\epsilon_1}\right)  \sum(\rho_{\boldsymbol{\omega} })^{\rm i}
+ \tfrac{1}{4} (f''_{\boldsymbol{\emptyset}} )^{\rm i}
\g_{\e_1}(\epsilon_1)_{\rm ij} (\rho''_{\boldsymbol{\omega} } )^{\rm j}
+ \tfrac{1}{4} (\rho''_{\boldsymbol{\omega} }  )^{\rm i}
\g_{\e_1}(\epsilon_1)_{\rm ij} (f''_{\boldsymbol{\emptyset}}  )^{\rm j}
\\
& + \tfrac{1}{4} (\rho''_{\boldsymbol{\omega} }  )^{\rm i}
\g_{\e_1}(\epsilon_1)_{\rm ij} (\rho''_{\boldsymbol{\omega} }  )^{\rm j},
\end{split}
\end{equation}
where
$$
\sum(\rho_{\boldsymbol{\omega} })^{\rm i} \equiv
\int\limits_\mathbb{R}\mbox{d}x\, \rho_{\mathbf{a},\boldsymbol{\omega} }(x|\epsilon_1).
$$
It is expected that the sum over all profiles and hence over all instanton densities
for a large $k$ approaches the path integral over the latter. This can be understood through the
analysis of number of configurations accessible for $k$ instantons, i.e., the dimension
of the configuration space $\Omega$ for a given $k$. Since the number of partitions
runs with $k$ like $p(k)\sim \exp(\pi \sqrt{2 k/3} - \log k)$,
which stems from the Hardy-Ramanujan theorem,
the sum over colored partitions runs with $k$ like ($k = \sum \Bbbk_\alpha$)\footnote{
$c(k,N)$ is defined as the number of possibilities for $k$ to be partitioned into $N$ nonnegative, integer
parts. This number equals $c(k,N) = (k+N-1)/k!(N-1)!$ .}
$$
\dim \Omega_k = \sum_{r=1}^{c(k,N)}\prod_{\alpha =1}^N p(|\Bbbk^r_\alpha|)
\stackrel{k\to \infty}{\sim}
\sum_{r=1}^{k^N/(N-1)!}\exp\left\{\pi\sum_{\alpha=1}^N\sqrt{\tfrac{2}{3} |\Bbbk^r_\alpha| }  \right\}.
$$
This estimation shows that the dimension grows very fast with $k$ such that the discreet
distribution of configurations approaches the continuous one. Therefore, the sum over densities may be
approximated by the functional integral over $\{\boldsymbol{\w}_i\}_{i\in \mathbb{N} }
\in \boldsymbol{\mathsf{c}}_0(\mathbb{R}_{\leq 0})^N$ i.e.,
the space of sequences monotonically convergent to zero. Hence,
\begin{equation}
\label{path_integral}
\mathcal{Z}^{\rm N_f=0, SU(N)}_{\rm inst}(\Lambda,\mathbf{a},\e_1,\e_2)
\stackrel{\substack{k\to\infty \\ \epsilon_2\to 0 }}{\sim}
\int\limits_{\boldsymbol{\mathsf{c}}_0(\mathbb{R}_{\leq 0})^N}
\prod\limits_{\substack{\a,i }}\mbox{d}\omega_{\a,i} \
\mbox{e}^{-\frac{1}{\e_2}\mathcal{H}[\r_{\mathbf{a},\boldsymbol{\omega}}] } ,
\end{equation}
where $\w_{\a,i} = x_{\a,i} - x^0_{\a,i}$.

\subsection{Saddle point equation}

We would like to determine the extremal density function which makes the weight in the
path integral in eq. \eqref{path_integral} the dominant contribution. In order to do this
let us consider two Young diagrams that differ in small fluctuations within each column, i.e.,
$$
\tilde \w_{\a,i} = \w_{\a,i} + \e_2 \d\w_{\a,i}
\quad
\Rightarrow
\quad
\r_{\mathbf{a},\tilde{\boldsymbol{ \w}} }(x)
= \r_{\mathbf{a},\boldsymbol{\w}}(x) + \e_2 \d\r_{\mathbf{a},\boldsymbol{\w}}(x) ,
$$
where\footnote{The signum function is defined as $\mbox{sgn}(x) \equiv |x|/x
=\sum\limits_{s\in\mathbb{Z}_2} s\theta(s x)$.}
\begin{equation}
\label{instanton_fluct}
\d\r_{\mathbf{a},\boldsymbol{\w}}(x)
= \sum_{\a=1}^N\sum_{i\geq 1}
\sbr{\mbox{sgn}(x-x_{\a,i} ) -\mbox{sgn}(x-x_{\a,i} - \e_1) }\d\w_{\a,i}.
\end{equation}
In this case the Hamiltonian \eqref{squeezed_hamiltonian_log_exposed} will suffer the change
$$
\mathcal{H}[\r_{\mathbf{a},\tilde{\boldsymbol{ \w}} } ] =
\mathcal{H}[\r_{\mathbf{a},\boldsymbol{\w}} ]
+ \e_2 \frac{\d \mathcal{H}[\r_{\mathbf{a},\boldsymbol{\w}}]}{\d \r^{\rm i}_{\mathbf{a},\boldsymbol{\w}} }
\d \r^{\rm i}_{\mathbf{a},\boldsymbol{\w}} + \ord{\e_2^2} .
$$
For some stationary point $\rho_{\ast\mathbf{a},\boldsymbol{\w}}$
\begin{equation}
\label{def_saddle_pt_eq}
\frac{\d \mathcal{H}[\r_{\ast\,\mathbf{a},\boldsymbol{\w}}]}{\d \r^{\rm i}_{\mathbf{a},\boldsymbol{\w}} }
\d \r^{\rm i}_{\mathbf{a},\boldsymbol{\w}}
\equiv
\int\limits_{\mathbb{R}}\mbox{d} x\
\frac{\d \mathcal{H}[\r_{\ast\,\mathbf{a},\boldsymbol{\w}}]}{\d \r_{\mathbf{a},\boldsymbol{\w}}(x) }
\d \r_{\mathbf{a},\boldsymbol{\w}}(x) = 0,
\quad \r_{\ast\,\mathbf{a},\boldsymbol{\w}}(x) = \r_{\mathbf{a},\boldsymbol{\w}_\ast}(x) ,
\end{equation}
where $\boldsymbol{\w}_\ast$ is the extremal of the colored Young diagram.
Hence, within the limit $\epsilon_2\to 0$ the only term that is left from eq, \eqref{path_integral}
reads
\begin{equation*}
\mathcal{Z}^{\rm N_f=0, SU(N)}_{\rm inst}(\Lambda,\mathbf{a},\e_1,\e_2)
\sim \exp\pbr{-\frac{1}{\e_2}\mathcal{H}[\r_{\ast\,\mathbf{a},\boldsymbol{\w}}](\hat\Lambda,\epsilon_1)  } ,
\end{equation*}
and by virtue of eq. \eqref{def_NS_limit} we obtain
\begin{equation}
\label{W_and_H}
\mathcal{W}^{\rm N_f=0,SU(N)}_{\rm inst}(\hat\Lambda,\mathbf{a},\epsilon_1)
= \mathcal{H}[\r_{\ast\,\mathbf{a},\boldsymbol{\w}}](\hat\Lambda,\epsilon_1).
\end{equation}
Thus, in order to obtain the form of the twisted superpotential we must find the explicit
form of the Hamiltonian at the extremal solving the saddle point equation \eqref{def_saddle_pt_eq}.
Its functional form reads
\begin{multline}
\label{saddle_point_eq}
\frac{N}{\e_1}\log\left(\frac{\hat\Lambda}{\epsilon_1}\right)
\int\limits_{\mathbb{R}} \mbox{d}x\,x\,\d\r'_{\mathbf{a},\boldsymbol{\w}}(x)
\\
+\tfrac{1}{4} \vpiint\limits_{\mathbb{R}^2}
\mbox{d} x\mbox{d} y \
\Big[ f''_{\mathbf{a},\boldsymbol{\emptyset} }(x)\pt_y\g_{\e_1}(x-y)\d \r'_{\mathbf{a},\boldsymbol{\w}}(y)
+ \d \r'_{\mathbf{a},\boldsymbol{\w}}(y)\pt_y\g_{\e_1}(y-x) f''_{\mathbf{a},\boldsymbol{\emptyset}}(x)
\\
+\r''_{\mathbf{a},\boldsymbol{\w} }(x)\pt_y\g_{\e_1}(x-y) \d \r'_{\mathbf{a},\boldsymbol{\w}}(y)
+ \d \r'_{\mathbf{a},\boldsymbol{\w}}(y)\pt_y\g_{\e_1}(y-x) \r''_{\mathbf{a},\boldsymbol{\w}}(x)
\Big] = 0.
\end{multline}
The above equation for critical density is directly related to the equation on critical
squeezed Young diagram. In order to find the latter one must express eq. \eqref{saddle_point_eq}
in terms of $x_{\alpha,i}$. After some algebra we find
\begin{multline}
0=\sum_{\a,i}
\sbr{2N \log\left(\frac{\hat\Lambda}{\epsilon_1}\right)
- \sum_{\b=1}^N\log\cbr{\frac{(a_\b -x_{\a ,i} -\e_1 )(x_{\a ,i} - a_\b)}{\epsilon_1^2} }
}\d\w_{\a,i}
\\
+\sum_{(\a,i)\neq(\b,j)}
\log\left(\frac{x_{\a,i} -  x_{\b,j} - \e_1}{x_{\a,i} -x_{\b,j} + \e_1} \right)
\d\w_{\a,i}
+\sum_{\a,i;\b,j}
\log\left(\frac{x_{\a,i}  -  x^0_{\b,j} + \e_1}{x_{\a,i} - x^0_{\b,j} - \e_1} \right)
\d\w_{\a,i}  .
\end{multline}
The ''off-diagonal'' sum comes from necessity of taking the principal values of integrals
in the last line of eq. \eqref{saddle_point_eq} which is due to
the singularity of $\gamma'_{\epsilon_1}(x) \sim \log\Gamma(x) $ at zero.
In order to put the sums on the same footing we add zero to the above equation in the form
that completes the ''off-diagonal'' sum with the missing diagonal term, i.e.,
$$
\sum_{\a,i}
\left[
\log\left(\frac{x_{\a,i} -  x_{\a,i} - \e_1}{x_{\a,i} -x_{\a,i} + \e_1} \right)  - \sqrt{-1}\,\pi
\right] \d\w_{\a,i}.
$$
The last step and the arbitrariness of the fluctuations $\d\omega_{\alpha,i}$
enable us to cast the saddle point equation into the form first
found in ref. \cite{Poghossian:2010pn}. Namely,
\begin{equation}
\label{saddle_point_eq_final}
(-1)^{N}(\hat\Lambda/\epsilon_1)^{2N}
\prod_{\b=1}^N\prod_{j \geq 1}\frac{(x_{\a,i}-x_{\b,j}-\e_1)(x_{\a,i}-x^0_{\b,j}+\e_1)}
{(x_{\a,i}-x_{\b,j}+\e_1)(x_{\a,i}-x^0_{\b,j}-\e_1)}
\prod_{j=1}^2\frac{\e_1}{x_{\a,i}-x^0_{\b,j}+\e_1} = -1  .
\end{equation}
The equation \eqref{saddle_point_eq_final} in fact represents an infinite
set of equations for the critical colored squeezed Young diagram.
As one may infer from the above formula, the infinite sequence of columns of the critical diagram with
a given color must be strictly decreasing in the sense of absolute value. In order to see this,
consider the factor in the denominator that takes the form $x_{\a,i}-x_{\b,j}+\e_1$ and
for $\b=\a$ choose the neighboring columns with $i=j-1$. For this specific choice just mentioned formula
amounts to $\w_{\a,j-1} - \w_{\a,j}$ abstract value of which, by definition of partitions, must be nonnegative.
In fact, it must be strictly positive, because it is a factor of the product of terms in the denominator.
Therefore, the critical diagram consists of columns each of which has the multiplicity at most one
and as such form a \emph{strictly monotone null sequence of negative real numbers},
i.e., $\{\boldsymbol{\w}_{\ast\,i}\}_{i\in\mathbb{N}}\in \boldsymbol{\mathsf{c}}_0(\mathbb{R}_{< 0})^N.$

\subsection{Solution to saddle point equation}

The saddle point equation \eqref{saddle_point_eq_final} can be solved iteratively
by means of the method first proposed by Poghossian \cite{Poghossian:2010pn}.
Noticing, that it depends on the only free parameter $(\hat\Lambda/\epsilon_1)^{2N}$,
the columns, that are sought quantities, can be expressed
in terms of powers of $(\hat\Lambda/\epsilon_1)^{2N}$. The advantage of this particular form
of the saddle point equation is that it does not change if we introduce an explicit cutoff length
for each of $N$ constituents of the colored Young diagram regardless of whether
this cutoff is colored or common for every $\alpha$.
Therefore, the first step is to introduce a common cutoff length
$L$ that starts from $1$ and increases along with each step of iteration,
such that $|\omega_{\alpha,i}|>0$ for $i\in\{1,\dots,L\}$ and assumes zero value elsewhere.
Furthermore, the ansatz one starts with takes the form
\begin{equation}
x^{(L)}_{\alpha,i} = x^0_{\a,i} + \d x^{(L)}_{\alpha,i},
\quad
\d x^{(L)}_{\alpha,i} \equiv x^{(L)}_{\alpha,i} - x^0_{\a,i} = \omega_{\ast\,\alpha, i}^{(L)}.
\end{equation}
The form of the above ansatz can be justified with the aid of the following argument. Let us first note
that by construction
\begin{equation}
\label{def_L_corrections}
\omega_\ast = \sum_{\a=1}^N \sum_{i \geq 1}\omega_{\ast\, \alpha,i}
= \sum_{i \geq 1}\left( \sum_{\a=1}^N \d x_{\alpha,i}\right)
= \lim_{L\to\infty}\sum_{i = 1}^L \sum_{\a=1}^N \d x^{(L)}_{\alpha,i}.
\end{equation}
On the other hand from eq. \eqref{squeezed_hamiltonian_log_exposed} we get
\begin{equation*}
-\w_\ast = (\hat\Lambda/\epsilon_1)^{2N}\frac{\mbox{d} \mathcal{H}[\rho_{\ast\, \mathbf{a},\boldsymbol{\omega}}] }
{\mbox{d} (\hat\Lambda/\epsilon_1)^{2N}}
= (\hat\Lambda/\epsilon_1)^{2N}\frac{\partial \mathcal{H}[\rho_{\ast\, \mathbf{a},\boldsymbol{\omega}}] }
{\partial (\hat\Lambda/\epsilon_1)^{2N}} ,
\end{equation*}
where we used the saddle point equation in the form given in eq. \eqref{def_saddle_pt_eq}.
Employing the definition of the Nekrasov-Shatashvili limit \eqref{def_NS_limit}
we can formally expand the right hand side of eq. \eqref{def_NS_limit}, and due to the established
relationship between the superpotential $\mathcal{W}_{\rm inst}$ and $\mathcal{H}$ in
eq. \eqref{W_and_H} we find\footnote{The relationship of the coefficients $W_i$ in the last formula with those introduced earlier in
eq. \eqref{W_Lambda_expansion} is $W_i/(i\,\e_1^{2Ni})=\mathcal{W}_i$.}
\begin{equation}
\label{relation_omega_W}
-\w_\ast = (\hat\Lambda/\epsilon_1)^{2N}
\frac{\partial \mathcal{W}_{\rm inst}(\hat\Lambda,\mathbf{a},\epsilon_1) }{\partial (\hat\Lambda/\epsilon_1)^{2N}}
=\sum_{i\geq 1} (\hat\Lambda/\epsilon_1)^{2N i}\, W_i(\mathbf{a},\e_1) .
\end{equation}
The above equation and eq. \eqref{def_L_corrections} imply
that $\sum\limits_\alpha \delta x^{(L)}_{\alpha,i} \sim \mathcal{O}((\hat\Lambda/\epsilon_1)^{2N i})$.
Hence,
\begin{equation}
\label{omega_coefficients}
\d x^{(L)}_{\alpha,i} =  \w^{(L)}_{\ast\,\alpha,i}\big(\hat\Lambda/\epsilon_1,\mathbf{a},\e_1\big)
= \sum_{j=i}^L(\hat\Lambda/\epsilon_1)^{2N j}\, \omega_{\alpha, i, j}(\mathbf{a},\e_1)\,  .
\end{equation}
Thus, the iterative solution of eq. \eqref{saddle_point_eq_final} amounts to
finding the coefficients $\omega_{\alpha, i, j}$. Their forms for the gauge symmetry group $SU(N)$ and $N=2$
with the above ansatz up to $j= 3$ are gathered in appendix \ref{app_speq_sols}.

The generic case when $L\to \infty$ can be examined through the saddle point equation
given in a different form. Making use of identities given in appendix \ref{app_ids}
the equation \eqref{saddle_point_eq_final} assumes yet another form
\begin{equation}
\label{saddle_point_eq_infinite}
\left(\frac{\hat\Lambda}{\epsilon_1}\right)^{2N}
\frac{Y(x_{\a,i} -\e_1 ) }{Y(x_{\a,i} + \e_1 )}
= (-1)^{N-1}  ,
\end{equation}
where the function $Y(z)$ is defined in eq. \eqref{app_def_Y}.
Since $x_{\a,i}$ are zeros of this function the saddle point equation in the form of eq. \eqref{saddle_point_eq_final}
is traded for the equations on zeros themselves.

\subsection{Mathieu eigenvalue -- contour integral representation}

The iterative solution of the saddle point equation \eqref{saddle_point_eq_final}
enables us to express the effective twisted superpotential in terms of the expansion
parameter $(\hat\Lambda/\e_1)^{2N}$ as a sum over columns of
the critical Young diagram. This can be formally accomplished by means of
the formula \eqref{relation_omega_W} and integrating it term by term with respect to
the expansion parameter. Explicitly
\begin{equation}
\label{W_in_cols}
\mathcal{W}_{\rm inst}^{\rm N_f=0, SU(N)}(\hat\Lambda,\mathbf{a},\epsilon_1)
= -\int\limits_0^{(\hat\Lambda/\e_1)^{2N}}\frac{\mbox{d} q}{q}\, \w_\ast(q)
= -\sum_{i\geq 1} \int\limits_0^{(\hat\Lambda/\e_1)^{2N}}
\mbox{d} q\, q^{i-1}\left(\sum_{\a=1}^N\sum_{j=1}^i\w_{\a,j,i}\right) .
\end{equation}
In the above expression we employed eq. \eqref{omega_coefficients} where the coefficients $\w_{\a,j,i}$
are introduced. For the gauge group with $N=2$ which is the case in question, the superpotential
was computed up to the third order in the expansion parameter. Namely, making use of the following formula ($\sum a_\a =0$)
$$
\mathcal{W}_{\rm inst}^{\rm N_f=0, SU(2)}(\hat\Lambda,a,\epsilon_1)
= \sum_{i\geq1} (\hat\Lambda/\e_1)^{4i}\, W_i(a,\e_1)/i\ ,
\qquad W_i = -\sum_{\a=1}^2\sum_{j=1}^i\w_{\a,j,i}\ .
$$
and the results in appendix \ref{app_speq_sols} the coefficients of the expansion written in the form that is
suitable for comparison read
\begin{eqnarray}
\nonumber
\frac{1}{\e_1^4}\,W^{\rm N_f=0, SU(2)}_1 &=& -\frac{2}{\epsilon_1^3 - 4a^2 \epsilon _1}\ ,
\\
\label{W_in_cols_coefficients}
\frac{1}{2\,\e_1^8}\,W^{\rm N_f=0, SU(2)}_2 &=& \frac{20 a^2+7 \epsilon _1^2}{4
   \left(\epsilon _1^2-4 a^2\right)^3
   \left(\epsilon _1^3-a^2 \epsilon
   _1\right)}\ ,
\\
\nonumber
\frac{1}{3\,\e_1^{12} }\,W^{\rm N_f=0, SU(2)}_3 &=& -\frac{4 \left(144 a^4+29 \left(8 a^2
   \epsilon _1^2+\epsilon
   _1^4\right)\right)}{3 \left(\epsilon
   _1^2-4 a^2\right)^5 \left(4 a^4
   \epsilon _1-13 a^2 \epsilon _1^3+9
   \epsilon _1^5\right)}\ .
\end{eqnarray}
As it may be observed these coefficients coincide
with those in appendix \ref{AppIII} ($W_i/(i\,\e_1^{2Ni})=\mathcal{W}_i$).
The above results can be compared with coefficients of the irregular classical block.
The relationship between coefficients of the latter with those of twisted superpotential
is given in eq. \eqref{coeffs_f_W_comparison}. In terms of $W_i$ this relationship reads
\begin{equation}
\label{f_critcolW_relationship}
\frac{1}{i\,\epsilon_1^{4 i}}\, W^{\rm N_f=0, SU(2)}_i(a,\e_1)
= -\frac{1}{\epsilon_1^{4 i}}\, f^{i}_{\frac{1}{4}-\frac{a^2}{\epsilon_{1}^{2}}}.
\end{equation}
Coefficients of the classical irregular block given in eq. \eqref{sec2_block_coeffs}
expressed in terms of instanton parameters $a,\,\e_1$ read
\begin{eqnarray}
\nonumber
f^{1}_{\frac{1}{4}-\frac{a^2}{\epsilon_{1}^{2}}} &=&\frac{2 \epsilon _1^2}{\epsilon _1^2-4 a^2}
,
\\[5pt]\label{f_coeffs_in_a_and_e1}
f^{2}_{\frac{1}{4}-\frac{a^2}{\epsilon_{1}^{2}}} &=&-\frac{\epsilon _1^6 \left(20 a^2+7 \epsilon _1^2\right)}
{4 \left(\epsilon _1^2-4 a^2\right)^3\left(\epsilon _1^2-a^2\right)}
,
\\[5pt]\nonumber
f^{3}_{\frac{1}{4}-\frac{a^2}{\epsilon_{1}^{2}}} &=&\frac{4 \epsilon _1^{10} \left(144 a^4+29 \left(8 a^2 \epsilon _1^2+\epsilon
   _1^4\right)\right)}{3 \left(\epsilon _1^2-4 a^2\right)^5 \left(4 a^4-13 a^2 \epsilon _1^2+9
   \epsilon _1^4\right)}
.
\end{eqnarray}
Heaving these coefficients in the explicit form it is now straightforward
to check the relationship with coefficients of the superpotential
stated in eq. \eqref{f_critcolW_relationship} where the latter are expressed in terms of the
critical Young diagram in eq. \eqref{W_in_cols_coefficients}. Although
just verified mathematical correspondence between the two contextually different
theories was performed up to the third order in expansion parameter $\hat\Lambda\sim \mu\,\mbox{e}^{\log(q)/(2N)}$,
it is by induction reasonable to assume that the range of validity extends
to all orders which was phrased earlier in eq. \eqref{coeffs_f_W_comparison}.
The last statement should be yet proved. Nevertheless,
the observed relationship, with aforementioned caveat, may be regarded as
an extension of the non-conformal limit of AGT correspondence found
in ref. \cite{Marshakov:2009} to the classical level.

The classical non-conformal AGT relation enables one to associate the eigenvalue of
the Mathieu operator expressed in terms of the classical block in eq. \eqref{eigenvalue}
with instanton parameters through the twisted superpotential. Let us recall at this point that
according to the Nekrasov-Shatashvili's conjecture put forward in \cite{NS:2009} (the Bethe/gauge correspondence)
for any $\mathcal{N}=2$ gauge theory there exists a quantum integrable system whose
Hamiltonians are given in terms of gauge independent functions of operators $\mathscr{O}_k(x)$
that form the so-called twisted chiral ring, and enumerate vacua of the relevant gauge theory.
These operators can be chosen as the traces of the lowest components of vector multiplet.
Their expectation values evaluated at the minimum of the (shifted) superpotential correspond to the spectrum
of Hamiltonians of the integrable system. Namely \cite{Nekrasov:Okounkov:2003}
\begin{equation*}
\mathscr{E}_k = \langle \mathscr{O}_k(x) \rangle = \langle \mbox{tr}\phi^k \rangle_{\mathbf{a}}\big|_{\e_2=0}
=\frac{1}{2}\int\limits_\mathbb{R}\mbox{d}x\, x^k f''_{\mathbf{a},\boldsymbol{\emptyset} }(x)
+ \frac{1}{2}\int\limits_\mathbb{R}\mbox{d}x\, x^k \rho''_{\mathbf{a},\boldsymbol{\w}_\ast }(x|\e_1),
\end{equation*}
where the first term on the right hand side is the classical part.
In the case under consideration the gauge group is $SU(2)$ and the
relevant quantum integrable system is the two-particle periodic Toda chain.
Its two Hamiltonians have the following spectrum: $\mathscr{E}_1 = 0$ and
\begin{equation}
\label{eigenvalue_from_gauge_th}
\begin{aligned}
\mathrm{E} = 2\mathscr{E}_2
=&\, 4 a^2 + 2\int\limits_\mathbb{R}\mbox{d}x\,\rho_{\mathbf{a},\boldsymbol{\w}_\ast }(x|\e_1)
= 4 a^2 -  4\epsilon_1\sum_\alpha\sum_{i\geq 1}\omega_{\ast\, \alpha, i}\big(\hat\Lambda/\epsilon_1\big)
\\
=&\, \epsilon_1 \hat\Lambda
\partial_{\hat\Lambda} \mathcal{W}^{\rm N_f=0, SU(2)}(\hat\Lambda,\mathbf{a},\epsilon_1)
= \epsilon^2_1\lambda\, ,
\end{aligned}
\end{equation}
where in the second line we accounted for the result \eqref{relation_omega_W} as well as
the perturbative part given in eq. \eqref{def_full_partition_func}
with the gamma function traded for the one for $g=0$ order of the expansion
in the limit $\epsilon_2\to 0$ given in eq. \eqref{gamma_expand_coeff}
that we signified as $\gamma_{\e_1}(x;\hat\Lambda)$. Explicitly,
the perturbative part reads
\begin{equation*}
\mathcal{W}^{\rm N_f=0, SU(2)}_{\rm pert}(\hat\Lambda,\mathbf{a},\epsilon_1)
= \sum_{\a,\b=1}^N \gamma_{\e_1}(a_\a - a_\b;\hat\Lambda)
= \frac{4}{\epsilon _1}\log\left( \frac{\hat\Lambda}{\epsilon _1} \right) \left( a^2+\frac{\epsilon^2 _1}{12}  \right).
\end{equation*}
It is worth noting that the relationship between the spectrum of the second Hamiltonian $E$, twisted superpotential $\mathcal{W}$ and
Mathieu eigenvalue $\lambda$ stated in eq. \eqref{eigenvalue_from_gauge_th} coincides with the one deduced from
the null vector decoupling in eq. \eqref{EnergyWKB}, which is the CFT side of classical AGT correspondence.

The formula \eqref{eigenvalue_from_gauge_th} can also be given yet another, concise form
in terms of the $Y_{\mathbf{a},\boldsymbol{\w}}(z)$ function which does not refer to the iterative expansion of $\mathcal{W}$.
Namely, noting that
\begin{displaymath}
\tfrac{1}{2}\int\limits_\mathbb{R}\mbox{d}x\, x^n \rho''_{\mathbf{a},\boldsymbol{\w}_\ast }(x|\e_1)
= -\oint\limits_C\frac{\mbox{d} z}{2\pi i}\left[(z+\e_1)^n - z^n \right]\partial_z\log\frac{Y(z)}{Y_0(z)},
\end{displaymath}
and
\begin{displaymath}
\tfrac{1}{2}\int\limits_{\mathbb{R}}\mbox{d}x\, x^n f''_{\mathbf{a},\boldsymbol{\emptyset} }(x)
= \oint\limits_C\frac{\mbox{d} z}{2\pi i} z^n \partial_z \log\frac{Y_0( z - \e_1)}{Y_0(z)},
\end{displaymath}
we obtain
\begin{equation}\label{konturowa}
\mathrm{E}
= -2 \oint\limits_C \frac{\mbox{d}z}{2\pi i}\, z^2 \partial_z
\log \frac{Y(z - \e_1)}{Y(z)},
\end{equation}
where the contour $C$ encloses all points $x^\ast_{\a,i}$ and $x^0_{\a,i}$ which
are zeros of $Y(z)$ and $Y_0(z)$, respectively.
The latter functions solve eq. \eqref{saddle_point_eq_infinite}.

\section{Conclusions}
\label{sec5}
In this paper we have postulated the existence of the classical irregular block
$f_{\delta}(\hat\Lambda/\epsilon_1)$.
Calculations presented in this work provide a convincing evidence that the classical
limit of the quantum irregular block (\ref{ClIrr}) exists yielding a consistent definition
of the function $f_{\delta}(\hat\Lambda/\epsilon_1)$. In particular, this hypothesis is strongly
supported by the tests described in subsection \ref{S22} (see fig.~\ref{fig:1}).

As an implication of the conjectured semiclassical asymptotic (\ref{ClAsymp}) the classical
irregular block enters the expression (\ref{eigenvalue}) for the Mathieu eigenvalue $\lambda$.
Indeed, we have checked that the formula (\ref{eigenvalue}) reproduces well known
week coupling (small $h^2$) expansion of $\lambda$.\footnote{Work is in progress in
order to verify wether calculations performed in section 3
pave the way for working out really new methods for calculating corresponding eigenfunctions.}
Hence, the existence of the function
$f_{\delta}(\hat\Lambda/\epsilon_1)$ seems to be well confirmed {\it a posteriori}, i.e.
by consequences of its existence. Let us stress that
although the classical limit of the null vector decoupling equation for the degenerate 3-point
irregular block was discussed before in \cite{Maruyoshi:2010} the expression
(\ref{eigenvalue}) has not  appeared in the literature so far.

The existence of the classical irregular block is also apparent from the semiclassical limit of the
``non-conformal" AGT relation. Two checks have been performed in the present work in order to show that
the function $f_{\delta}(\hat\Lambda/\epsilon_1)$ corresponds to the instanton twisted superpotential
${\cal W}_{\rm inst}^{{\rm N_f}=0, {\rm SU(2)}}$ of the ${\cal N}=2$ $SU(2)$ pure gauge SYM theory.
One of these checks (see section 4) is highly non-trivial. It employs a representation of
${\cal W}_{\rm inst}^{{\rm N_f}=0, {\rm SU(2)}}$ as a critical value of the gauge theory ``free energy''.
The criticality condition or equivalently the saddle point
equation takes the form of the Bethe-like equation. This equation can be solved by a power
expansion in $\hat\Lambda/\epsilon_1$ (see appendix~\ref{app_speq_sols}). Solution to this equation (or
in fact a system of equations) describes a shape of the two critical Young diagrams extremizing
the ``free energy''. Finally, the latter evaluated on the ``critical configuration'' and expanded in
$\hat\Lambda/\epsilon_1$ yields $f_{\delta}(\hat\Lambda/\epsilon_1)$
provided that certain relations between parameters are assumed. As a by-product, the identity (\ref{classAGT})
and the formula (\ref{EnergyWKB}) imply another new expression for the Mathieu eigenvalue $\lambda$,
namely, as a sum of columns' lengths 
in critical Young diagrams, cf.~(\ref{eigenvalue_from_gauge_th}).\footnote{
See also appendix \ref{app_WKB}, where has been shown an exact matching between this result and
that obtained by means of the WKB analysis.}
Moreover, as was noticed in \cite{Poghossian:2010pn}, once critical columns' 
lengths are known in a
closed form such sum can be rewritten in terms of
the contour integral with integrand built out with certain special functions.
We have re-derived this statement on the gauge theory side within the formalism of profile functions
of the Young diagrams, cf.~(\ref{konturowa}). The latter and (\ref{eigenvalue_from_gauge_th})
imply contour integral representation of $\lambda$.

Let us interpret results of this paper within the context of the triple correspondence
\begin{equation}\label{triple}\boxed{
2d\;{\rm CFT}\;\Big/\;2d\;{\cal N}=2\;SU(N)\,{\rm SYM}\;
\Big/\;N-{\rm particle}\;{\rm QIS}}
\end{equation}
mentioned in the introduction. In our case we have $N\!=\!2$. On the
conformal field theory side we have found the formula for the Mathieu eigenvalue $\lambda$
expressed in terms of the classical irregular block. Its analogue on the gauge theory side
is given by ${\cal W}_{\rm inst}^{{\rm N_f}=0, {\rm SU(2)}}$ and realized here as v.e.v.
of certain gauge theory operator (see section 4). Then, according to the Bethe/gauge correspondence
$\lambda$ is nothing but an eigenvalue of the corresponding integrable system 2-particle Hamiltonian
(see appendix \ref{app_WKB}). An interesting further line of research is to extent this observation
going beyond $N=2$. In such a case the AGT duality is extended
to the correspondence between the $2d$ conformal Toda theory and the $4d$ ${\cal N}=2$ $SU(N)$
gauge theories \cite{Wyllard:2009hg}. Then, in order to get examples of (\ref{triple}) one has to
study classical limit of the $W_N$--symmetry conformal blocks.\footnote{Interestingly, there
is also a need for the classical $W_N$--blocks in the study of the entanglement entropy within
the ${\rm AdS}_3/2d\;{\rm CFT}$ holography \cite{Perlmutter:2013paa}.}

\section*{Acknowledgments}
The support of the Polish National Center of Science,
scientific project No. N N202 326240, is gratefully acknowledged.

\newpage

\appendix

\section{Appendices}
\subsection{Expansion coefficients of \texorpdfstring{$2d$}{2d} CFT and gauge theory functions}
\label{App0}

\textbf{Gram matrix.}

\medskip\noindent
$n=1$: $\lbrace L_{-1}|\,\nu_\Delta\,\rangle \rbrace$,
\begin{eqnarray*}
G^{n=1}_{c,\Delta} = \langle L_{-1}\nu_\Delta\,|\,L_{-1}\nu_\Delta\rangle
=\langle\nu_\Delta\,|\,L_{1}L_{-1}\nu_\Delta\rangle = 2\Delta,
\end{eqnarray*}

\bigskip\noindent
$n=2$: $\lbrace L_{-2}|\,\nu_\Delta\,\rangle,L_{-1} L_{-1}|\,\nu_\Delta\,\rangle\rbrace$,
\begin{equation*}
G^{n=2}_{c,\Delta} =
\begin{pmatrix}
\langle L_{-2}\nu_\Delta\,|\,L_{-2}\nu_\Delta\rangle  &  \langle L_{-1}^2\nu_\Delta\,|\,L_{-2}\nu_\Delta\rangle \\
\langle L_{-2}\nu_\Delta\,|\,L_{-1}^{2}\nu_\Delta\rangle & \langle L_{-1}^{2}\nu_\Delta\,|\,L_{-1}^{2}\nu_\Delta\rangle  \\
\end{pmatrix}
=
\begin{pmatrix}
    \frac{c}{2}+4\Delta  & 6\Delta \\
    6\Delta & 4\Delta(2\Delta+1) \\
\end{pmatrix},
\end{equation*}

\bigskip\noindent
$n=3$: $\lbrace L_{-3}|\,\nu_\Delta\,\rangle,
L_{-2}L_{-1}|\,\nu_\Delta\,\rangle, L_{-1}L_{-1}L_{-1}|\,\nu_\Delta\,\rangle\rbrace$,
\begin{equation*}
G^{n=3}_{c,\Delta}=
  \begin{pmatrix}
    2c+6\Delta  & 10\Delta & 24\Delta \\
    10\Delta & \Delta(c+8\Delta+8) & 12\Delta(3\Delta+1) \\
    24\Delta & 12\Delta(3\Delta+1) & 24\Delta(\Delta+1)(2\Delta+1) \\
  \end{pmatrix},
\end{equation*}

\noindent
$\ldots\;\;\;\;$.

\bigskip\noindent
\textbf{Coefficients of the quantum irregular block.}
\begin{eqnarray*}
&& \left[G^{1}_{c,\Delta}\right]^{(1) (1)}=\frac{1}{2\Delta},
\\[13pt]
&&\left[G^{2}_{c,\Delta}\right]^{(1^2) (1^2)}=\frac{c+8 \Delta }{4 \Delta Ê(2 c \Delta +c+2 \Delta Ê(8 \Delta -5))},
\\[13pt]
&&\left[G^{3}_{c,\Delta}\right]^{(1^3) (1^3)}=
\frac{(11 c-26) \Delta +c (c+8)+24 \Delta ^2}{24 \Delta Ê\left((c-7) \Delta +c+3 \Delta ^2+2\right)
 Ê \left(2 (c-5) \Delta +c+16 \Delta ^2\right)},
\\
&&\ldots\;\;\;\; .
\end{eqnarray*}

\noindent\bigskip
\textbf{Coefficients of the quantum 4-point block on the sphere.}
\begin{eqnarray*}
{\cal F}^{\,1}_{c,\Delta}\!\left[_{\Delta_{4}\;\Delta_{1}}^{\Delta_{3}\;\Delta_{2}}\right]
&=&\frac{(\Delta+\Delta_3-\Delta_4)(\Delta+\Delta_2-\Delta_1)}{2\Delta},
\\
{\cal F}^{\,2}_{c,\Delta}\!\left[_{\Delta_{4}\;\Delta_{1}}^{\Delta_{3}\;\Delta_{2}}\right]
&=&
\Big[
\left(4\Delta(1+2\Delta)\right)^{-1}
(\Delta-\Delta_1 +\Delta_2)
(1+\Delta-\Delta_1+\Delta_2)
\\
&\times &
(\Delta+\Delta_3 -\Delta_4)
(1+\Delta+\Delta_3-\Delta_4)
\\
&+&
\left(
\Delta-\Delta^{2}-\Delta_1-\Delta_2
+3(\Delta_2 -\Delta_1)^2 - 2\Delta (\Delta_1 +\Delta_2)
\right)
\\
&\times &
\left(
\Delta-\Delta^{2}-\Delta_3 -\Delta_4
+3(\Delta_3 -\Delta_4)^2 - 2\Delta (\Delta_3 +\Delta_4)
\right)
\Big]
\\
&\times &
\left[
2(1+2\Delta)^2 \left( c-\frac{4\Delta(5-8\Delta)}{2+4\Delta}\right)
\right]^{-1},\\
\ldots &&.
\end{eqnarray*}

\noindent\bigskip
\textbf{Coefficients of the classical 4-point block on the sphere.}
\begin{eqnarray*}
f^{\,1}_{\delta}\!\left[_{\delta_{4}\;\delta_{1}}^{\delta_{3}\;\delta_{2}}\right]
&=&
\frac{(\delta + \delta_3 -\delta_4)(\delta + \delta_2 - \delta_1)}{2\delta},
\\
f^{\,2}_{\delta}\!\left[_{\delta_{4}\;\delta_{1}}^{\delta_{3}\;\delta_{2}}\right]
&=&
\Big[16\delta^3 (4\delta +3)\Big]^{-1}
\Big[
13\delta^5 +\delta^4 \left(18\delta_2 -14\delta_1 +18\delta_3 -14\delta_4 + 9\right)
\\
&+&
\delta^3 \Big(
\delta_{1}^{2}+\delta_{2}^{2}-2\delta_1 (\delta_2 +6\delta_3 -10\delta_4 +6)
\\
&+&
4\delta_2 (5\delta_3 -3\delta_4 +3) + (\delta_3 -\delta_4)(\delta_3 -\delta_4 +12)
\Big)
\\
&-&
3\delta^2
\Big(\delta_{1}^{2}(2\delta_3 +2\delta_4 -1)
+ 2\delta_1 (\delta_{3}^{2}+\delta_{4}^{2} +2\delta_3 +\delta_2 -2\delta_2 \delta_3 - 2\delta_4 (\delta_2+\delta_3 +1))
\\
&+&
\delta_{2}^{2}(2\delta_3 + 2\delta_4 -1)
+2\delta_2 (\delta_3 -\delta_4 -2)(\delta_3 -\delta_4)-(\delta_3 -\delta_4)^2
\Big)
\\
&+&
5\delta (\delta_1 -\delta_2)^2 (\delta_3 -\delta_4)^2 - 3(\delta_1 -\delta_2)^2 (\delta_3 -\delta_4)^2
\Big],
\\
\ldots&&\;.
\end{eqnarray*}

\noindent\bigskip
\textbf{Coefficients of the quantum 1-point block on the torus.}
\begin{eqnarray*}
{\cal F}_{c,\Delta}^{\tilde\Delta,1}&=&
\frac{\left(\tilde\Delta-1\right)\tilde\Delta}{2\Delta }+1,
\\[10pt]
{\cal F}_{c,\Delta}^{\tilde\Delta,2}&=&
\left[4 \Delta
\left(2 c \Delta +c+16 \Delta ^2-10 \Delta \right)\right]^{-1}
\\
&&
\Big[\left(8 c \Delta +3 c+128 \Delta ^2+56 \Delta \right)
   \tilde\Delta^2+\left(-8 c \Delta -2 c-128 \Delta
   ^2\right) \tilde\Delta
\\
&+&
   (c+8 \Delta ) \tilde\Delta^4+(-2 c-64 \Delta ) \tilde\Delta^3+16 c \Delta
   ^2+8 c \Delta +128 \Delta ^3-80 \Delta ^2\Big].
\\
\ldots\;&&.
\end{eqnarray*}

\noindent\bigskip
\textbf{Coefficients of the classical 1-point block on the torus.}
\begin{equation*}
f^{\,\tilde\delta, 1}_{\delta} \;=\; \frac{\tilde{\delta }^2}{2 \delta }\,,
\;\;\;\;\;\;\;\;\;\;\;\;\;
f^{\,\tilde\delta, 2}_{\delta} \;=\;
\frac{\tilde\delta^2 [24\delta^2 \left(4\delta+1\right)
+ \tilde\delta^2 \left(5\delta-3\right) - 48\tilde\delta\delta^2]}{16 \delta^3 \left(4 \delta +3\right)}\,,
\;\;\;\;\;\;\;\;\;\;\;\;\;
\ldots\;.
\end{equation*}

\noindent\bigskip
\textbf{Coefficients of the SU(2) pure gauge Nekrasov's instanton function.}
\begin{eqnarray*}
\mathcal{Z}_{1}^{\rm N_f=0, SU(2)} &=& \frac{2}{\epsilon_1\epsilon_2\left((\epsilon_1+\epsilon_2)^2-4a^2\right)},
\\
\mathcal{Z}_{2}^{\rm N_f=0, SU(2)} &=& \frac{-8 a^2+8 \epsilon _1^2+8 \epsilon _2^2+17 \epsilon _1 \epsilon _2}{\epsilon _1^2 \epsilon
   _2^2 \left(\left(\epsilon _1+\epsilon _2\right){}^2-4 a^2\right) \left(\left(2 \epsilon
   _1+\epsilon _2\right){}^2-4 a^2\right) \left(\left(\epsilon _1+2 \epsilon _2\right){}^2-4
   a^2\right)},
\\
\mathcal{Z}_{3}^{\rm N_f=0, SU(2)} &=&
\Big[2 \Big(\epsilon _1^2 \left(594 \epsilon _2^2-104 a^2\right)+\epsilon _1 \left(363 \epsilon
   _2^3-188 a^2 \epsilon _2\right)
\\
&&+\;
8 \left(4 a^4-13 a^2 \epsilon _2^2+9 \epsilon _2^4\right)+72
   \epsilon _1^4+363 \epsilon _2 \epsilon _1^3\Big)\Big]
\\
&&
\Big[3 \epsilon _1^3 \epsilon _2^3 \left(-2
   a+\epsilon _1+\epsilon _2\right) \left(2 a+\epsilon _1+\epsilon _2\right) \left(-2 a+2 \epsilon
   _1+\epsilon _2\right) \left(-2 a+3 \epsilon _1+\epsilon _2\right)
\\
&&
\left(2 a+3 \epsilon
   _1+\epsilon _2\right) \left(2 \left(a+\epsilon _1\right)+\epsilon _2\right)
\\
&&
\left(-2 a+\epsilon
   _1+2 \epsilon _2\right) \left(-2 a+\epsilon _1+3 \epsilon _2\right) \left(2 a+\epsilon _1+3
   \epsilon _2\right) \left(2 \left(a+\epsilon _2\right)+\epsilon _1\right)\Big]^{-1},
\\
\ldots\;&&\;.
\end{eqnarray*}

\noindent\bigskip
\textbf{Coefficients of the SU(2) pure gauge twisted superpotential.}
\label{AppIII}
\begin{eqnarray*}
\mathcal{W}_{1}^{\rm N_f=0, SU(2)} &=& \frac{2}{\epsilon _1^3-4 a^2 \epsilon _1},
\\
\mathcal{W}_{2}^{\rm N_f=0, SU(2)} &=&-\frac{20 a^2+7 \epsilon _1^2}{4 \left(\epsilon _1^2-4 a^2\right){}^3 \left(\epsilon _1^3-a^2
   \epsilon _1\right)},
\\
\mathcal{W}_{3}^{\rm N_f=0, SU(2)} &=&\frac{4 \left(144 a^4+232 a^2 \epsilon _1^2+29 \epsilon _1^4\right)}{3 \left(\epsilon _1^2-4
   a^2\right){}^5 \left(4 a^4 \epsilon _1-13 a^2 \epsilon _1^3+9 \epsilon _1^5\right)}.
\\
\ldots &&\;.
\end{eqnarray*}

\subsection{Ward identities}
\label{AppI}

In order to derive the Ward identity (\ref{MT2}) first we will need to prove
the following relation:
\begin{equation}\label{WI1}
\langle\Delta',\Lambda^2 |V_{+}(z)L_{0}|\tilde\Delta,\Lambda^2\rangle
\;=\;\frac{1}{2}\left(\frac{\Lambda}{2}\frac{\partial}{\partial\Lambda}
+\Delta'+\tilde\Delta -\Delta_+ - z\frac{\partial}{\partial z}\right)\Psi(\Lambda, z).
\end{equation}
Indeed, using
$
L_{0}|\,\Delta,\Lambda^2\,\rangle=
\left(\Delta+\frac{\Lambda}{2}\partial_\Lambda\right)
|\,\Delta,\Lambda^2\,\rangle
$
one gets
\begin{eqnarray*}
\langle\,\Delta',\Lambda^2\, |\,V_{+}(z)(L_{0}-\tilde\Delta)\,|\,\tilde\Delta,\Lambda^2\,\rangle
&=&\langle\,\Delta',\Lambda^2\, |\,V_{+}(z)\,\frac{\Lambda}{2}\,\frac{\partial}{\partial\Lambda}\,|\,\tilde\Delta,\Lambda^2\rangle
\\[8pt]
&&\hspace{-100pt}=\;\;\frac{\Lambda}{2}\,\frac{\partial}{\partial\Lambda}\,\Psi(\Lambda, z)
-\langle\,\Delta',\Lambda^2\, |\,(L_{0}-\Delta') V_{+}(z)\,|\,\tilde\Delta,\Lambda^2\rangle.
\end{eqnarray*}
Hence, taking into account that
$
\left[ L_0 , V_{+}(z)\right]=\left(z\partial_z+\Delta_+\right)V_{+}(z)
$
one finds
\begin{eqnarray*}
\frac{\Lambda}{2}\,\frac{\partial}{\partial\Lambda}\,\Psi(\Lambda, z)
&=&\langle\,\Delta',\Lambda^2\, |\,V_{+}(z)(L_{0}-\tilde\Delta)\,|\,\tilde\Delta,\Lambda^2\,\rangle
+\langle\,\Delta',\Lambda^2\, |\,(L_{0}-\Delta') V_{+}(z)\,|\,\tilde\Delta,\Lambda^2\,\rangle
\\[8pt]
&&\hspace{-60pt}=\;-\left(\Delta'+\tilde\Delta -\Delta_+ - z\frac{\partial}{\partial z}\right)\Psi(\Lambda, z)
+ 2 \langle\,\Delta',\Lambda^2\, |\,V_{+}(z)L_{0}\,|\,\tilde\Delta,\Lambda^2\,\rangle.
\end{eqnarray*}
Next, let us consider the matrix element
$\langle\,\Delta', \Lambda^2\,|\,T(w)V_{+}(z)\,|\,\tilde\Delta,\Lambda^2\,\rangle$
with 
$$
T(w)\;=\;\sum\limits_{n\in\mathbb{Z}}w^{-n-2}L_{n},
$$
i.e.:
\begin{eqnarray*}
&&\langle\,\Delta', \Lambda^2\,|
\,T(w)V_{+}(z)\,|\,\tilde\Delta,\Lambda^2\,\rangle
\;=\;
\sum\limits_{n=-\infty}^{n=+\infty}\frac{1}{w^{n+2}}\,\langle\,\Delta', \Lambda^2\,|
\,L_n V_{+}(z)\,|\,\tilde\Delta,\Lambda^2\,\rangle
\\
&&\hspace{55pt}=
\sum\limits_{n=-\infty}^{n=-2}\frac{1}{w^{n+2}}\,\langle\,\Delta', \Lambda^2\,|
\,L_n V_{+}(z)\,|\,\tilde\Delta,\Lambda^2\,\rangle
+\frac{1}{w}\,\langle\,\Delta', \Lambda^2\,|\,L_{-1} V_{+}(z)\,|\,\tilde\Delta,\Lambda^2\,\rangle
\\
&&\hspace{55pt}+
\sum\limits_{n=0}^{\infty}\frac{1}{w^{n+2}}\,
\,\langle\,\Delta', \Lambda^2\,|
\,L_n V_{+}(z)\,|\,\tilde\Delta,\Lambda^2\,\rangle.
\end{eqnarray*}
>From (\ref{Gstate3}) and $L_{n}^{\dagger}=L_{-n}$ we have
$
\sum_{n=-\infty}^{n=-2} w^{-n-2}\,\langle\,\Delta', \Lambda^2\,|
\, L_n V_{+}(z)\,|\,\tilde\Delta,\Lambda^2\,\rangle=0.
$
Then,
\begin{eqnarray*}
\langle\,\Delta', \Lambda^2\,|
\,T(w)V_{+}(z)\,|\,\tilde\Delta,\Lambda^2\,\rangle
&=&\frac{1}{w}\,\langle\,\Delta', \Lambda^2\,|\,L_{-1} V_{+}(z)\,|\,\tilde\Delta,\Lambda^2\,\rangle
\\
&+&
\sum\limits_{n=0}^{\infty}\frac{1}{w^{n+2}}\,
\,\langle\,\Delta', \Lambda^2\,|
\,L_n V_{+}(z)\,|\,\tilde\Delta,\Lambda^2\,\rangle
\\
&&\hspace{-50pt}\;=
\frac{1}{w}\,\Lambda^2\Psi(\Lambda, z)+\sum\limits_{n=0}^{\infty}\frac{1}{w^{n+2}}\,
\,\langle\,\Delta', \Lambda^2\,|
\,\left[L_n, V_{+}(z)\right]\,|\,\tilde\Delta,\Lambda^2\,\rangle
\\
&&\hspace{-50pt}+\;\frac{1}{w^2}\,\langle\,\Delta', \Lambda^2\,|
\,V_{+}(z) L_0 \,|\,\tilde\Delta,\Lambda^2\,\rangle
+\frac{1}{w^3}\,\langle\,\Delta', \Lambda^2\,|\,V_{+}(z) L_1 \,|\,\tilde\Delta,\Lambda^2\,\rangle,
\end{eqnarray*}
where again the conditions (\ref{Gstate1})-(\ref{Gstate3}) defining the Gaiotto state have been used.
Now, using
$
\left[L_n , V_{+}(z)\right] = z^{n}\left(z
\partial_z + (n+1)\Delta_{+}
\right)V_{+}(z)
$
one gets
\begin{eqnarray*}
\langle\,\Delta', \Lambda^2\,|
\,T(w)V_{+}(z)\,|\,\tilde\Delta,\Lambda^2\,\rangle
&=&\left(\frac{\Lambda^2}{w}+\frac{\Lambda^2}{w^3}\right)\Psi(\Lambda, z)+
\frac{1}{w^2}\,\langle\,\Delta', \Lambda^2\,|
\,V_{+}(z) L_0 \,|\,\tilde\Delta,\Lambda^2\,\rangle
\\
&+&\sum\limits_{n=0}^{\infty}\frac{1}{w^{n+2}}\,
z^n \left(z\frac{\partial}{\partial z}+(n+1)\Delta_{+}\right)\Psi(\Lambda, z).
\end{eqnarray*}
Let us note that for $|\frac{z}{w}|<1$ we have
\begin{eqnarray*}
\sum\limits_{n=0}^{\infty}\frac{z^{n+1}}{w^{n+2}}
&=&\frac{z}{w^2}\sum\limits_{n=0}^{\infty}\left(\frac{z}{w}\right)^n
= \frac{z}{w^2}\frac{1}{1-\frac{z}{w}}=\frac{z}{w(w-z)},
\\
\sum\limits_{n=0}^{\infty}\frac{z^n}{w^{n+2}}(n+1)&=&\frac{\partial}{\partial z}\left(\frac{z}{w(w-z)}\right)
=\frac{1}{(w-z)^2}.
\end{eqnarray*}
Then,
\begin{eqnarray*}
\langle\,\Delta', \Lambda^2\,|
\,T(w)V_{+}(z)\,|\,\tilde\Delta,\Lambda^2\,\rangle
&=&\left[\left(\frac{\Lambda^2}{w}+\frac{\Lambda^2}{w^3}\right)
+\frac{\Delta_+}{(w-z)^2}+\frac{z}{w(w-z)}\frac{\partial}{\partial z}\right]\Psi(\Lambda, z)
\\
&+&\frac{1}{w^2}\,\langle\,\Delta', \Lambda^2\,|\,V_{+}(z) L_0 \,|\,\tilde\Delta,\Lambda^2\,\rangle.
\end{eqnarray*}
Finally, an application of (\ref{WI1}) yields (\ref{MT2}).

Having (\ref{MT2}) one can prove (\ref{L2}). Indeed, using
(\ref{hatL}), (\ref{MT2}) and Cauchy's contour integral formula one gets
\begin{eqnarray*}
\langle\,\Delta', \Lambda^2\,|
\,\widehat{L}_{-2}(z)V_{+}(z)\,|\,\tilde\Delta,\Lambda^2\,\rangle &=&
\frac{1}{2\pi i}\oint\limits_{C_z}dw (w-z)^{-1}
\langle\,\Delta', \Lambda^2\,|
\,T(w)V_{+}(z)\,|\,\tilde\Delta,\Lambda^2\,\rangle
\\
&=&
\frac{1}{2\pi i}\oint\limits_{C_z}dw (w-z)^{-1}\times
{\rm r.h.s.}\;{\rm of}\;(\ref{MT2})
\\
&=& {\rm r.h.s.}\;{\rm of}\;(\ref{L2}).
\end{eqnarray*}

\subsection{Mathieu eigenvalue from WKB}
\label{app_WKB}

In this appendix we perform computations for the spectrum of the
Mathieu operator by means of the WKB method along the line of refs.
\cite{NS:2009,Fateev:2009aw,Mironov:Morozov:2010,Maruyoshi:2010}.

The Seiberg-Witten (SW) theory \cite{Seiberg:1994rs} with $SU(2)$ gauge group is
defined by the prepotential $\mathcal{F}_{SW}(a,\hat\Lambda)$ which is determined by
the set of quantities $(a_{\alpha},\partial\mathcal{F}_{SW}/\partial a_{\alpha})$ i.e., moduli and their duals and the elliptic curve
\begin{equation}
\label{app_WKB_SWcurve}
y^2 = (z^2-\hat\Lambda^4)(z - u),
\qquad u\equiv \hat\Lambda^2 \left(1 -2 \frac{\vartheta^4_2(0|\tau)}{\vartheta^4_3(0|\tau)} \right),
\end{equation}
where $\vartheta^4_i(0|\tau)$, $i=1,2$ are Jacobi functions and $\tau$ is a complexified coupling constant that parametrizes
the modular half-plane for the one-dimensional complex tori. $u$ is the parameter on the moduli space of the theory.
The latter has three singular points: two branch points $u=\hat\Lambda^2,\,(\tau=0)$, $u=-\hat\Lambda^2,\,(\tau=\pm 1)$
and a singular point at $u=\infty,\,(\tau=i\infty)$. The latter corresponds to the perturbative
region of the moduli space (asymptotic freedom). The modulus and its dual with appropriate monodromies about the singular points
are found by means of the curve \eqref{app_WKB_SWcurve} and SW differential which takes the form
$$
\lambda = \frac{\sqrt{2}}{2\pi}\frac{z-u}{y(z)}\mbox{d}z
= \frac{\sqrt{2}}{2\pi}\sqrt{\frac{z-u}{z^2-\hat\Lambda^4}}\mbox{d}z .
$$
The mentioned functions are determined by integrals of $\lambda$ with appropriate cycles, i.e., ($\sum a_\alpha = 0$)
$$
a = \oint\limits_{A} \lambda , \qquad a_D = \oint\limits_{B} \lambda ,
\qquad a_D \equiv \frac{\partial\mathcal{F}_{SW}(a,\hat\Lambda)}{\partial a}.
$$
Integration along the contour $A$ is the one that circumvents a branch cut between the points $[-\hat\Lambda^2,\hat\Lambda^2]$,
whereas the integration over contour $B$ starts from the point on one sheet, passes on the second sheet through the branch cut between
$u$ and $\infty$ and returns to the starting point on the first sheet through the branch cut $[-\hat\Lambda^2,\hat\Lambda^2]$. Hence,
the contour integrals can be given the form
\begin{equation}
\begin{aligned}
\label{vev_vevD}
a_{\hat\Lambda^2}(u) =& \frac{\sqrt{2}}{\pi}\int\limits_{-\hat\Lambda^2}^{\hat\Lambda^2}
\sqrt{\frac{z-u}{z^2-\hat\Lambda^4}}\mbox{d}z
=\frac{\sqrt{2 u}}{\pi}\int\limits_{-\hat\Lambda^2/u}^{\hat\Lambda^2/u}
\sqrt{\frac{z-1}{z^2-(\hat\Lambda^2/u)^2}}\mbox{d}z
\equiv a(\hat\Lambda^2/u),
\\[5pt]
(a_{\hat\Lambda^2})_D(u)
=&\frac{\sqrt{2}}{\pi}\int\limits_{\hat\Lambda^2}^u \sqrt{\frac{z-u}{z^2-\hat\Lambda^4}}\mbox{d}z
= \frac{\sqrt{2u}}{\pi}\int\limits^{1}_{\hat\Lambda^2/u} \sqrt{\frac{z-1}{z^2-(\hat\Lambda^2/u)^2}}\mbox{d}z
\equiv a_D(\hat\Lambda^2/u) .
\end{aligned}
\end{equation}
In a parametrization $z = \hat\Lambda^2\cos(\theta) $, $\pi\lambda
= \sqrt{2u - 2\hat\Lambda^2 \cos(\theta) }\mbox{d}\theta = p(\theta;u,\hat\Lambda)\mbox{d}\theta$, and
\begin{equation}
\begin{aligned}
\label{a_momentum}
a(\hat\Lambda^2/u) =& \frac{1}{2\pi}\oint\limits_A p(\theta;u,\hat\Lambda)\mbox{d} \theta
=\frac{1}{\pi}\int\limits_0^{\pi} p(\theta;u,\hat\Lambda)\mbox{d} \theta ,
\\
a_D(\hat\Lambda^2/u) =& \frac{1}{2\pi}\oint\limits_B p(\theta;u,\hat\Lambda)\mbox{d} \theta
= \frac{1}{\pi}\int\limits_\pi^{\theta_u} p(\theta;u,\hat\Lambda)\mbox{d} \theta ,
\qquad \theta_u\equiv \arccos(u/\hat\Lambda^2) .
\end{aligned}
\end{equation}
Equations \eqref{vev_vevD} are integral representations of the hypergeometric fuctions, namely
\begin{subequations}
\begin{eqnarray}
a(u,\upsilon)&=&\frac{\sqrt{2 u}}{\pi}\int\limits_{-\upsilon}^{\upsilon} \sqrt{\frac{z-1}{z^2-\upsilon^2}}\mbox{d}z
= \sqrt{2 u}\, \sqrt{1+\upsilon}\,
F_{2,1}\left(-\tfrac{1}{2},\tfrac{1}{2},1;\tfrac{2 \upsilon}{1+\upsilon}\right) ,
\\
a_D(u,\upsilon)&=&\frac{\sqrt{2 u}}{\pi}\int\limits_{\upsilon}^{1} \sqrt{\frac{z-1}{z^2-\upsilon^2}}\mbox{d}z
= \tfrac{i}{2^{3/2}} \sqrt{2 u}\, (1-\upsilon)
F_{2,1}\left(\tfrac{1}{2},\tfrac{1}{2},2;-\tfrac{1-\upsilon}{2\upsilon}\right) ,
\end{eqnarray}
\end{subequations}
where $\upsilon \equiv \hat\Lambda^2/u$. The SW prepotential $\mathcal{F}_{SW}(a,\hat\Lambda)$
corresponds to the $\epsilon_1\epsilon_2\to 0$ limit of the
exponentiated Nekrasov partition function. It is related to the classical two-particle Toda periodic chain or
sine-Gordon model action of which takes the form
\begin{equation}
\label{app_WKB_action}
S[\theta(t_2),\theta(t_1),t_2,t_1]=\int\limits^{t_2}_{t_1}\mbox{d}t
\left(\tfrac{1}{2}\dot\theta^2 - \hat\Lambda^2 \cos\theta \right)
= \int\limits^{\theta(t_2)}_{\theta(t_1)} p(\theta;\textrm{E},\hat\Lambda) \mbox{d}\theta - (t_2-t_1)\textrm{E} ,
\end{equation}
where
\begin{equation}
\label{app_WKB_momentum}
p(\theta;\textrm{E},\hat\Lambda) \equiv \sqrt{2\textrm{E} - 2\hat\Lambda^2 \cos\theta} .
\end{equation}
Canonical quantization of the system in eq. \eqref{app_WKB_action} results in the
following form of the time independent Schroedinger equation ($\Psi(t,\theta)=\mbox{e}^{-i \textrm{E} t/\epsilon_1}\psi(\theta)$)
\begin{equation}
\label{app_WKB_schroed}
\left[
-\frac{\epsilon^2_1}{2}\frac{\mbox{d}^2}{\mbox{d}\theta^2} + \hat\Lambda^2\cos\theta - \textrm{E} \right]\psi(\theta) = 0 ,
\end{equation}
with the energy given in eq. \eqref{eigenvalue_from_gauge_th}.
This equation can be solved with the aid of the \emph{exact WKB method}. Namely,
the WKB ansatz takes the form
\begin{equation}
\label{app_WKB_ansatz}
\psi_{\mbox{\tiny WKB}}(\theta) = \exp\left\{\frac{i}{\epsilon_1}
\int\limits^\theta_{\theta_0} P(\vartheta;E,\hat\Lambda) \mbox{d}\vartheta \right\}.
\end{equation}
The turning points are at $\theta^{(\rm tp)}_{1,2} = \pm\arccos(\textrm{E}/\hat\Lambda)$.
Inserting the above wave function into the eq. \eqref{app_WKB_schroed} we obtain equation for
$P(\vartheta;E,\hat\Lambda)$. In general it can be solved for
\begin{equation}
P(\theta;u,\hat\Lambda)\, \mbox{d}\theta
= \sum_{m\geq 0} \epsilon^m_1\, p_m(\theta;u,\hat\Lambda),
\end{equation}
order by order. The first term $p_0(\theta;u,\hat\Lambda) \equiv p(\theta;u,\hat\Lambda)$
corresponds to quasi-classical solution. The modulus $a$ related to this solution
is related to the ''classical'' momentum through
\begin{displaymath}
a(\hat\Lambda^2/ u) =\frac{1}{2\pi} \oint_A\, p_0(\theta;u,\hat\Lambda)\, \mbox{d}\theta .
\end{displaymath}
Its quantum counterpart corresponds to modulus on the ''quantum'' moduli space and reads
\begin{equation}
\label{app_WKB_deformed_a}
\mathsf{a}[a(\hat\Lambda^2/ u) ]= \frac{1}{2\pi}\oint_A P(\theta;u,\hat\Lambda)\, \mbox{d}\theta
= \frac{1}{2\pi}\sum_{m\geq 0} \epsilon^m_1\, \oint_A\, p_m(\theta;u,\hat\Lambda)\, \mbox{d}\theta .
\end{equation}
The above corrections to momentum can be obtained by means of the following relation
\begin{equation}
p_m(\theta;u,\hat\Lambda) = \mathcal{O}_m(\gamma,\mathcal{E},
\partial_\gamma,\partial_ {\mathcal{E}}) p_0(\theta;u,\hat\Lambda),
\end{equation}
such that
\begin{equation}
\mathsf{a}[a(\hat\Lambda^2/ u) ]
=\sum_{m\geq 0} \epsilon^m_1\, \mathcal{O}_m(\gamma,\mathcal{E},
\partial_\gamma,\partial_ {\mathcal{E}}) a(\gamma/\mathcal{E}),
\end{equation}
where $\mathcal{E} \equiv 2u,\ \gamma\equiv 2\hat\Lambda^2$. We have found these operators up to
third order in $\epsilon_1$. The result reads
\begin{subequations}
\begin{eqnarray}
\mathcal{O}_1(\gamma,\mathcal{E},\partial_\gamma,\partial_ {\mathcal{E}})
&\equiv&
\frac{\gamma }{2^2\cdot 3}\,\partial_\gamma \partial_ {\mathcal{E}} ,
\\
\mathcal{O}_2(\gamma,\mathcal{E},\partial_\gamma,\partial_ {\mathcal{E}})
&\equiv&
\frac{\gamma}{2^5\cdot 3^2\cdot 5}
\Big[
5 \gamma\,\partial^2_\mathcal{E}\,\partial^2_\gamma
-2 \mathcal{E}\,\partial^3_\mathcal{E}\,\partial_\gamma
\Big] ,
\\\nonumber
\mathcal{O}_3(\gamma,\mathcal{E},\partial_\gamma,\partial_ {\mathcal{E}})
&\equiv&
\frac{\gamma}{2^{7}\cdot 3^6\cdot 5\cdot 7}
\Big[
463\,\gamma^2\,\partial^3_\mathcal{E}\, \partial^3_\gamma
-118\,\mathcal{E}\gamma\,\partial^4_\mathcal{E}\,\partial^2_\gamma
\\
&&\qquad+\left(20\mathcal{E}^2 +16 \gamma ^2\right)\,\partial^5_\mathcal{E}\, \partial_\gamma
+16\, \mathcal{E} \gamma\, \partial^6_\mathcal{E}
\Big] .
\end{eqnarray}
\end{subequations}
In order to find the energy for the full quantum system we have to invert \eqref{app_WKB_deformed_a}
so that $\textrm{E} = \textrm{E}(\textsf{a}(a))$. The result for the quantum modulus takes the form
\begin{multline}
\mathsf{a}\left(a(\gamma/\mathcal{E})\right)
= \sqrt{\mathcal{E}}
-\gamma^2 \left(
\frac{1}{16\,\mathcal{E}^{3/2}}
+\frac{\epsilon_1^2}{64\,\mathcal{E}^{5/2}}
+\frac{\epsilon_1^4}{256\,\mathcal{E}^{7/2}}
+\frac{\epsilon_1^6}{1024\,\mathcal{E}^{9/2}}
\right)
\\
- \gamma^4\left(
\frac{15}{1024\,\mathcal{E}^{7/2}}
+\frac{35\,\epsilon_1^2}{2048\,\mathcal{E}^{9/2}}
+\frac{273\,\epsilon_1^4}{16384\,\mathcal{E}^{11/2}}
+\frac{33\,\epsilon_1^6}{2048\,\mathcal{E}^{13/2}}
\right)
\\
-\gamma^6\left(
\frac{105}{16384\,\mathcal{E}^{11/2}}
+\frac{1155\,\epsilon_1^2}{65536\,\mathcal{E}^{13/2}}
+\frac{5005\,\epsilon _1^4}{131072\,\mathcal{E}^{15/2}}
+\frac{42185\, \epsilon_1^6}{524288\,\mathcal{E}^{17/2}}
\right)
+ \mathcal{O}(\gamma^{8}).
\end{multline}
Inverting the above series we obtain the energy up to the third order in $\hat\Lambda^4$,
\begin{multline}
2\textrm{E}(a) =
a^2
+\hat\Lambda^4\left(
\frac{1}{2\, a^2}
+\frac{\epsilon _1^2}{8\, a^4}
+\frac{\epsilon _1^4}{32\, a^6}
+\frac{\epsilon _1^6}{128\, a^8}
\right)
+\hat\Lambda^8 \left(
\frac{5}{32\, a^6}
+\frac{21\, \epsilon_1^2}{64\, a^8}
+\frac{219\, \epsilon_1^4}{512\, a^{10}}
+\frac{121\, \epsilon_1^6}{256\, a^{12}}
\right)
\\
+ \hat\Lambda^{12} \left(
\frac{9}{64\, a^{10}}
+\frac{55\, \epsilon_1^2}{64\, a^{12}}
+\frac{1495\, \epsilon_1^4}{512\, a^{14}}
+\frac{4035\,\epsilon _1^6}{512\, a^{16}}
\right)
+ \mathcal{O}(\hat\Lambda^{16}).
\end{multline}
As one may check, all the coefficients of $\hat\Lambda^{4n}$ in the above formula for energy
are Maclaurin expansions of coefficients of $-\e_1\omega_\ast(a,\Lambda^{4}/\epsilon^4_1)$ in $\epsilon_1$.
Explicitly,
\begin{equation}
\label{app_WKB_Esg}
\begin{aligned}
2\textrm{E}(a) =&
a^2
-\epsilon_1\hat\Lambda^4\sum_{\a=1}^2 \w_{\ast\,\a,1,1}
-\epsilon_1\hat\Lambda^8\sum_{\a=1}^2\sum_{j=1}^2 \w_{\ast\,\a,j,2}
-\epsilon_1\hat\Lambda^{12}\sum_{\a=1}^2\sum_{j=1}^3 \w_{\ast\,\a,j,3}
\\
=& a^2 - \epsilon_1\omega_{\ast} ,
\end{aligned}
\end{equation}
where the coefficients $\w_{\a,i,j}$ are given in eqs. \eqref{app_speq_sols_L1}-\eqref{app_speq_sols_L3}.
This result should be related to the two particle periodic Toda chain (pToda). The system is defined either by means
of the canonical Poisson-commuting Hamiltonians with potential $U(x_1,x_2) = 2\hat\Lambda^2 \cosh(x_1 - x_2)$
or by the Hamiltonians obtained as coefficients of the characteristic polynomial of Lax operator (for details see \cite{NS:2009})
with potential $U(x_1,x_2) = 2\hat\Lambda^2 \cosh(i x_1 - i x_2)$. In both cases canonical quantization
leads to the Schroedinger equation which can be cast in the form of Mathieu equation, where
the relationship between parameters $E_{\rm Toda}$, $\hat\Lambda^2$ and Mathieu parameters $\lambda$ and $h^2$
reads
$$
\lambda = \frac{2E_{\rm Toda}}{\e^2_1} ,\quad h^2 = \frac{\hat\Lambda^2}{\e^2_1} .
$$
The relation between the spectrum of the sin-Gordon operator and Toda operator is $4 E_{\rm SG}= E_{\rm Toda} $.
Hence, multiplying eq. \eqref{app_WKB_Esg} by two we get the formula given in eq. \eqref{eigenvalue_from_gauge_th}.

\subsection{Young diagrams, their profiles and Nekrasov partition function}
\label{app_instNekrasov}

In this appendix we introduce basic notions concerning partitions
and their use in the context of instanton partition function for pure $\mathcal{N}=2$
gauge theory with $U(N)$ symmetry group.

Let $\Bbbk^s(n)$ denote a partition
of a number $n\in \mathbb{N}$ into $\ell$ positive integers $k^s_i(n)$, such that
for any $i,j\in\{1,..,\ell\}$ and $\ell\geq i>j$, $k^s_i\leq k^s_j$. The upper index $s$ indicates that the partition of $n$
may be done in $p(n)$ distinct ways. Thus $s^{\rm th}$ partition of $n$ for some $s\in \{1,\dots,p(n)\}$
is a finite ordered sequence of nonincreasing positive integers, which for any
$n\in \mathbb{N}$ and $s\in \{1,\dots,p(n)\}$ may be represented
by an ordered $\ell$-tuple of numbers
\begin{equation}
\label{def_partition}
\Bbbk^s(n) = \big(k^s_1(n),k^s_2(n),\dots,k^s_\ell(n)\big),
\quad
n\equiv |\Bbbk^s(n)| = \sum_{i=1}^\ell k^s_i(n).
\end{equation}
The number $\ell = \ell(\Bbbk^s(n))$ is called a length of partition $\Bbbk^s(n)$ and
it may vary from 1 to $\ell_{\rm max} = n$.
Let $\mathcal{P}_n$ denote a set of all partitions of $n\in\mathbb{N}$,
i.e., $\mathcal{P}_n \equiv \{\Bbbk^s(n)\}_{s=1}^{p(n)}$.
Note, that $\ell(\mathcal{P}_n) = p(n)$.
If we also include the null partition, i.e., $\mathcal{P}_0\equiv \pbr{k^1(0)}$  ($p(0)=1$), where $\ \Bbbk^1(0) = (0)$,
then the space of all partitions of each element of $\mathbb{N}_0\equiv\mathbb{N}\cup\{0\}$ termed
\emph{Young graph} and denoted by $\mathbb{Y}$ is
a disjoint union of sets $\mathcal{P}_n$, that is
$$
\mathbb{Y} = \bigsqcup_{n\in\mathbb{N}_{0} }\mathcal{P}_n .
$$
\begin{figure}[htb]
  \centering
\centering
\begin{tikzpicture}[scale=.35]
\begin{scope}[shift={(-5,10)}]
\node (a) at (-4.5,-10) {(a)} ;
\draw[>=latex,->,help lines] (-12.25,0.25) -- (-12.25,-11) node [left] at (-12.5,-11){$i$};
\draw[>=latex,->,help lines] (-12.25,0.25) -- (-3.25,0.25) node [right] at (-3.3,0.25){$j$};
\begin{scope}
\draw[clip] (-4,0) |- (-5,-1)  |- (-7,-3) |- (-8,-4) |- (-9,-5) |- (-10,-7) |- (-11,-9) |- (-12,-10) |- (-4,0);
\draw (-4,-10) grid (-12,0);
\draw (-12,0) grid (-4,-10);
\draw (-4,0) |- (-5,-1)  |- (-7,-3) |- (-8,-4) |- (-9,-5) |- (-10,-7) |- (-11,-9) |- (-12,-10) ;
\end{scope}
\draw[help lines,dashed] (-7,-4) to (-3.5,-4) ;
\draw[<->,help lines] (-3.7,0) to (-3.7,-4) node [right] at (-3.7,-2) {$\tilde k_j$} ;
\draw[help lines,dashed] (-7,-3) to (-7,-10.7);
\draw[<->,help lines] (-12,-10.7) to (-7,-10.7) node [below] at (-9.5,-10.9) {$k_i$};
\end{scope}
%----------------------------
\draw[->] (-7,3) -- (-3,3) node[above] at (-5,3.3) {\small rotate $135^\circ$};
%---------------------------
\begin{scope}[rotate=45,shift={(4,-1)}]
\node (b) at (-1.3,-1.5) {(b)} ;
    \draw[>=latex,->,help lines] (-0.25,-.25) -- (11,-.25) node [left] at (11.7,-.5){$i$};
    \draw[>=latex,->,help lines] (-.25,-.25) -- (-.25,9) node [right] at (-.6,9.7){$j$};
   % \draw[-,thick] (0,8) -- (0,11);
    %\draw[-,thick] (10,0) -- (11,0);
    \draw[clip] (0,8) -| (1,7) -| (3,5) -| (4,4) -| (5,3) -| (7,2) -| (9,1) -| (10,0) -| (0,8);
    \draw (0,0) grid (10,8);
    \draw[-] (0,8) -| (1,7) -| (3,5) -| (4,4) -| (5,3) -| (7,2) -| (9,1) -| (10,0);
\end{scope}
\end{tikzpicture}
\caption{
\label{app_fig_1}
The Young diagram for the partition $\Bbbk=(8,7,7,5,4,3,3,2,2,1)$ of $|\Bbbk|=42$
  with quantities defined in the text and the relationship between the
  two different conventions: English (panel (\textit{a}) and Russian (panel (\textit{b}) ).
}
\end{figure}
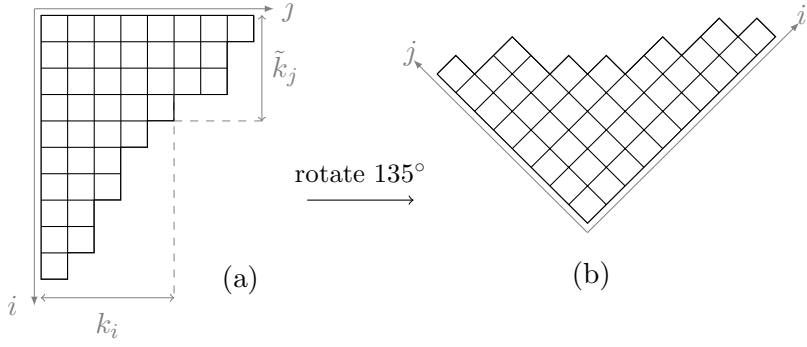
\noindent
The partition $\Bbbk^s(n)$ can be represented graphically by means of the Young diagram,
whose columns contain $k^s_i(n)$ unenumerated boxes for each $1\leq i\leq\ell$ and $n>0$.

Let $k\in\mathbb{N}_0$ be the instanton number (charge). In order to extend the notion of partitions
to the colored ones we assume that this number
can be partitioned into a sum of $N$ \emph{nonnegative} integers $n_\alpha$ for $\a = 1,\dots,N$.
Therefore, in opposite to the earlier partitioning the maximal length of the latter is constrained to $N$.
However, each of $n_\a$ can be partitioned as before i.e., in an unconstrained way ($\ell_{\a\,\mathrm{max}}=n_\a$).
Let $\mathbf{k}^s(k)$ be a $s^{\rm th}$ \emph{colored partition} of $k$.
The space of all colored partitions termed \emph{colored Young graph} and its elements are
defined as follows\footnote{In what follows we ascribe
Greek letters to color indexes $\a = 1, \dots, N$ and Latin
letters to coordinates of the box within a Young diagram, $(i,j)\in \Bbbk^s(n) $.}
\begin{equation}
\mathbb{Y}^N \equiv \bigtimes_{i=1}^N \mathbb{Y} ,
\quad
\mathbf{k}^s(k) \in \mathbb{Y}^N,
\quad
\mathbf{k}^s(k) \equiv \big(\Bbbk^{s_1}(n_1),\Bbbk^{s_2}(n_2),\dots,\Bbbk^{s_N}(n_N)\big),
\end{equation}
where for a given $k\in\mathbb{N}_0$
\begin{equation}
|\mathbf{k}^s(k)|\equiv k = \sum_{\a = 1}^N n_\a = \sum_{\a = 1}^N |\Bbbk^{s_\a}(n_\a) |
=\sum_{\a = 1}^N \sum_{i_\a=1}^{\ell_\a} k^{s_\a}_{i_\a}(n_\a)  .
\end{equation}
for any $ s_\a\in \{1,\dots,p(n_\a)\}$ with $\a = 1,\dots,N$.
Note that any $\mathbf{k}^s(k)$ is uniquely specified by the set of numbers $s_\a$ and $n_\a$,
such that for a fixed $n_\a$ for all $\a$ we have $\prod p(n_\a)$ number of
different $\mathbf{k}^s(k)$'s.
The above equation represents a specific decomposition of $k$ into a sum of parts $n_\a$.
However, this decomposition is not unique. In fact, there is a $\mbox{c}(n,N) = (n+N-1)!/(N-1)!n!$ possible
ways to do this, where $\mbox{c}(n,N)$ is a number of constrained partitions (so called \emph{compositions}) of
$n\in \mathbb{N}_{0}$ into a sum of exactly $N$ parts.
Let $\mathcal{S}(k,N)$ denote a \emph{shell} in $\mathbb{Y}^N$, that is a
set of points defined as follows
$\mathcal{S}(k,N) \equiv \pbr{\mathbf{k}\in\mathbb{Y}^N \big| |\mathbf{k}| = k }$,
then the number of points in the shell amounts to
$$
\#\mathcal{S}(k,N)  = \sum_{r=1}^{\mbox{c}(k,N)}\prod_{\a = 1}^{N} p(n_\a^r) .
$$
This enables us to decompose the sum in the instanton part of
Eq. \eqref{def_z_inst} into a sum over distinct shells
and the sum over points in a given shell, namely
\begin{equation}
\sum_{\mathbf{k}\in \mathbb{Y}^N} \La^{2N|\mathbf{k}|}
Z_{\mathbf{k}}(\mathbf{a},\e_1,\e_2)
= \sum_{k\geq 0}\La^{2N k}\sum_{\mathbf{k}\in \mathcal{S}(k,N)}
Z_{\mathbf{k}}(\mathbf{a},\e_1,\e_2) .
\end{equation}
Sum over shells amounts to
\begin{equation}
\sum_{\mathbf{k}\in \mathcal{S}(k,N)} Z_{\mathbf{k}}(\mathbf{a},\e_1,\e_2)
= \sum_{r=1}^{c(k,N)}\prod_{\a=1}^N\sum_{s_\a=1}^{p(n^r_\a)}
Z_{\rm inst}(\mathbf{a},\{\Bbbk^{s_\a}(n^r_\a)\},\e_1,\e_2 ) .
\end{equation}

Instead of regarding $s^{\rm th}$ partition $\Bbbk^s(n)$ as a $\ell$-tuple of non-increasing, positive integers
one can extend this notion to an infinite sequence of non-increasing, \emph{nonnegative} integers $k_i$
whose length is $\ell(\Bbbk)=\ell$, and for $i>\ell$, $k_i = 0$. Moreover, we follow the practice to keep implicit
both, the label $s$ that positions a partition in $\mathcal{P}_n$ and the number $n$ the relevant partition corresponds to.
Within this extended notion, the partition is the following sequence
$$
\Bbbk=\{k_i\}_{i\in\mathbb{N}}, \quad \mbox{where}\quad k_i\in \mathbb{N}_{0},
\quad \underset{i,j\in\mathbb{N}}{\bigforall}\  i>j \Rightarrow k_i\leq k_j ,
\quad |\Bbbk| = \sum_{i\geq 1} k_i = n .
$$
Note that this implies that for $i\to \infty,\, k_i\to 0$.
Form this point of view, $\mathbb{Y}$ is a subset of nonnegative, integer null sequences $\boldsymbol{\mathsf{c}}_0(\mathbb{Z}_{\geq 0})$.
For any partition (Young diagram) $\Bbbk\in \mathbb{Y}$ there is a dual partition $\tilde{\Bbbk}\in \mathbb{Y}^D$.
It is related to the original one through the transposition,
i.e., $\tilde{\Bbbk} =\Bbbk^T$ ($\tilde{\Bbbk}^T=\Bbbk$) and $|\tilde{\Bbbk}| =|\Bbbk|$.
Hence, $\mathbb{Y} \cong \mathbb{Y}^D=\mbox{Reverse}(\mathbb{Y})$, where the last map is a specific permutation
that reverses the ordering o partitions within each $\mathcal{P}_n$.
Extension to a colored partition is straightforward and for any
$\mathbf{k}\in\mathbb{Y}^N$ it takes the following form
$$
\mathbf{k}=(\Bbbk_1,\dots,\Bbbk_N)
= \{k_{\a,i}\}_{\substack{\a\in\{1,..,N\}\\i\in\mathbb{N}}} \quad \text{and}
\quad k = |\mathbf{k}| = \sum_{\a=1}^N|\Bbbk_\a|
=\sum_{\a=1}^N\sum_{i\geq 1} k_{\a,i} .
$$

Notions of partition theory play the crucial role in counting of instanton configurations.
As it has been shown by Nekrasov \cite{Nekrasov:2002qd} and independently
by Nakajima and Yoshioka \cite{Nakajima:Yoshioka:2005} the instanton part of the partition
function for the $\mathcal{N}=2$ nonabelian gauge theory with $SU(N)$ symmetry, when computed on the
so called $\Omega$-background \cite{Nekrasov:Okounkov:2003}, appears to be a sum over all colored Young diagrams.
The instanton charge $k$ corresponds to the total number of boxes
that are distributed into $N$ parts (compositions) each of which can be further partitioned, thus
forming a colored Young diagram as described above. The form of Nekrasov's instanton partition function
for a pure $SU(N)$ gauge theory reads \cite{Nekrasov:Okounkov:2003}
\begin{equation}
\label{app_Nekrasov_inst_part}
\mathcal{Z}_{\rm inst}(\hat\Lambda,\mathbf{a},\epsilon_1,\epsilon_2)
= \sum_{\mathbf{k}\in \mathbb{Y}^N} \hat\Lambda^{2Nk} Z_{\mathbf{k}}(\mathbf{a},\epsilon_1,\epsilon_2),
\end{equation}
where
\begin{subequations}
\begin{equation}
\label{nekrasov_1}
Z_{\mathbf{k}}(\mathbf{a},\e_1,\e_2)
= \prod_{\a,\b=1}^N \prod_{i,j\geq 1}
\frac{a_\a - a_\b +\e_1 (i-1) + \e_2(-j)}{a_\a - a_\b + \e_1(i-\tilde{k}_{\b,j}-1) + \e_2(k_{\a,i}-j)} ,
\end{equation}
where $\hat\Lambda$ is an infrared effective scale which is related to the complexified $SU(N)$ gauge
coupling $\tau = 4\pi i/g^2 + \theta/2\pi$ through $\hat\Lambda^{2N}\sim\mu^{2N}_{UV}\exp\{ 2\pi i\tau\}$.
$\mu_{UV}$ is the scale where the classical (microscopic) theory is defined.
The deformation parameters are in general $\e_{1,2}\in \mathbb{C}$ and either $\Re(\e_1)>0, \Re(\e_2)<0$ or
$\Re(\e_1)<0, \Re(\e_2)>0$. For the sake of definiteness in what follows we assume $\Re(\e_1)>0, \Re(\e_2)<0$.
The contribution \eqref{nekrasov_1} to eq. \eqref{app_Nekrasov_inst_part} may
be cast into three equivalent forms, which we quote following ref. \cite{Nekrasov:Okounkov:2003},
namely
\begin{align}
\nonumber
\lefteqn{Z_{\mathbf{k}}(\mathbf{a},\e_1,\e_2) }
\\
\label{nekrasov_2}
&= \prod_{\a,\b=1}^N \prod_{i,j\geq 1}
\frac{a_{\a\b} +\e_1 (-i) + \e_2(j-1)}{a_{\a\b} + \e_1(\tilde{k}_{\a,j}-i) + \e_2(j-k_{\b,i}-1)}
\\
\label{nekrasov_3}
&= \frac{1}{\e_2^{2Nk}}\prod_{(\a,i)\neq(\b,j)}
\frac{\Gamma\cbr{b_{\a\b}+\n(j-i+1)+k_{\a i}-k_{\b j}}
\Gamma\cbr{b_{\a\b}+\n(j-i)}}{\Gamma\cbr{b_{\a\b}+\n(j-i)+k_{\a i}-k_{\b j}}
\Gamma\cbr{b_{\a\b}+\n(j-i+1)} }
\\
\label{NakajimaYoshioka}
&=\prod_{\a,\b=1}^N \prod_{s\in \Bbbk_\a}\frac{1}{a_{\a\b} - \e_1 L_\b(s) + \e_2 \big(A_\a(s)+1\big)}
\prod_{s\in\Bbbk_\b}\frac{1}{a_{\a\b}+\e_1\big( L_\a(s)+1\big) - \e_2 A_\b(s)} ,
\end{align}
\end{subequations}
where
$$
a_{\a\b}\equiv a_\a - a_\b,\quad b_{\a\b}\equiv a_{\a\b}/\e_2 , \quad \n\equiv -\e_1/\e_2 ,
$$
and $A_\a(s)\equiv k_{\a,i}-j,\ L_\a(s) \equiv \tilde k_{\a,j}-i $ are arm-length and leg-length, respectively.
$s\equiv (i,j)$ represents a box in a Young diagram $\Bbbk_\a$, where $1\leq i \leq \ell_\a = \tilde k_{\a,1}$ and
$1\leq j \leq k_{\a,i}$. The forms \eqref{nekrasov_1} -- \eqref{nekrasov_3} although defined
in terms of infinite product are finite and well defined due to a finiteness of the relevant Young diagrams.
In what follows we are concerned with yet another form, namely the contribution expressed
in terms of the profiles of Young diagrams.
This form can be obtained by means of the $\zeta$-function regularization techniques that enable to deal with
divergent expressions like the numerator or the denominator of eq. \eqref{nekrasov_1} when considered separately.
The contribution  \eqref{nekrasov_1} to the instanton partition function
\eqref{app_Nekrasov_inst_part} along with the perturbative part introduced in eq. \eqref{def_full_partition_func}
can be cast into the following form
\begin{multline}
\label{app_pert_inst_full_form}
\La^{2Nk}\exp\pbr{-\sum_{\a,\b=1}^N \g_{\e_1,\e_2}(a_{\a\b} ;\La)}Z_{\mathbf{k}}(\mathbf{a},\e_1,\e_2)
\\
= \exp\left\{-\frac{\mbox{d}}{\mbox{d} s}\bigg|_{s=0}\frac{\La^s}{\Gamma(s)}\int\limits^\infty_0 \mbox{d} t\ t^{s-1}
\frac{1}{(\mbox{e}^{\e_1 t}-1)(\mbox{e}^{\e_2 t}-1 ) }\right.
\\
\left. \times\sum_{\a,\b=1}^N
\Bigg[
\mbox{e}^{a_{\a\b} t}
+ (\mbox{e}^{-\e_1 t}-1)(\mbox{e}^{\e_2 t}-1 )\sum_{i,j\geq 1}\cbr{\mbox{e}^{t [ a_{\a\b} + \e_1(i-\tilde{k}_{\b j}) + \e_2(k_{\a i}-j ) ]}
- \mbox{e}^{t [ a_{\a\b} + \e_1 i - \e_2 j  ]}}
\Bigg] \right\},
\end{multline}
where the perturbative part is defined in eq. \eqref{def_full_partition_func} with the $\e_1\e_2$-parametric gamma function
assumes a form (recall that we take $\Re \e_1>0,\,\Re\e_2<0$)
\begin{equation}
\label{app_def_gamma}
\gamma_{\e_1,\e_2}(x;\hat\Lambda) \equiv
\frac{\mbox{d}}{\mbox{d} s}\bigg|_{s=0}\frac{\hat\Lambda^s}{\Gamma(s)}\int\limits^\infty_0 \mbox{d} t\ t^{s-1}
\frac{\mbox{e}^{-xt}}{(\mbox{e}^{\e_1 t}-1)(\mbox{e}^{\e_2 t}-1 ) }  ,\quad \Re x>0,\,\Re s>2 .
\end{equation}
The above gamma function satisfies the following second order functional partial difference equation
$$
\Delta_{\e_1}\Delta_{\e_2}\gamma_{\e_1,\e_2}(x;\hat\Lambda) = \log\left(\frac{\hat\Lambda}{x+\e_1+\e_2}\right),
$$
where $\Delta_{\e_i}f(x)\equiv f(x+\e_i)-f(x)$ for $i=1,2$.
For convenience we introduce the following notation
\begin{equation}
\label{app_notation_pqe}
e_\a(t) \equiv \mbox{e}^{a_\a t} , \quad p(t) \equiv \mbox{e}^{-\e_1 t} ,\quad q(t) \equiv \mbox{e}^{\e_2 t} \ ,
\qquad \Re(\e_1)>0,\ \Re(\e_2)<0 .
\end{equation}
With this notation the formula in the last line of eq. \eqref{app_pert_inst_full_form} reads
\begin{multline}
\label{eq_parameters_pqe}
-\frac{\mbox{d}}{\mbox{d} s}\bigg|_{s=0}\frac{\La^s}{\Gamma(s)}\int\limits^\infty_0 \mbox{d} t\ t^{s-1}
\frac{1}{(p^{-1}-1)(q-1 ) }
\\
\times\sum_{\a,\b=1}^N e_\a e_\b^{-1}
\Bigg\{
1+ (p-1)(q -1)\sum_{i,j\geq 1}\cbr{p^{\tilde{k}_{\b j}-i} q^{k_{\a i}-j} - p^{-i} q^{-j}}
\Bigg\} ,
\end{multline}
The above series can be transformed into the finite form, namely ($\tilde\ell_\a = \ell(\Bbbk^T_\a) = k_{\a,1}$)
\begin{multline}
\label{app_to_profile}
1+(p-1)(q-1) \sum_{i,j\geq 1}\cbr{p^{\tilde{k}_{\b j}-i} q^{k_{\a i}-j} - p^{-i} q^{-j}}
\\
\begin{aligned}
=& 1+(p-1)(q-1)\sum_{i=1}^{\ell_\a} \sum_{j=1}^{\tilde{\ell}_\b}p^{-i} q^{-j}
\cbr{p^{\tilde{k}_{\b j}} -1}\cbr{q^{k_{\a i}} - 1}
\\
&+(p-1)\sum_{i=1}^{\ell_\a} p^{-i}\cbr{q^{k_{\a i}} - 1}
+(q-1)\sum_{j=1}^{\tilde{\ell}_\b} q^{-j}\cbr{p^{\tilde{k}_{\a j} } - 1}
\\
=&\sbr{1+(p-1)\sum_{i=1}^{\ell_\a}p^{-i} \cbr{q^{k_{\a i}} - 1}}
\sbr{1+(q-1) \sum_{j=1}^{\tilde{\ell}_\b} q^{-j} \cbr{p^{\tilde{k}_{\b j}} -1} } .
\end{aligned}
\end{multline}
In what follows it proves useful to define the following quantity
\begin{equation}
\label{def_fpq}
\vf_{\Bbbk_\a}(p,q) \equiv 1+(p-1)\sum_{i=1}^{\ell_\a}p^{-i} \cbr{q^{k_{\a i}} - 1} .
\end{equation}
Note, that in the above formula we can extend the upper sum limit $\ell_\a$ up to infinity which
leaves its form intact. Multiplying eq. \eqref{def_fpq} by $e_\a$, summing over color index and rewriting it
back in the original form as in eq. \eqref{app_notation_pqe} we obtain
\begin{multline*}
\sum_{\a=1}^N e_\a \vf_{\Bbbk_\a}(p,q)
\\
= \sum_{\a=1}^N \mbox{e}^{a_\a t}
+\sum_{\a=1}^N\sum_{i\geq 1}\cbr{\mbox{e}^{[a_\a +\e_1(i - 1) + \e_2 k_{\a i} ] t }
-\mbox{e}^{[a_\a +\e_1 i + \e_2 k_{\a i} ] t } - \mbox{e}^{[a_\a +\e_1 (i-1) ] t } + \mbox{e}^{[a_\a +\e_1 i ] t }}  .
\end{multline*}
which can be written as
\begin{subequations}
\begin{equation}
\label{profile_to_phi}
\frac{1}{2} \int\limits_\mathbb{R}\mbox{d} x\ f_{\mathbf{a},\mathbf{k}}''(x|\e_1,\e_2)\mbox{e}^{ x t}
=  \sum_{\a=1}^N e_\a(t) \vf_{\Bbbk_\a}\big(p(t),q(t)\big) ,
\end{equation}
where
\begin{multline}
\label{app_def_profile}
 f_{\mathbf{a},\mathbf{k}}(x|\e_1,\e_2)
 = \sum_{\a =1}^N |x-a_\a|
 +\sum_{\a =1}^N \sum_{i\geq1}\Big( \left|x - x_{\a,i}\right|
 -\left|x - x_{\a,i} - \e_1\right|
 \\
  - \left|x-x^0_{\a,i}\right|+\left|x- x^0_{\a,i}-\e_1\right|
 \Big ) ,
\end{multline}
is the \emph{profile function} of the colored Young diagram $\Bbbk_\a$.
In analogy to the above, one defines for the second factor of the last line of eq. \eqref{app_to_profile}
the following form of the profile
\begin{equation}
\label{profileT_to_phiT}
\frac{1}{2} \int\limits_\mathbb{R}\mbox{d} x\ f_{\mathbf{a},\mathbf{k}^T}''(x|\e_2,\e_1)\mbox{e}^{ - x t}
=  \sum_{\a=1}^N e_\a^{-1}(t) \vf_{\Bbbk^T_\a}(q(t),p(t)) ,
\end{equation}
and
\begin{multline}
f_{\mathbf{a},\mathbf{k}^T}(x|\e_2,\e_1)
 = \sum_{\a =1}^N |x + a_\a|
 +\sum_{\a =1}^N \sum_{i\geq1}\Big( \left|x -\tilde x_{\a,i}\right|
 -\left|x - \tilde x_{\a,i} - \e_2\right|
 \\
  - \left|x-\tilde x^0_{\a,i}\right|+\left|x- \tilde x^0_{\a,i}-\e_2\right|
 \Big ).
\end{multline}
\end{subequations}
This form of the profile function equals to the one in eq. \eqref{app_def_profile} as it takes its form from subjecting the former to the
transposition of the Young diagram and subsequent reversion of the order of pair of parameters $(\e_1,\e_2)\to(\e_2,\e_1)$ . It
stems from the fact that in case of the arbitrary Young diagram the following identity holds true
\begin{equation}
\label{app_sum_over_boxes}
\sum_{(i,j)\in \Bbbk}p^i q^j = \sum_{i=1}^{\ell} \sum_{j=1}^{k_i} p^i q^j
= \sum_{j=1}^{\tilde{\ell}} \sum_{i=1}^{\tilde{k}_j} p^i q^j = \sum_{(j,i)\in \Bbbk^T}p^i q^j.
\end{equation}
It entails that $\vf_{\Bbbk_\a}\big(p(t),q(t)\big) = \vf_{\Bbbk^T_\a}(q(t),p(t))$
and from eq. \eqref{app_notation_pqe} it
results that $p^{-1}(t)=p(-t)$ which holds true also for $q$ and $e_\a$. Hence, we obtain the equality
$f_{\mathbf{a},\mathbf{k}}(x|\e_1,\e_2) = f_{\mathbf{a},\mathbf{k}^T}(x|\e_2,\e_1)$.
\begin{figure}[htb]
\centering
\begin{tikzpicture}[scale=1.3]
\draw[->] (-3,0) -- (3,0) node[right] {$x$};
\draw[->] (0,0) -- (0,3.5) node[above] {$f_{a_\alpha,\Bbbk_\alpha}(x|\hbar,-\hbar)$};
\draw [very thick,domain=-3:3] plot (\x, { abs(-(5/2) + \x) - abs(-(9/4) + \x) + abs(-2 + \x)
- abs(-(7/4) + \x) + abs(-(5/4) + \x) - abs(-1 + \x) + abs(-(1/2) + \x) - abs(-(1/4) + \x)
+ abs(\x) - abs(1/4 + \x) + abs(1/2 + \x) - abs(1 + \x) + abs(3/2 + \x) - abs(7/4 + \x) + abs(2 + \x) }) ;
\draw [black,domain=-3:3] plot (\x,{abs(\x)});
\draw[color=orange,fill] (-2, 2) circle (.05) (-3/2, 2) circle (.05) (-5/4, 9/4)
circle (.05) (-1/2, 2) circle (.05) (0, 2) circle (.05) (1/2, 2) circle (.05) (3/4, 9/4)
circle (.05) (5/4, 9/4) circle (.05) (3/2, 5/2) circle (.05) (2, 5/2) circle(.05) (5/2, 5/2) ;
\draw (0, 0) -- (0, -1/10) node [below] {\small $a_\alpha$};
\draw (-2, 0) -- (-2, -1/10) node [below] {\small $x_{\alpha,1}$};
\draw[dashed,help lines] (-2,2) to (-2,0);
\draw (-3/2, 0) -- (-3/2, -1/10) node [below] {\small $x_{\alpha,2}$};
\draw[dashed,help lines] (-3/2,2) to (-3/2,0);
\draw (-5/4, 0) -- (-5/4, -1/10) ;
\draw[dashed,help lines] (-5/4,2) to (-5/4,0);
\draw (-1/2, 0) -- (-1/2, -1/10) ;
\draw[dashed,help lines] (-1/2,2) to (-1/2,0);
\draw (1/2, 0) -- (1/2, -1/10) ;
\draw[dashed,help lines] (1/2,2) to (1/2,0);
\draw (3/4, 0) -- (3/4, -1/10) ;
\draw[dashed,help lines] (3/4,9/4) to (3/4,0);
\draw (5/4, 0) -- (5/4, -1/10) ;
\draw[dashed,help lines] (5/4,9/4) to (5/4,0);
\draw (3/2, 0) -- (3/2, -1/10) ;
\draw[dashed,help lines] (3/2,5/2) to (3/2,0);
\draw (2, 0) -- (2, -1/10)  node [below] {\small $x_{\alpha,\ell_\alpha}$} ;
\draw[dashed,help lines] (2,5/2) to (2,0);
\draw[clip,scale=.355,rotate=45] (0,8) -| (1,7) -| (3,5) -| (4,4) -| (5,3) -| (7,2) -| (9,1) -| (10,0) -| (0,8);
\draw[help lines,scale=.355,rotate=45] (0,0) grid (10,8);
\end{tikzpicture}
\caption{This is the profile function (thick line) for the colored Young diagram of fig. \ref{app_fig_1} with the columns projected
onto $\mathbb{R}$ axis.}
\end{figure}
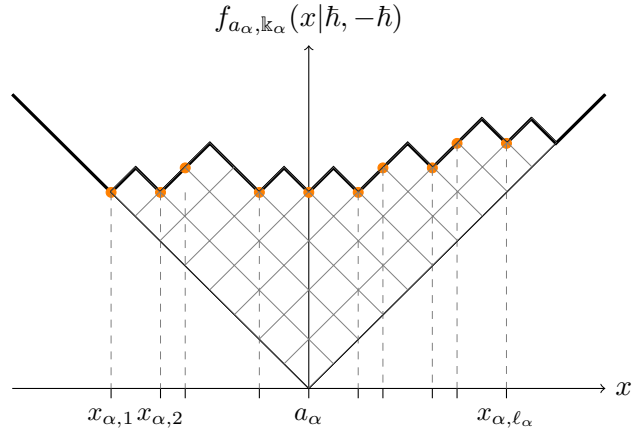
Using the two expressions for the profile functions we obtain the exponent of right hand side of eq. \eqref{app_pert_inst_full_form}
in the sought form, namely
\begin{multline}
\label{Nekrasov_exp_in_profile}
-\frac{\mbox{d}}{\mbox{d} s}\bigg|_{s=0}\frac{\La^s}{\Gamma(s)}\int\limits^\infty_0 \mbox{d} t\ t^{s-1}
\frac{1}{(p^{-1}-1)(q-1 ) }\sum_{\a ,\b=1}^N e_\a e_\b^{-1} \vf_{\Bbbk_\a}(p,q)
\vf_{\Bbbk^T_\b}(q,p)
\\
= -\frac{1}{4}\vpiint\limits_{\mathbb{R}^2}\mbox{d} x\mbox{d} y\ f''_{\mathbf{a},\mathbf{k}}(x|\e_1,\e_2)
\g_{\e_1 ,\e_2}(x-y;\La) f''_{\mathbf{a},\mathbf{k}^T}(y|\e_2,\e_1) .
\end{multline}
The partition function for the $\mathcal{N}=2$ pure gauge theory with $SU(N)$ symmetry takes now the form
\begin{multline}
\label{Nekrasov_profile_rep}
\mathcal{Z}(\La,\mathbf{a},\e_1,\e_2)
= \sum_{f_{\mathbf{a},\mathbf{k}}\in\mathscr{F}}
\exp\pbr{-\frac{1}{4}\vpiint\limits_{\mathbb{R}^2}\mbox{d} x\mbox{d} y\ f''_{\mathbf{a},\mathbf{k}}(x|\e_1,\e_2)
\g_{\e_1 ,\e_2}(x-y;\La) f''_{\mathbf{a},\mathbf{k}^T}(y|\e_2,\e_1)} ,
\end{multline}
Due to the fact that the profile function equals its (parameter) reverse (Young diagram) transpose, the exponent
in the above formula can be cast into yet two equivalent forms with exponents that have schematic form
$(f''_{\mathbf{k}^T})^{\rm i}(\e_2,\e_1)\g_{\rm i j}(\hat\Lambda) (f''_{\mathbf{k}})^{\rm j}(\e_1,\e_2)$
and $(f''_{\mathbf{k}})^{\rm i}(\e_1,\e_2)\g_{\rm i j}(\hat\Lambda) (f''_{\mathbf{k}})^{\rm j}(\e_1,\e_2)$.
The last two correspond to eqs. \eqref{nekrasov_2} and \eqref{nekrasov_3}, respectively multiplied by the perturbative part.
This statement can be phrased yet in another form. Let us define the following quantity
\begin{equation}
\label{def_N}
N_{\a\b}(p,q)\equiv \frac{\vf_{\Bbbk_\a}(p,q)\vf_{\Bbbk^T_\b}(q,p) - 1}{(p^{-1}-1)(q-1 )}.
\end{equation}
Then the left hand side of eq. \eqref{Nekrasov_exp_in_profile} can be put in the form
\begin{multline*}
-\frac{1}{4}\vpiint\limits_{\mathbb{R}^2}\mbox{d} x\mbox{d} y\ f''_{\mathbf{a},\mathbf{k}}(x|\e_1,\e_2)
\g_{\e_1 ,\e_2}(x-y;\La) f''_{\mathbf{a},\mathbf{k}^T}(y|\e_2,\e_1)
\\
= -\sum^N_{\a,\b=1}\g_{\e_1,\e_2}(a_{\a\b};\hat\Lambda)
-\frac{\mbox{d}}{\mbox{d} s}\bigg|_{s=0}\frac{\hat\Lambda^s}{\Gamma(s)}\int\limits^\infty_0 \mbox{d} t\ t^{s-1}
\sum_{\a ,\b=1}^N e_\a e_\b^{-1} N_{\a\b}(p,q).
\end{multline*}
The quantity under the second integral with the use of eq. \eqref{app_sum_over_boxes}
can be brought to the following equivalent forms
\begin{subequations}
\begin{align}
\label{app_N_1}
\sum_{\a ,\b=1}^N e_\a e_\b^{-1} N_{\a\b}(p,q)
=& - \sum_{\a ,\b=1}^N e_\a e_\b^{-1} \sum_{i,j\geq 1}\cbr{p^{\tilde{k}_{\b j}-i+1} q^{k_{\a i}-j} - p^{-i+1} q^{-j}}
\\
\label{app_N_2}
=&  -\sum_{\a ,\b=1}^N e_\a e_\b^{-1} \sum_{i,j\geq 1} \cbr{p^{-\tilde k_{\a j}+i} q^{-k_{\b i} + j -1 }-p^i q^{j-1} }
\\
\label{app_N_3}
=& \sum_{(\a,i)\neq(\b,j)} e_\a e^{-1}_\b \frac{p-1}{q-1}\cbr{q^{k_{\a i}-k_{\b j}} p^{j-i}- p^{j-i}}
\\
\label{app_N_4}
=& - \sum_{\a ,\b=1}^N e_\a e_\b^{-1} \left(\sum_{s\in\Bbbk_\a} p^{L_\b(s)+1}q^{A_\a(s)}
+\sum_{s\in \Bbbk_\b}p^{-L_\a(s)}q^{-A_\b(s)-1}\right) .
\end{align}
\end{subequations}
The ordering of the above equations corresponds to the one in eqs. \eqref{nekrasov_1} -- \eqref{NakajimaYoshioka}.
The definition of $A_\a$ and $L_\a$ can be found below the latter equation.
The ''off-diagonal'' sum in eq. \eqref{app_N_3} is understood as follows
$$
\sum_{(\a,i)\neq(\b,j)} A_{\a,i;\b,j}
= \sum_{\a=1}^N \sum_{i\neq j} A_{\a,i;\a,j} + \sum^N_{\a\neq\b}\sum_{i,j}A_{\a,i;\b,j}.
$$

\subsection{Product identities and \texorpdfstring{$Y$}{Y} functions}

\label{app_ids}

The saddle point equation can be given yet another form. First, let us
note that the following identities hold
\begin{subequations}
\begin{equation}
\label{id_prod1}
\prod_{\a=1}^N\prod_{j = 1}^{L}\frac{x- x^0_{\a,j}+\e_1}{x- x^0_{\a,j}}
= \prod_{\a=1}^N \frac{x- x^0_{\a,1}+\e_1}{x- x^0_{\a,1}-\e_1(L-1)}.
\end{equation}
Shifting the above equation in $x\to x-\e_1$ and multiplying thus obtained result by the eq. \eqref{id_prod1}
leads to the following identity
\begin{equation}
\label{app_empty_diag_ids}
\prod_{\a=1}^N\prod_{j = 1}^{L}\frac{x- x^0_{\a,j}+\e_1}{x- x^0_{\a,j}-\e_1}
\prod_{j=1}^2\frac{\e_1}{x- x^0_{\a,j}+\e_1}
= \prod_{\a=1}^N \prod_{j=1}^2\frac{\e_1}{x- x^0_{\a,j}-\e_1(L-1)}.
\end{equation}
\end{subequations}
As the first step to the iterative solution of eq. \eqref{saddle_point_eq_final} one assumes
that there exists some $L\in\mathbb{N}$ at which a sequence of columns terminates, i.e., $\w_{\a,i}=0$ for $i>L$.
In this case the only change in the saddle point equation \eqref{saddle_point_eq_final} is
\begin{multline*}
\prod_{\b=1}^N\prod_{j \geq 1}\frac{(x_{\a,i}-x_{\b,j}-\e_1)(x_{\a,i}-x^0_{\b,j}+\e_1)}
{(x_{\a,i}-x_{\b,j}+\e_1)(x_{\a,i}-x^0_{\b,j}-\e_1)}
\prod_{j=1}^2\frac{\e_1}{x_{\a,i}-x^0_{\b,j}+\e_1}
\\
=\prod_{\b=1}^N\prod_{j = 1}^{L}\frac{(x_{\a,i}-x_{\b,j}-\e_1)(x_{\a,i}-x^0_{\b,j}+\e_1)}
{(x_{\a,i}-x_{\b,j}+\e_1)(x_{\a,i}-x^0_{\b,j}-\e_1)}
\prod_{j=1}^2\frac{\e_1}{x_{\a,i}-x^0_{\b,j}+\e_1}   .
\end{multline*}
Due to identity in eq. \eqref{app_empty_diag_ids}
it is possible to rewrite the right hand side of the above equation in the form
\begin{displaymath}
\prod_{\b=1}^N\prod_{j = 1}^{L}\frac{x_{\a,i}-x_{\b,j}-\e_1}{x_{\a,i}-x_{\b,j}+\e_1}
\prod_{j=1}^2\frac{\e_1}{x_{\a,i}- x^0_{\b,j}-\e_1(L-1)}
= \e_1^{2N} \prod_{\b=1}^N\prod_{j \geq 1}\frac{x_{\a,i}-x_{\b,j}-\e_1}{x_{\a,i}-x_{\b,j}+\e_1} .
\end{displaymath}
Hence, the saddle point equation takes the infinite product form ($\xi_{\a,i} \equiv x_{\a,i}/\e_1$)
\begin{equation}
\label{app_speq_short}
\hat\Lambda^{2N}\prod_{\b=1}^N\prod_{j \geq 1}\frac{x_{\a,i}-x_{\b,j}-\e_1}{x_{\a,i}-x_{\b,j}+\e_1}
= (\hat\Lambda/\e_1)^{2N}
\prod_{\b=1}^N\prod_{j \geq 1}\frac{\xi_{\a,i}-\xi_{\b,j}-1}{\xi_{\a,i}-\xi_{\b,j}+1}
= (-1)^{N-1},
\end{equation}
It is now possible to release the condition for the finiteness of the number
of columns such that $\{\boldsymbol{\w}_i\}_{i=1}^L\to \{\boldsymbol{\w}_i\}_{i\in \mathbb{N}} $,
which leave the form of the equation intact. As one may notice, this form resembles TBA equation.

Before we proceed, let us rewrite the saddle point equation using the following function
$$
U(z) \equiv \prod_{\b=1}^N\prod_{j \geq 1}\frac{z /\e_1-1 -\xi_{\b,j}}{z/\e_1+1-\xi_{\b,j}},
\qquad
(\hat\Lambda/\e_1)^{2N} U(x_{\a,i}) = (-1)^{N-1}.
$$
Note, that numerator and denominator if taken separately are divergent.
For the study of non-iterative solution of eq. \eqref{app_speq_short}
it is convenient to work with well defined functions. Therefore, following Poghossian \cite{Poghossian:2010pn}
we bring the saddle point equation into yet another form.
Namely, let us cast $U(z)$ function back into the form of eq. \eqref{saddle_point_eq_final}
with the cutoff $L$ for safety reasons, i.e.,
\begin{multline}
\label{app_deriv_Y_step0}
U_L(z) = \frac{\e^{2N}_1}{P(z)P(z+\e_1)}
\prod_{\b=1}^N\prod_{j = 1}^{L}\frac{(z-x_{\b,j}-\e_1)(z-x^0_{\b,j}+\e_1)}
{(z-x_{\b,j}+\e_1)(z-x^0_{\b,j}-\e_1)}
\\
= \frac{\e^{2N}_1}{P(z)P(z+\e_1)}
\prod_{\b=1}^N\prod_{j = 1}^L \frac{\left( 1 - \frac{z - \e_1}{x_{\b,j}} \right) \left( 1 - \frac{z + \e_1}{x^0_{\b,j}} \right)}
{\left( 1 - \frac{z + \e_1}{x_{\b,j}} \right) \left( 1 - \frac{z - \e_1}{x^0_{\b,j}} \right) },
\end{multline}
where $P(z) \equiv \prod (z - a_\a)$.
Each factor of the above product may be transformed as follows
\begin{multline*}
\prod_{\b=1}^N\prod_{j = 1}^L\left( 1 - \frac{z}{x_{\b,j}} \right)\mbox{e}^{z/x^0_{\b,j}}\mbox{e}^{-z/x^0_{\b,j}}
=\prod_{\b=1}^N\mbox{e}^{-(z/\e_1)\sum\limits^L_{j = 1} (\e_1/x^0_{\b,j}) }
\prod_{j = 1}^L\left( 1 - \frac{z}{x_{\b,j}} \right)\mbox{e}^{z/x^0_{\b,j}}
\\
= \mbox{e}^{-\frac{z}{\e_1}N\left(\gamma + \sum\limits^L_{j = 1}\frac{1}{j} \right) }
\prod_{\b=1}^N\mbox{e}^{(z/\e_1)\psi_L(a_\b/\e_1) }
\prod_{j = 1}^L\left( 1 - \frac{z}{x_{\b,j}} \right)\mbox{e}^{z/x^0_{\b,j}},
\end{multline*}
where
$$
\psi_L(z+1) \equiv -\gamma + \sum_{i = 1}^L \left( \frac{1}{i} - \frac{1}{z+i} \right) \xrightarrow{L\to\infty}
\psi(z+1) = \frac{\mbox{d}}{\mbox{d}z}\log\Gamma(z+1) .
$$
Inserting this result in the appropriate form into the eq. \eqref{app_deriv_Y_step0} we get
\begin{displaymath}
U_L(z)=\frac{\e^{2N}_1}{P(z)P(z+\e_1)}\prod_{\b=1}^N\frac{\mbox{e}^{2(z/\e_1)\psi_L(a_\b/\e_1) }}
{\mbox{e}^{2(z/\e_1)\psi_L(a_\b/\e_1) }}
\prod_{j =1}^L
\frac{
\left( 1 - \frac{z - \e_1}{x_{\b,j}} \right) \mbox{e}^{(z-\e_1)/x^0_{\b,j}}
\left( 1 - \frac{z + \e_1}{x^0_{\b,j}} \right) \mbox{e}^{(z+\e_1)/x^0_{\b,j}}
}
{
\left( 1 - \frac{z + \e_1}{x_{\b,j}} \right) \mbox{e}^{(z+\e_1)/x^0_{\b,j}}
\left( 1 - \frac{z - \e_1}{x^0_{\b,j}} \right) \mbox{e}^{(z-\e_1)/x^0_{\b,j}}
}.
\end{displaymath}
Each factor in the numerator and the denominator are now separately well defined.
Within the limit  $L\to\infty$ these functions resemble inverse of the Gamma function
which is known to be entire function. Define the following quantities
\begin{equation}
\label{app_def_Y}
\begin{split}
Y(z) \equiv& \prod_{\b=1}^N\mbox{e}^{-(z/\e_1)\psi( a_\b/\e_1 ) }
\prod_{j \geq 1}\left( 1 - \frac{z}{x_{\b,j}} \right)\mbox{e}^{z/x^0_{\b,j}},
\\
Y_0(z) \equiv& \prod_{\b=1}^N\mbox{e}^{-(z/\e_1)\psi( a_\b/\e_1 ) }
\prod_{j \geq 1}\left( 1 - \frac{z}{x^0_{\b,j}} \right)\mbox{e}^{z/x^0_{\b,j}}.
\end{split}
\end{equation}
These are either entire functions with zeros at points $x_{\b,j},\, x^0_{\b,j}$ respectively.
We can now safely send $L\to\infty$ so that the result reads
\begin{equation}
\label{app_U_in_Y}
U(z) = \frac{\e^{2N}_1}{P(z)P(z+\e_1)}
\frac{Y(z-\e_1) Y_0(z+\e_1)}{Y(z+\e_1)Y_0(z-\e_1) }
= \frac{Y(z - \e_1)}{Y(z+\e_1)} .
\end{equation}
With the aid of these functions the saddle point equation \eqref{app_speq_short}
can now be expressed in yet another enlightening form given in eq. \eqref{saddle_point_eq_infinite}.

\subsection{The result of iterative solution of the saddle point equation for SU(2)}
\label{app_speq_sols}

In this appendix we present results of iterative solution of the saddle point
equation \eqref{saddle_point_eq_final}. This is done up to $L=3$, where $L$
is a cutoff for number of columns within each colored Young diagram.
Quantities obtained this way are coefficients of expansion in $(\hat\Lambda/\epsilon_1)^{2N}$
of colored column in an extreme Young diagram, namely
$$
\omega_{\ast\, \alpha,i} = \sum_{j\geq i} \omega_{\alpha,i,j}\mathfrak{q}^{j} ,
\qquad
\mathfrak{q}\equiv (\hat\Lambda/\epsilon_1)^{2N} .
$$
The sum over colors within a given order of $\mathfrak{q}$ expansion yields coefficients of the twisted superpotential.
The saddle potint equation given in eq. \eqref{saddle_point_eq_final} which is to be soleved iteratively in this appendix may be
rewritten in a more useful form, namely
\begin{multline}
\mathfrak{q} \epsilon_1 ^{2 N} \prod _{\alpha =1}^N \prod _{j=1}^L
\left(x^{(L)}_{\alpha ,i}-x^{0}_{\beta ,j}+\epsilon_1 \right)
\left(x^{(L)}_{\alpha ,i}-x^{(L)}_{\beta ,j}-\epsilon_1 \right)
\\
+\prod_{\alpha =1}^N \prod _{j=1}^2 \left(x^{(L)}_{\alpha ,i}-x^{0}_{\beta
,j}+\epsilon_1 \right) \prod _{j=1}^L \left(x^{(L)}_{\alpha ,i}-x^{0}_{\beta
,j}-\epsilon_1 \right) \left(x^{(L)}_{\alpha ,i}-x^{(L)}_{\beta ,j}+\epsilon_1
\right)=0 .
\end{multline}

\begin{itemize}
\begin{subequations}
\item $N=2, L=1$

\begin{equation}
\label{app_speq_sols_L1}
\omega _{1,1,1} =- \frac{\epsilon_1^3}{2 a (2 a+\epsilon_1 )},
\quad
\omega _{2,1,1} = -\frac{\epsilon _1^3}{2 a \left(2 a-\epsilon _1\right)} .
\end{equation}

\item $N=2, L=2$

\begin{align}
\omega _{1,1,2}=& -\frac{\epsilon _1^5 \left(-8 a^5+4 a^4 \epsilon _1+22 a^3 \epsilon _1^2+3 a^2
   \epsilon _1^3+3 a \epsilon _1^4+\epsilon _1^5\right)}{8 a^3 \left(2 a-\epsilon _1\right){}^2
   \left(a+\epsilon _1\right) \left(2 a+\epsilon _1\right){}^3},
\\
\omega _{1,2,2} =& -\frac{\epsilon _1^5}{8 a \left(a+\epsilon _1\right) \left(2 a+\epsilon_1\right){}^2} ,
\\
\omega _{2,1,2} =& \frac{\epsilon _1^5 \left(8 a^5+4 a^4 \epsilon _1-22 a^3 \epsilon _1^2+3 a^2
   \epsilon _1^3-3 a \epsilon _1^4+\epsilon _1^5\right)}{8 a^3 \left(a-\epsilon _1\right) \left(2
   a-\epsilon _1\right){}^3 \left(2 a+\epsilon _1\right){}^2}
\\
\omega _{2,2,2}=& -\frac{\epsilon _1^5}{8 a \left(a-\epsilon _1\right) \left(2 a-\epsilon
   _1\right){}^2} .
\end{align}

\item $N=2, L=3$

In this case the solutions are for $\alpha = 1$
\begin{align}
\nonumber
\omega _{1,1,3} =& -\frac{\epsilon _1^7 \left(896 a^{12}-1600 a^{11} \epsilon _1-2656 a^{10}
   \epsilon _1^2\right)}{96 a^5 \left(a-\epsilon _1\right){}^2
   \left(\epsilon _1-2 a\right){}^4 \left(a+\epsilon _1\right) \left(2 a+\epsilon _1\right){}^5
   \left(2 a+3 \epsilon _1\right)}
   \\ \nonumber
   & -\frac{\epsilon _1^7 \left(7344 a^9 \epsilon _1^3+3000 a^8 \epsilon _1^4-7788 a^7 \epsilon _1^5+7102 a^6
   \epsilon _1^6\right)}{96 a^5 \left(a-\epsilon _1\right){}^2
   \left(\epsilon _1-2 a\right){}^4 \left(a+\epsilon _1\right) \left(2 a+\epsilon _1\right){}^5
   \left(2 a+3 \epsilon _1\right)}
   \\
   &-\frac{\epsilon _1^7 \left(-2999 a^5 \epsilon _1^7-677 a^4 \epsilon _1^8-144 a^3 \epsilon _1^9-96 a^2 \epsilon_1^{10}
   +30 a \epsilon _1^{11}+18 \epsilon _1^{12}\right)}{96 a^5 \left(a-\epsilon _1\right){}^2
   \left(\epsilon _1-2 a\right){}^4 \left(a+\epsilon _1\right) \left(2 a+\epsilon _1\right){}^5
   \left(2 a+3 \epsilon _1\right)} ,
\\[5pt]
\omega _{1,2,3} =& -\frac{2 \epsilon _1^7 \left(-2 a^3-a^2 \epsilon _1+5 a \epsilon _1^2+\epsilon
   _1^3\right)}{3 a \left(\epsilon _1-2 a\right){}^2 \left(a+\epsilon _1\right) \left(2 a+\epsilon
   _1\right){}^4 \left(2 a+3 \epsilon _1\right)},
\\
\omega _{1,3,3} =& -\frac{\epsilon _1^7}{96 a \left(2 a+3 \epsilon _1\right) \left(2 a^2+3 a
   \epsilon _1+\epsilon _1^2\right){}^2} ,
\end{align}
and for $\a = 2$
\begin{align}
\nonumber
\omega _{2,1,3}=& -\frac{\epsilon _1^7 \left(896 a^{12}+1600 a^{11} \epsilon _1-2656 a^{10}
   \epsilon _1^2\right)}{96 a^5 \left(2 a-\epsilon
   _1\right){}^5 \left(a+\epsilon _1\right){}^2 \left(2 a+\epsilon _1\right){}^4 \left(2 a^2-5 a
   \epsilon _1+3 \epsilon _1^2\right)}
   \\ \nonumber
   &-\frac{\epsilon _1^7 \left(-7344 a^9 \epsilon _1^3+3000 a^8 \epsilon _1^4+7788 a^7 \epsilon _1^5+7102 a^6
   \epsilon _1^6\right)}{96 a^5 \left(2 a-\epsilon
   _1\right){}^5 \left(a+\epsilon _1\right){}^2 \left(2 a+\epsilon _1\right){}^4 \left(2 a^2-5 a
   \epsilon _1+3 \epsilon _1^2\right)}
   \\
   & -\frac{\epsilon _1^7 \left(2999 a^5 \epsilon _1^7-677 a^4 \epsilon _1^8+144 a^3 \epsilon _1^9-96 a^2 \epsilon
   _1^{10}-30 a \epsilon _1^{11}+18 \epsilon _1^{12}\right)}{96 a^5 \left(2 a-\epsilon
   _1\right){}^5 \left(a+\epsilon _1\right){}^2 \left(2 a+\epsilon _1\right){}^4 \left(2 a^2-5 a
   \epsilon _1+3 \epsilon _1^2\right)} ,
\\[10pt]
\omega _{2,2,3} =& \frac{2 \epsilon _1^7 \left(2 a^3-a^2 \epsilon _1-5 a \epsilon _1^2+\epsilon
   _1^3\right)}{3 a \left(\epsilon _1-2 a\right){}^4 \left(2 a+\epsilon _1\right){}^2 \left(2 a^2-5
   a \epsilon _1+3 \epsilon _1^2\right)} ,
\\
\label{app_speq_sols_L3}
\omega _{2,3,3} =& -\frac{\epsilon _1^7}{96 a \left(2 a-3 \epsilon _1\right) \left(2 a^2-3 a
   \epsilon _1+\epsilon _1^2\right){}^2} .
\end{align}
\end{subequations}
\end{itemize}

\providecommand{\href}[2]{#2}
\begingroup\raggedright

\endgroup

\end{document}